\documentclass[useAMS,usenatbib]{mn2e}

\usepackage{epsfig,color,pifont,array}
\usepackage[colorlinks=true,allcolors=blue]{hyperref}

\DeclareMathAlphabet{\mathpzc}{OT1}{pzc}{m}{it}
\usepackage{bm}
\usepackage{mathabx}
\usepackage{multirow}
\usepackage{breqn}

\usepackage{natbib}
\bibpunct{(}{)}{;}{a}{}{,}
\usepackage{etoolbox}

\makeatletter

\patchcmd{\NAT@citex}
  {\@citea\NAT@hyper@{%
     \NAT@nmfmt{\NAT@nm}%
     \hyper@natlinkbreak{\NAT@aysep\NAT@spacechar}{\@citeb\@extra@b@citeb}%
     \NAT@date}}
  {\@citea\NAT@nmfmt{\NAT@nm}%
   \NAT@aysep\NAT@spacechar\NAT@hyper@{\NAT@date}}{}{}

\patchcmd{\NAT@citex}
  {\@citea\NAT@hyper@{%
     \NAT@nmfmt{\NAT@nm}%
     \hyper@natlinkbreak{\NAT@spacechar\NAT@@open\if*#1*\else#1\NAT@spacechar\fi}%
       {\@citeb\@extra@b@citeb}%
     \NAT@date}}
  {\@citea\NAT@nmfmt{\NAT@nm}%
   \NAT@spacechar\NAT@@open\if*#1*\else#1\NAT@spacechar\fi\NAT@hyper@{\NAT@date}}
  {}{}

\makeatother

\newcommand\Tstrut{\rule{0pt}{2.6ex}}         % = `top' strut
\newcommand\Bstrut{\rule[-1.5ex]{0pt}{2.6ex}}   % = `bottom' strut
\newcommand\Dstrut{\rule[-3.0ex]{0pt}{2.6ex}}   % = double `bottom' strut

\def\spose#1{\hbox to 0pt{#1\hss}}
\def\ltsimm{\mathrel{\spose{\lower 3pt\hbox{$\sim$}}
        \raise 2.0pt\hbox{$<$}}}
\def\gtsimm{\mathrel{\spose{\lower 3pt\hbox{$\sim$}}
        \raise 2.0pt\hbox{$>$}}}

\def\cm{{\rm\thinspace cm}}

\def\s{{\rm\thinspace s}}

\def\g{{\rm\thinspace g}}

\def\erg{{\rm\thinspace erg}}
\def\Hz{{\rm\thinspace Hz}}

\def\ster{{\rm\thinspace ster}}
\def\ergps{\hbox{${\rm\erg\s^{-1}\,}$}}

\def\pcm{\hbox{${\rm\cm^{-1}\,}$}}
\def\pcm2{\hbox{${\rm\cm^{-2}\,}$}}
\def\pcm3{\hbox{${\rm\~{\rm cm^{-3}}\,}$}}
\def\ergpscm3Hz{\hbox{${\rm\ergps\~{\rm cm^{-3}}\Hz^{-1}\,}$}}
\def\ergpscm3Hzster{\hbox{${\rm\ergps\~{\rm cm^{-3}}\Hz^{-1}\ster^{-1}\,}$}}
\def\gpcm3{\hbox{${\rm\g\~{\rm cm^{-3}}\,}$}}
\def\ergpcm2{\hbox{${\rm\erg\cm^{-2}\,}$}}
\def\ergpcm3{\hbox{${\rm\erg\~{\rm cm^{-3}}\,}$}}
\def\phpscm2{\hbox{${\rm photons\s^{-1}\cm^{-2}\,}$}}

\title [Filaments in wind-cloud interactions (I)]{Filament formation in wind-cloud interactions. I. Spherical clouds in uniform magnetic fields}

\author[W.~E.~Banda-Barrag\'{a}n, E.~R.~Parkin, C.~Federrath, R.~M.~Crocker, G.~V.~Bicknell]
{W.~E.~Banda-Barrag\'{a}n\thanks{E-mail: wlady.bsc@gmail.com}, E.~R.~Parkin, C.~Federrath, R.~M.~Crocker, and G.~V.~Bicknell
  \\Research School of Astronomy and Astrophysics,
  Australian National University, Canberra, ACT 2611, Australia}

\begin{document}
%\begin{onecolumn}

\date{Accepted ... Received ...; in original form ...}

\pagerange{\pageref{firstpage}--\pageref{lastpage}} \pubyear{2015}

\maketitle

\label{firstpage}

\begin{abstract}
{Filamentary structures are ubiquitous in the interstellar medium, yet their formation, internal structure, and longevity have not been studied in detail. We report the results from a comprehensive numerical study that investigates the characteristics, formation, and evolution of filaments arising from magnetohydrodynamic interactions between supersonic winds and dense clouds. Here we improve on previous simulations by utilising sharper density contrasts and higher numerical resolutions. By following multiple density tracers, we find that material in the envelopes of the clouds is removed and deposited downstream to form filamentary tails, while the cores of the clouds serve as footpoints and late-stage outer layers of these tails. Aspect ratios $\gtrsim12$, subsonic velocity dispersions $\sim0.1-0.3$ of the wind sound speed, and magnetic field amplifications $\sim100$ are found to be characteristic of these filaments. We also report the effects of different magnetic field strengths and orientations. The magnetic field strength regulates vorticity production: sinuous filamentary towers arise in non-magnetic environments, while strong magnetic fields inhibit small-scale perturbations at boundary layers making tails less turbulent. Magnetic field components aligned with the direction of the flow favour the formation of pressure-confined flux ropes inside the tails, whilst transverse components tend to form current sheets. Softening the equation of state to nearly isothermal leads to suppression of dynamical instabilities and further collimation of the tail. Towards the final stages of the evolution, we find that small cloudlets and distorted filaments survive the break-up of the clouds and become entrained in the winds, reaching velocities $\sim0.1$ of the wind speed.}
\end{abstract}


%\maketitle-

\section{Introduction}
\label{sec:Introduction}
Magnetic structures with elongated, tail-shaped morphologies are ubiquitous in the Universe. They can be the result of a variety of dynamic processes occurring in both the interstellar medium (ISM) and the intergalactic medium (IGM)\footnote{Outside the Milky Way boundaries, it is also possible to observe (magneto)tails emerging when small galaxies move through the IGM in gravitationally-bound galactic aggregations (see \citealt{Recchi2007} for wind-clump simulations of dwarf galaxies, and \citealt{2008ApJ...677..993D} and \citealt{2010NatPh...6..520P} for some recent numerical studies on cluster magnetic fields).}. Shock waves, gravitational forces, turbulence, and magnetically-driven events can together or separately be involved in the formation of filamentary structures (see \citealt*{1967SoPh....1..254B,1979ApJS...41...87S,1981ASSL...82.....A,1986Sci...231..907A,1996ApJ...470L..49R,2000ApJ...540..797W,2001PASA...18..431B}; \citealt{2008ApJ...684.1384R,2011ApJ...731...13N,2012ApJ...746L..21W}; \citealt*{2012ApJ...751...69S}; \citealt{2013MNRAS.436.3707L,2014ApJ...780...72E,2014ApJS..213....8T}, and references therein for discussions on cosmic filaments formed in different environments). In this and subsequent papers in this series, however, we focus our analysis exclusively on those filaments that arise from wind-clump interactions. Structures of this kind are found at all scales in the ISM and range from the relatively small cometary tails found in the Solar System (e.g., \citealt{2000Icar..148...52B,2008ApJ...677..798B}), through the complex optical, \textit{X}-ray, and infrared filamentary shells observed in supernova remnants (e.g., \citealt{1996ApJ...456..225H,2009ApJ...693.1883S,2009ApJ...697..535P,2010Ap&SS.330..123D,2011Ap&SS.331..521V,2013ApJ...774..120M}), to the large-scale $\rm H$$\,\alpha$- and $\rm H\,${\scriptsize{I}}-emitting filaments detected in some galaxies (e.g., \citealt{1998ApJ...493..129S}; \citealt*{1999ApJ...523..575L}; \citealt*{2002ApJ...576..745C}; \citealt{2005MNRAS.363..216C}). Despite having different sizes and being observed at various wavelengths, all of these structures are believed to share a common origin, namely the interaction of fast-moving, low-density winds with ISM inhomogeneities (i.e., clumps). The radio threads observed in the Galactic Centre of the Milky Way (\citealt*{1984Natur.310..557Y}; \citealt{1985AJ.....90.2511M,2000AJ....119..207L}; \citealt*{2004ApJS..155..421Y}) may also be exemplars of such filaments. Both thermal and non-thermal emissions can be expected from these interactions as both are connected with the emergence of shock waves and instabilities in cosmic plasmas (see \citealt{1993ARA&A..31..373D}; \citealt*{1994ApJ...432..194J}; \citealt{1994ApJ...433..757M,2012SSRv..173..369H} for discussions on emission processes involving winds and clumps).\par

Winds are known to play a major role in shaping the ISM, altering its dynamics, and changing its physical and chemical properties (see \citealt{1995ApJ...439..365S,1995ApJ...439..381S} for studies on supernova remnants, or \citealt{2000MNRAS.314..511S,2001ApJ...556..121K} for models of starburst galaxies). Winds expand and interact with clumps leaving behind imprints of their passage in the form of shocked gas (e.g., \citealt{1985ApJ...298..316W,2001ApJ...552..175K}), excited atomic or molecular species (e.g., \citealt{2002Sci...296.2350W,2007ApJ...664..890N}), and topologically-altered magnetic fields (e.g., \citealt{2001ApJ...548L..69B,2009AdSpR..44..433S,2012SSRv..166..231R}). Typical wind sources in the ISM include isolated stars, star clusters, star-forming regions, and explosive events associated with dying stars (e.g., supernovae, afterglows of gamma-ray bursts, among others). As the wind moves away from its source, it encounters a plurality of differently-sized clumps in the surrounding inhomogeneous environments. These clumps could be collections of solid bodies, conglomerates of stars, or entire regions permeated with gas and dust clouds. Both the wind and the surrounding inhomogeneities undergo dramatic physical and chemical changes when they interact. For example, solid objects sublimate when immersed in stellar winds (e.g., \citealt{2014ApJ...794...14M}), stars lose mass and magnetic energy from their outer atmospheres to a prevailing external wind (e.g., \citealt{2003ApJ...598..325Y,2013ApJ...776...13B}), and atomic and dense clouds are disrupted by the ram pressure exerted by outflowing material (e.g., \citealt*{1986Sci...232..185B,1996ApJ...473..365J}). Another effect that has been seen in simulations of wind-swept clouds is that they can be accelerated by the net force resulting from the excess pressure on the upstream side of the cloud pushing dense material downstream (e.g., \citealt*{1978ApJ...219L..23M}; \citealt{2000ApJ...543..775G,2005MNRAS.362..626M}).\par

On the other hand, the wind itself can also be altered during these interactions and it often evolves from purely adiabatic to radiative expansion phases (e.g., \citealt*{1975ApJ...200L.107C}; \citealt{1978ApJ...220..742W,2008ARA&A..46...89R}). The transition into a highly-efficient cooling regime occurs when lateral and reverse shocks inject additional kinetic energy into the wind and excite atomic and molecular species as a result (e.g., \citealt{1997ApJ...475..194K,1997ApJ...485..263K}). Some significant effects observed in winds when they encounter clouds in their trajectories also include ageing, deceleration, and (de)magnetisation (see \citealt*{1994ApJ...420..213K}; \citealt{2014MNRAS.444..971A}). The importance of studying wind-cloud systems in the context of this work, then, lies in three main points: a) the wind-swept clouds may be distorted into tail-shaped structures (i.e., filaments) by disruptive processes and magnetohydrodynamic instabilities; b) winds encountering clumpy regions in the ISM can trigger shocks that may produce detectable thermal and non-thermal emission in these filaments; and c) advective and compressive processes combined with turbulence can radically change the topology and strength of magnetic fields and ultimately lead to the appearance of magnetohydrodynamic waves and the occurrence of highly energetic processes, such as magnetic reconnection (see e.g., \citealt{1996ApJ...473..365J}, \citealt*{1999ApJ...517..242M}; \citealt{1999ApJ...517..700L,2015ASSL..407..311L}). In this and subsequent papers we systematically investigate these three aspects in detail with the help of MHD numerical models of wind-cloud systems in which the supersonic motion of a hot wind produces filaments as it interacts with clouds of either uniform or fractal geometry.\par

The aim of this first paper is to provide insights into the processes that lead to the formation of filaments in a non-turbulent environment, i.e., a medium with uniform magnetic fields permeated by spherical, pressure-bound clouds. We shall extend the analysis to fractal clouds and turbulent environments in a subsequent paper (hereafter Paper II). In addition, we advance the overall understanding of the magnetohydrodynamics involved in the interaction of a wind-cloud system by performing three-dimensional simulations using a set of previously-unexplored, more realistic initial conditions. The remainder of this paper is organised as follows. In Section \ref{sec:ProblemDefinition} we review the current literature and contextualise the present work. In Section \ref{sec:Method} we include a description of the numerical methods, initial and boundary conditions, time-scales, and diagnostics that we employ for our study. In Sections \ref{sec:CloudDisruption}, \ref{sec:FilamentFormation}, and \ref{sec:Entrainment} we present our results. In Section \ref{sec:CloudDisruption} we describe the processes that lead to the disruption of clouds immersed in winds. In Section \ref{sec:FilamentFormation} we include an overall description of filament formation and comparisons between different initial configurations (e.g., non-magnetised and magnetised environments, adiabatic and quasi-isothermal equations of state, and different strengths and orientations of the magnetic field). We utilise 2D slices and 3D volume renderings to illustrate the structure, acceleration, and survival of filaments against dynamical instabilities, as well as the evolution, in the magnetotails, of the magnetic energy and the plasma $\beta$ (the ratio of thermal pressure to magnetic energy density). In Section \ref{sec:Entrainment} we analyse entrainment processes of clouds and filaments in winds. In Section \ref{sec:FutureWork} we discuss the limitations of our study and the work to be pursued in the future. In Section \ref{sec:Summary} we summarise our findings and conclusions.

\section{Problem Definition}
\label{sec:ProblemDefinition}
A wind-cloud system constitutes an idealised scenario in which an initially static, isolated cloud or a collection of clouds interact with a wind represented by a velocity field contained within a finite volume (an alternative approach is to consider that the wind is actually static and the cloud is a bullet moving through it with a certain velocity, e.g., ballistically). Because of the intrinsically non-linear character of the equations describing the evolution of wind-cloud interactions, these systems can only be studied analytically in simplified cases (see the pioneering work by \citealt{1975ApJ...195..715M}), and, in general, they need to be studied with numerical simulations. In purely hydrodynamic studies, where source terms are neglected, the wind-cloud system is often characterised by three numbers:\par

\noindent 1) The adiabatic index of the gas 

\begin{equation}
\gamma=\frac{c_p}{c_v},
\label{eq:PolytropicIndex}
\end{equation}

\noindent where $c_p$ and $c_v$ are the specific heat capacities at constant pressure and volume, respectively.\par

\noindent 2) The Mach number of the wind

\begin{equation}
{\cal M_{\rm w}}=\frac{|\bm{v_{\rm w}}|}{c_{\rm w}},
\label{eq:MachNumber}
\end{equation}

\noindent where $|{\bm{v_{\rm w}}}|\equiv v_{\rm w}$ and $c_{\rm w}=\sqrt{\gamma \frac{P_{\rm th}}{\rho_{\rm w}}}$ are the speed and adiabatic sound speed of the wind, respectively.\par

\noindent 3) The density contrast

\begin{equation}
\chi=\frac{\rho_{\rm c}}{\rho_{\rm w}}
\label{eq:DensityContrast}
\end{equation}

\noindent between the cloud, $\rho_{\rm c}$, and wind material, $\rho_{\rm w}$ (\citealt{1996ApJ...473..365J}). Note that in Equations (\ref{eq:MachNumber}) and (\ref{eq:DensityContrast}): a) we utilise normalised quantities in code units, and b) we assume an ideal single-fluid approximation characterised by a constant polytropic index, $\gamma$, and a uniform mean molecular weight, $\bar{\mu}$. The thermal pressure, $P_{\rm th}$, can be obtained from the gas temperature using thermodynamic relations. If the Mach number of the wind is much higher than unity, \cite{1994ApJ...420..213K} and \cite{Nakamura:2006bc} demonstrated that Mach scaling is applicable, so that the evolution solely depends upon the density contrast. This parametrisation indicates that adiabatic simulations are scale-free and therefore independent of any absolute dimensions or primitive inputs.\par

When additional source terms, e.g., cooling or heating, are added to the basic hydrodynamic model, the scaling of the simulations is restricted to a one-parameter scaling (see the discussion on scaling in section 3.2 of \citealt{2007ApJS..173...37S}). In such cases, however, simulations are specifically designed for a pre-defined problem, and results are generally not transferable to other situations. In numerical simulations where magnetic fields are incorporated, a fourth parameter is included:\par

\noindent 4) The so-called plasma beta\footnote{Note that the factor $\frac{1}{\sqrt{4\pi}}$ has been subsumed into the definition of magnetic field. The same normalisation applies henceforth.}

\begin{equation}
\beta=\frac{P_{\rm th}}{P_{\rm mag}}=\frac{P_{\rm th}}{\frac{1}{2}|\bm{B}|^2},
\label{eq:Beta}
\end{equation}

\noindent a dimensionless number that relates the thermal pressure, $P_{\rm th}$, to the magnetic pressure, $P_{\rm mag}=\frac{1}{2}|\bf{B}|^2$, in a medium, needs to be specified for the system. Sometimes, the Alfv\'{e}nic Mach number, 

\begin{equation}
{\cal M_A}=\frac{v_{\rm w}}{v_A}=\frac{v_{\rm w}}{\frac{|\bm{B}|}{\sqrt{\rho_{\rm w}}}},
\label{eq:AlfvenMach}
\end{equation}

\noindent is used instead. Here, $v_A$ is the Alfv\'{e}n speed in the wind. Additionally, if the magnetic field is uniformly distributed in the simulation domain, such as in the models presented here, a set of additional parameters describing the topology of the field needs to be added as an input to the set-ups (e.g., in three-dimensional models two direction angles are reported). If more complex magnetic fields are implemented, alternative quantities, such as the maximum field strength, the average plasma beta, or parameters associated with magnetic turbulent cascades are often used as problem descriptors (see Paper II).\par

\subsection{Previous Work}
\label{subsec:PreviousWork}
Over the last two decades, a considerable amount of two-dimensional (planar 2D or axisymmetric 2.5D) and three-dimensional (3D) hydrodynamic (HD) and magnetohydrodynamic (MHD) simulations of wind-cloud interactions have been performed. \cite{1993ApJ...407..588M} studied how dynamic instabilities affect pressure-bound and gravitationally-bound clouds as they move subsonically through a background gas in a two-phase medium. \cite{1994ApJ...432..194J} employed a two-fluid numerical model to identify particle acceleration sites in cosmic bullets and showed that they can be radio sources. \cite*{1995ApJ...439..237S} studied wind-accelerated, radiating clouds and found that radiation losses enhance the ablation of small-scale perturbations and prolong cloud lifetimes. Later, \citealt{1996ApJ...473..365J} modelled cylindrical clouds threaded by aligned and transverse magnetic fields with different Alfv\'{e}nic Mach numbers and found that stretching, folding, and compression of field lines are the dominant effects for field amplification. \cite{1999ApJ...517..242M} studied the exchange of kinetic and magnetic energy in two-dimensional wind-cloud interactions and showed that the magnetic pressure at the leading edge of the cloud can exceed the wind ram pressure and become dynamically important in clouds with high density contrasts with respect to the wind.\par

\begin{table*}\centering
\caption{Comparison between the parameter space explored by previous authors and our present work. Column 1 contains the references. Column 2 provides the number of dimensions considered in their simulations and whether the models reported are purely hydrodynamic (HD), magnetohydrodynamic (MHD), or both (M/HD). Column 3 indicates the type of geometry employed to describe clouds, i.e., spherical (Sph), cylindrical (Cyl) or fractal (Fra). Column 4 indicates the resolutions ($R_{\rm x}$) used for the simulations in terms of the number of cells ($x$) per cloud radius. Columns 5, 6, 7, and 8 summarise the polytropic indices ($\gamma$), density contrasts ($\chi$), Mach numbers (\textbf{$\cal M_{\rm w}$}), and initial plasma betas ($\beta$) reported in the references. Finally, column 9 indicates the topological structure of the field (when relevant), which could be tangled (Ta), turbulent (Tu), or aligned (Al), transverse (Tr), and oblique (Ob) with respect to the direction of the wind velocity.}
\begin{tabular}{*9c}
\hline
\textbf{(1)} & \textbf{(2)} & \textbf{(3)} & \textbf{(4)} & \textbf{(5)} & \textbf{(6)} & \textbf{(7)} & \textbf{(8)} & \textbf{(9)}\Tstrut\\
\textbf{Reference} & \textbf{Type} & \textbf{Cloud} & \textbf{Resolution} & \textbf{$\gamma$} & \textbf{$\chi$} & \textbf{$\cal M_{\rm w}$} & \textbf{$\beta$} & \textbf{Topology}\Bstrut \\ \hline 
\cite{1993ApJ...407..588M}  & 2D HD & Cyl & $R_{25}$ & $1.67$ & $500$, $10^3$ & $0.25$ - $1$ & $\infty$ & --\Tstrut \\ 
\cite{1994ApJ...432..194J}  & 2D HD  & Cyl & $R_{43}$ & $1.67$ & $30$, $100$ & $3$, $10$ & $\infty$ & --\\
\cite{1995ApJ...439..237S}  & 2D HD  & Cyl/Sph & $R_{128}$ - $R_{270}$ & $1.67$ & $10$ - $2000$ & $10$ & $\infty$ & --\\
\cite{1996ApJ...473..365J}  & 2D M/HD & Cyl & $R_{50}$, $R_{100}$ & $1.67$ & $10$, $40$, $100$ & $10$ & $1$ - $256$, $\infty$ & Al, Tr\\ 
\cite{1999ApJ...517..242M} & 2D MHD & Cyl & $R_{26}$ & $1.67$ & $10$, $100$ & $1.5$,$10$ & $4$ & Ob  \\ 
\cite{1999ApJ...527L.113G} & 3D M/HD & Sph & $R_{26}$ & $1.67$ & $100$ & $1.5$ & $4$, $100$, $\infty$ & Tr  \\ 
\cite{2000ApJ...543..775G}  & 3D MHD & Sph & $R_{26}$ & $1.67$ & $100$ & $1.5$ & $4$, $100$ & Tr  \\
\cite{2005RMxAA..41...45R}  & 3D HD & Sph & $R_{25}$ & $1.0$ & $50$ & $2.6$ & $\infty$ & --  \\
\cite{2005MNRAS.361.1077P} & 2D HD & Cyl & $R_{<32}$ & $1.0$ & $\leq 350$ & $1$, $20$ & $\infty$ &  --  \\
\cite{2009ApJ...703..330C}& 3D HD & Fra & $R_{6}$ - $R_{38}$ & $1.67$ & $630$ - $1260$ & $4.6$ & $\infty$ &  -- \\
\cite{2014arXiv1409.6719M} & 3D MHD & Sph & $R_{32}$ & $1.67$ & $50$ & $1.5$ & $0.1$ - $10$ &  Ta \\
Paper I & 3D M/HD& Sph & $R_{128}$ & $1.67$, $1.1$ & $10^3$ & $4$, $4.9$ & $10$, $100$, $\infty$ & Al, Tr, Ob \\
Paper II & 3D MHD & Fra & $R_{128}$ & $1.67$ & $10^3$ & $4$ & $25$ - $100$ & Tu\Bstrut \\
\hline
\end{tabular}
\label{Table1}
\end{table*}

Additionally, \cite{1999ApJ...527L.113G,2000ApJ...543..775G} explored 3D scenarios of a single cloud immersed in a magnetised wind in which the field was oriented perpendicular to the wind velocity. They showed that the growth of dynamical instabilities at the leading edge of the cloud is increased, owing to an enhanced magnetic pressure caused by the effective trapping of field lines in surface deformations. \cite*{2005RMxAA..41...45R} explored the effects of photoionisation on wind-swept clouds and reported that strong ionising fields can radically reduce the fragmentation of clouds by creating an interposing layer of photo-evaporated gas around them. Later, \cite{2005MNRAS.361.1077P} analysed the case of non-magnetised winds interacting with multiple embedded sources of mass injection in 2D and concluded that a collection of such clouds can act as a barrier for the wind if the mass injection rate in them is higher than the wind mass flux. \cite{2008ApJ...674..157C,2009ApJ...703..330C} studied the 3D HD interaction of star-formation-driven winds with spherical and fractal clouds. They found that long filamentary tails can result from such interactions and survive acceleration aided by their ability to radiate. More recently, \cite{2014arXiv1409.6719M} performed another set of 3D simulations adding a constant magnetic field to the wind and a tangled, force-free magnetic field to the cloud. Their results suggested that an internal tangled field can suppress the disruption of the cloud and lead to fragments co-moving with their surroundings. A list of publications related to wind-cloud interactions is provided in Table \ref{Table1}.\par

Additional publications that are relevant to the study of filaments arising from wind-cloud systems include those related to the study of shock-cloud interactions in which a shock, injected from one side of the simulation volume, impacts a cloud (or clouds) immersed in a pre-shocked medium initially at rest. The wind-cloud problem may actually be seen as a particular case of the shock-cloud problem in which the clumpy gas interacts with the flow behind a blast wave shock rather than with the shock itself (see Section $9$ in \citealt{1994ApJ...420..213K} for a thorough discussion). Wind-cloud systems, however, may also be found in other scenarios in which an initial shock-driven crush is not necessarily involved, such as clouds forming and falling through thermally-unstable outflows and gaseous disks immersed in accelerating winds (see Section $5$ of \citealt{1995ApJ...439..237S} for further details). Despite foreseeable differences in the time-scales involved in the evolution of wind-cloud and shock-cloud systems, the main aspects of the physics entailed in the cloud disruption and gas entrainment in both problems are similar. Thus, a brief summary of the literature on shock-cloud interactions is warranted. We will compare the contributions of each author with our conclusions in Section \ref{sec:CloudDisruption}, so that in this section we limit ourselves to solely providing details of their configurations and highlights of their work. \par

Early semi-analytical studies of shock-cloud interactions include the works by \cite{1975ApJ...195...53C,1976ApJ...207..484W,1982MNRAS.201..833N,1983MNRAS.203...67H}; and \cite{1985ApJ...291..523H}. Later, the advent of novel computational algorithms and advanced tools allowed more sophisticated numerical models. \cite{1992ApJ...390L..17S,1994ApJ...420..213K}; and \cite{1995ApJ...454..172X}, for example, described the adiabatic evolution, in 2D and 3D, of an interstellar cloud being impacted by a planar shock. Different cloud geometries and orientations were tested, but only non-radiative clouds with uniform density profiles were considered in these studies. In particular, \cite{1994ApJ...420..213K} showed that convergence in adiabatic HD simulations is achieved at resolutions of $120$ cells per cloud radius. Later, \cite{Nakamura:2006bc} introduced a mathematical function to prescribe smoothed density profiles in the clouds. Other HD simulations reported in the literature include studies of the propagation of a shock wave in the presence of multiple clouds (see \citealt*{2002ApJ...576..832P,2005AA...443..495M}; \citealt{2012MNRAS.425.2212A}). Although less frequent than HD models, MHD simulations have also been reported in the past. \cite{1994ApJ...433..757M} introduced the first adiabatic, axisymmetric shock-cloud simulations including magnetic fields. Later, \cite{2005ApJ...619..327F,2006AA...457..545O,2008ApJ...680..336S} studied the dynamic evolution of shocked clouds inserted in uniform fields in 2D, 2.5D, and 3D simulations, respectively.\par

Simulations in 2D and 2.5D have historically been used as simplifications of otherwise computationally-expensive 3D models. Nonetheless, 2D models constrain the cloud geometries and magnetic field topologies that can be employed, and this reduces the number of scenarios that can be tested computationally. For instance, turbulent flows can only be studied in 3D. More recently, \cite*{2013HEDP....9..132L,2013ApJ...774..133L} simulated 3D cases where the magnetic field was self-contained within the clouds, and \cite{2014MNRAS.444..971A} reported 2D adiabatic simulations of a shock interacting with multiple magnetised clouds. Several shock-cloud and wind-cloud simulations reported in the literature have incorporated source terms into their mathematical description of the systems. The effects of optically-thin radiative cooling (see \citealt*{2002AA...395L..13M}; \citealt{2004AA...424..817M,2004ApJ...604...74F,2005ApJ...619..327F,2005AA...443..495M,2010ApJ...722..412Y,2013ApJ...774..133L,2013ApJ...766...45J}), thermal conduction \citep{2005MNRAS.362..626M,2006AA...457..545O,2008ApJ...678..274O,2013MNRAS.430.2864M}, photo-evaporation (\citealt{2005AA...443..495M,2006ApJ...643..186T}), and self-gravity \citep{2004ApJ...604...74F} have been considered in the past. The turbulent destruction in 2D shock-cloud interactions has also been studied by \cite{2009MNRAS.394.1351P}; \cite*{2010MNRAS.405..821P}; and \cite{2011Ap&SS.336..239P} using a non-Eulerian, sub-grid compressible turbulence model.\par

\subsection{Current Work}
\label{subsec:CurrentWork}
Notwithstanding the significant progress made towards the understanding of the processes leading to the disruption of clouds by shocks and winds in the ISM, the detailed mechanisms that lead to the formation of filamentary tails from these interactions have not been analysed thoroughly. For instance, how is the cloud disruption process associated with the formation of filaments? How do filaments evolve in time in non-magnetised and magnetised cases? How long can (magneto)tails survive against the wind ram pressure and plasma instabilities in the ISM? What is the internal structure of these filaments?\par

Filamentary tails have been observed in some previous simulations, but not every wind-cloud interaction has been capable of producing long-lived structures. Thus, which initial conditions really favour the formation of (magneto)tails? How does the initial magnetic field topology affect the evolution of wind-cloud interactions and the resulting tail morphology? What is the fate of the dense gas originally in the cloud cores and of their associated filamentary tails? Could high-density cores provide the required footpoints for tails to form?\par

Large differences in the Courant time of wind and cloud material can make scenarios with very high density contrasts computationally expensive, so that most previous studies considered cases in which the density contrasts between ambient and cloud material ranged from $10-100$. Realistically, however, clouds can be $10^3-10^6$ times denser than low-density winds in the ISM. Higher density contrasts can be influential in the development of fluid instabilities and disruption time-scales. Our study is the first to thoroughly probe this physically relevant parameter range at high resolution.

\section{Method}
\label{sec:Method}

\subsection{Simulation code}
\label{subsec:SimulationCode}
To study the filamentary structures arising from wind-cloud interactions and provide answers to the questions discussed in Section \ref{subsec:CurrentWork}, we solve the equations of ideal magnetohydrodynamics using the {\sevensize PLUTOv4.0} code \citep{Mignone:2007,2012ApJS..198....7M} in a 3D cartesian coordinate system $(X_1,X_2,X_3)$. The relevant system of equations for mass, momentum, energy conservation, and magnetic induction are:

\begin{figure}
\begin{center}
  \begin{tabular}{c}
     \resizebox{70mm}{!}{\includegraphics{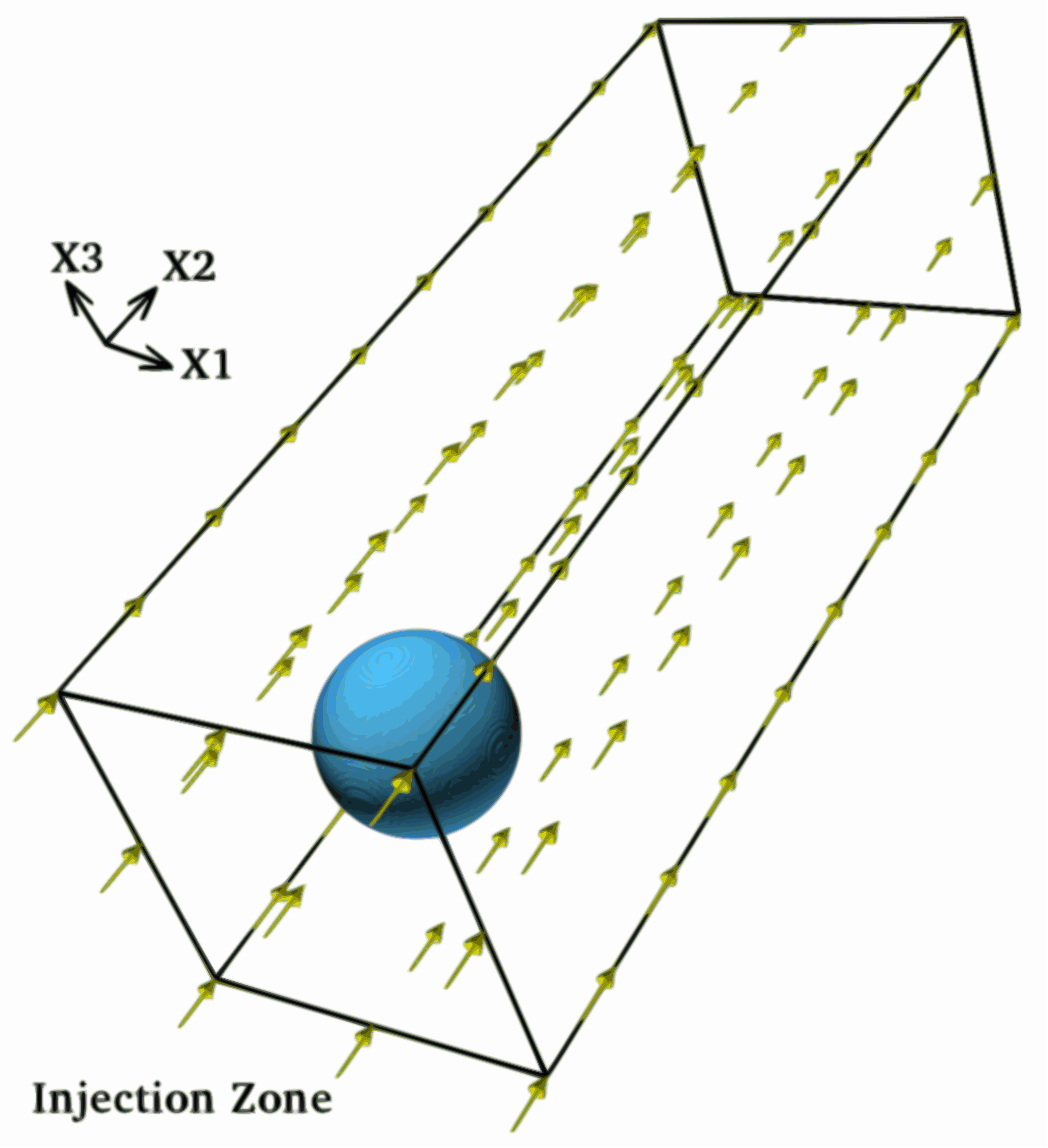}}\\
  \end{tabular}
  \caption{Simulation set-up for spherical clouds with smoothed density profiles. The wind velocity field is represented by arrows and the cloud density ($\rho C_{\rm cloud}$) is shown is represented by the sphere.}
  \label{Figure1}
\end{center}
\end{figure}

\begin{equation}
\frac{\partial \rho}{\partial t}+\bm{\nabla\cdot}\left[{\rho \bm{v}}\right]=0,
\end{equation}

\begin{equation}
\frac{\partial \left[\rho \bm{v}\right]}{\partial t}+\bm{\nabla\cdot}\left[{\rho\bm{v}\bm{v}}-{\bm{B}\bm{B}}+{\bm{I}}P\right]=0,
\end{equation}

\begin{equation}
\frac{\partial E}{\partial t}+\bm{\nabla\cdot}\left[\left(E+P\right)\bm{v}-\bm{B}\left(\bm{v}\bm{\cdot B}\right)\right]=0,
\end{equation}

\begin{equation}
\frac{\partial \bm{B}}{\partial t}-\bm{\nabla\times}\left(\bm{v}\bm{\times B}\right)=0,
\end{equation}

\noindent where $\rho$ is the mass density, $\bm{v}$ is the velocity, $\bm{B}$ is the magnetic field, $P=P_{\rm th}+P_{\rm mag}$ is the total pressure (thermal plus magnetic: $P_{\rm mag}=\frac{1}{2}|\bm{B}|^2$), $E=\rho\epsilon+\frac{1}{2}\rho\bm{v^2}+\frac{1}{2}|\bm{B}|^2$ is the total energy density, $\epsilon$ is the specific internal energy, and $T$ the gas temperature.  To close the above system of conservation laws, we use an ideal equation of state, i.e., 

\begin{equation}
P_{\rm {th}}=P_{\rm th}(\rho,\epsilon)=\left(\gamma-1\right)\rho\epsilon, 
\end{equation}

\noindent assuming a polytropic index $\gamma=\frac{5}{3}$ for adiabatic simulations and $\gamma=1.1$ for quasi-isothermal simulations. We also include additional advection equations of the form:

\begin{equation}
\frac{\partial\left[\rho C_{\alpha}\right]}{\partial t}+\bm{\nabla\cdot}\left[{\rho C_{\alpha} \bm{v}}\right]=0,
\label{eq:tracer}
\end{equation}

\noindent where $C_{\alpha}$ represents a set of three Lagrangian scalars used to track the evolution of gas initially contained in the cloud as a whole ($\alpha=\rm cloud/filament$), in its core ($\alpha=\rm core/footpoint$), and in its envelope ($\alpha=\rm envelope/tail$). Initially we define $C_{\alpha}=1$ for the whole cloud, the cloud core and the cloud envelope, respectively, and $C_{\alpha}=0$ everywhere else. This configuration allows us to follow the evolution of distinct parts of the cloud separately as they are swept up by the wind, as well as carefully examine the internal structure of the filaments that form downstream.\par

To solve the above system of hyperbolic conservation laws, we configure the {\sevensize PLUTO} code to use the \verb#HLLD# approximate Riemann solver of \cite{Miyoshi:2005} jointly with the constrained-transport upwind scheme of \cite{Gardiner:2005,Gardiner:2008} (used to preserve the solenoidal condition, $\bm{\nabla\cdot B}=0$). Equivalent algorithms are employed to solve the equations in the purely hydrodynamic simulation, i.e., the \verb#HLLC# approximate Riemann solver of \cite*{Toro:1994}, and the corner-transport upwind method of \cite{Colella:1990} and \cite{Saltzman:1994}. In order to achieve the required stability, the Courant-Friedrichs-Lewy (CFL) number was assigned a value of $0.3$ in all these cases.\par

The numerical resolutions and initial conditions used in our models and characterised by high density contrasts and supersonic wind speeds, have not been considered in previous three-dimensional studies. Despite being adequate to describe more realistic models of the ISM, the combination of these initial conditions may be challenging for some numerical solvers as a result of high-Mach-number flows near contact discontinuities and sharp density jumps. We address these technical difficulties by adding numerical diffusion to those cells affected by the high-Mach number problem.

\subsection{Initial and Boundary Conditions}
\label{subsec:Initial and Boundary Conditions}
For these simulations we consider a two-phase ISM composed of a single spherical cloud surrounded by a hot tenuous wind. The cloud is initially static and immersed in a wind represented by a uniform velocity field. The simulation volume consists of a rectangular prism with diode boundary conditions (i.e., outflow is allowed and inflow is prevented) on seven of its sides plus an inflow boundary condition (i.e., an injection zone) on the remaining side. The injection zone, located on the bottom left ghost zone of the computational domain (see Figure \ref{Figure1}), ensures that the flow of wind material is continuous over time.\par

All the simulations reported here utilise Cartesian $(X_1,X_2,X_3)$ coordinates and cover a physical spatial range $-2r_{\rm c}\leq X_1\leq2r_{\rm c}$, $-2r_{\rm c}\leq X_2\leq10r_{\rm c}$, and $-2r_{\rm c}\leq X_3\leq 2r_{\rm c}$, where $r_{\rm c}$ is the radius of the cloud. In the standard model, the grid resolution is $(N_{\rm X_{1}}\times N_{\rm X_{2}}\times N_{\rm X_{3}})=(512\times1536\times512)$, i.e., there are $128$ cells covering the cloud radius  ($R_{\rm 128}$) and $64$ cells covering the core radius. Other resolutions are explored in Paper III. The cloud is initially centred in the position $(0,0,0)$ of the simulation domain and has a density distribution that smoothly decreases as the distance from its centre increases (see \citealt{2000ApJ...531..366K,Nakamura:2006bc}). The function describing the radial density gradient is

\begin{figure}\centering
\includegraphics[scale=1]{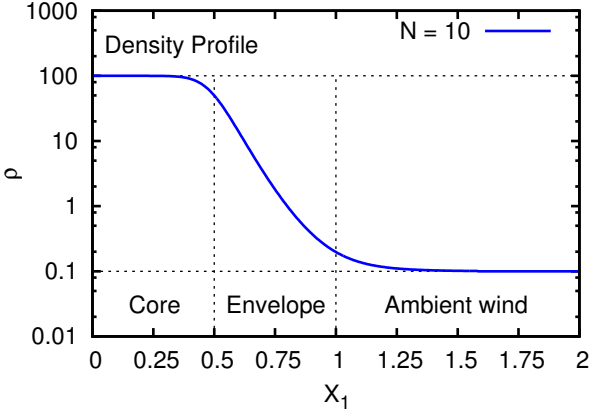}
\caption{Density profile for $N=10$ along the radial direction (see Equation \ref{eq:DensityProfile}).}
\label{Figure2}
\end{figure}

\begin{equation}
\rho(r)=\rho_{\rm w} + \frac{(\rho_{\rm c}-\rho_{\rm w})}{1+\left(\frac{r}{r_{\rm core}}\right)^N},
\label{eq:DensityProfile}
\end{equation}

\noindent where $\rho_c=100$ is the density at the centre of the cloud, $\rho_w=0.1$ is the density of the wind, $r_{\rm core}=0.5$ is the radius of the cloud core, and $N$ is an integer that determines the steepness of the curve describing the density gradient (see Figure \ref{Figure2}). The density profile given in Equation (\ref{eq:DensityProfile}) extends to infinity, so that we impose a boundary for the cloud by selecting an appropriate exponent $N$ and a cut-off radius. We truncate the density function at $r_{\rm cut}=1.58$, at which $\rho(r_{\rm cut})=1.01\rho_{\rm w}$ to ensure a smooth transition into the background gas, but we define the boundary of the cloud at $r_{\rm c}=1$, at which $\rho(r_{\rm c})=2.0\rho_{\rm w}$ (i.e., for $N=10$) for all the simulations reported here. A density gradient similar to that described by Equation (\ref{eq:DensityProfile}) is expected in clouds populating the ISM. In molecular clouds, for example, the dense $\rm H_2$ cores are surrounded by colder atomic $\textsc{h\,i}$ shells and thence by low-density, photo-ionised $\textsc{h\,ii}$ envelopes (a schematic view of typical molecular clouds can be found in Figure 3 of \citealt*{2009ApJ...698..350H}).\par

\begin{table}\centering
\caption{Simulation parameters for different models. In column 1, HD refers to the purely hydrodynamic scenario, while MHD-Al, MHD-Tr and MHD-Ob are magnetohydrodynamic models with magnetic fields aligned, transverse and oblique to the wind velocity, respectively. MHD-Ob-S and MHD-Ob-I include oblique fields with an increased strength and a reduced polytropic index, respectively. The initial conditions are reported in columns $2-5$.}
\begin{tabular}{c c c c c c}
\hline
\textbf{(1)} & \textbf{(2)} & \textbf{(3)} & \textbf{(3)} & \textbf{(4)} & \textbf{(5)}\Tstrut\\
\textbf{Model}  & $\gamma$ & $\cal M$ & $\chi$ & $\beta$ & \textbf{Topology}\Bstrut \\ \hline
HD & $1.67$ & $4$ & $10^3$ & $\infty$ & --\Tstrut \\
MHD-Al & $1.67$ & $4$ & $10^3$ & $100$ & Aligned\\
MHD-Tr & $1.67$ & $4$ & $10^3$ & $100$ & Transverse\\
MHD-Ob & $1.67$ & $4$ & $10^3$ & $100$ & Oblique\\
MHD-Ob-S & $1.67$ & $4$ & $10^3$ & $10$ & Oblique\\
MHD-Ob-I & $1.10$ & $4.9$ & $10^3$ & $100$ & Oblique\Bstrut\\ \hline
\end{tabular}
\label{Table2}
\end{table}

This study comprises six numerical simulations in total (see Table \ref{Table2}). Model HD, the only model without magnetic fields, serves as a comparison between filament formation mechanisms in HD and MHD configurations. Models MHD-Al and MHD-Tr study the formation of magnetotails in environments where the field has the same initial strength ($\beta=100$) and it is aligned (Al), i.e.,

\begin{equation}
\bm{B}=\bm{B_{\rm Al}=B_{\rm 2}}=\sqrt{\frac{2P_{\rm th}}{\beta}}~\bm{{\rm \hat{e}_{\rm 2}}},
\label{eq:AlField}
\end{equation}

\noindent or transverse (Tr), i.e.,

\begin{equation}
\bm{B}=\bm{B_{\rm Tr}=B_{\rm 1}}=\sqrt{\frac{2P_{\rm th}}{\beta}}~\bm{{\rm \hat{e}_{\rm 1}}},
\label{eq:TrField}
\end{equation}

\noindent with respect to the wind velocity, respectively. MHD-Ob is our standard model in which the magnetic field has the same strength as before (i.e., $\beta=100$) and is defined as follows

\begin{equation}
\bm{B}=\bm{B_{\rm Ob}=B_{\rm 1}+B_{\rm 2}+B_{\rm 3}},
\label{eq:ObliqueField}
\end{equation}

\noindent i.e., a 3D field obliquely oriented with respect to the wind direction with components of identical magnitude, i.e.,

\begin{equation}
|\bm{B_{\rm 1}|=|B_{\rm 2}|=|B_{\rm 3}}|=\sqrt{\frac{2P_{\rm th}}{3\beta}}.
\label{eq:BComponents}
\end{equation}

\noindent Note that the cloud is in thermal pressure equilibrium with the ambient medium at the beginning of the calculation ($P_{\rm th}=0.1$).\par

Model MHD-Ob-S uses the same initial field with an oblique topology, but it explores the evolution of magnetotails in an environment with a slightly stronger initial magnetic field ($\beta=10$). In addition, model MHD-Ob-I studies the quasi-isothermal ($\gamma=1.1$) evolution of a wind-cloud system. Note that in this model, the Mach number needs to be altered (to ${\cal M_{\rm w}}=4.9$) in order to keep the wind speed and dynamic time-scales constant. In all our MHD  simulations, we use the magnetic vector potential $\bm{A}$, where $\bm{B}=\bm{\nabla\times A}$, to initialise the field and ensure that the initial field has zero divergence. 

\subsection{Diagnostics}
\label{subsec:Diagnostics}
To study the formation and evolution of filaments, a series of diagnostics, involving geometric, kinetic, and magnetic quantities, can be estimated from our simulated data. Following previous authors \citep{1994ApJ...420..213K,Nakamura:2006bc,2008ApJ...680..336S} we define the volume-averaged value of a variable $\cal F$ by:

\begin{equation}
[~{\cal F}_{\alpha}~]=\frac{\int {\cal F}C_{\alpha}dV}{\int C_{\alpha}dV},
\label{eq:AveragedF}
\end{equation}

\noindent denoted by square brackets, and the mass-weighted volume average of the variable $\cal G$ by:

\begin{equation}
\langle~{\cal G}_{\alpha}~\rangle=\frac{\int {\cal G}\rho C_{\alpha}dV}{\int \rho C_{\alpha}dV},
\label{eq:IntegratedG}
\end{equation}

\noindent denoted by angle brackets. In Equations (\ref{eq:AveragedF}) and (\ref{eq:IntegratedG}), $V$ is the volume and $C_{\alpha}$ are the advected scalars defined in Section \ref{subsec:SimulationCode}. Note that the denominators in Equations (\ref{eq:AveragedF}) and (\ref{eq:IntegratedG}) are the total cloud volume, $\cal V_{\rm cl}$, and cloud mass, $M_{\rm cl}$, respectively, which are both, in general, functions of time. Taking Equation (\ref{eq:AveragedF}), we define functions describing the average density, $[~\rho_{\alpha}~]$; and the average plasma beta, $[~\beta_{\alpha}~]$. Taking Equation (\ref{eq:IntegratedG}), we define the averaged cloud extension, $\langle~X_{{\rm j},\alpha}~\rangle$; its rms along each axis, $\langle~X^2_{{\rm j},\alpha}~\rangle$; the averaged velocity, $\langle~v_{{\rm j},\alpha}~\rangle$; and its rms along each axis, $\langle~v^2_{{\rm j},\alpha}~\rangle$. Note that $j=1,2,3$ specifies the direction along $X_1$, $X_2$, and $X_3$, respectively. The initial values of the above quantities are used to normalise their averaged values and retain the scalability of our results. Velocity measurements are the exemption to this as they are normalised with respect to either the wind sound speed, $c_{\rm w}$, or the wind speed, $v_{\rm w}$.\par 

In order to quantify changes in the shape of the cloud, the effective radii along each axis

\begin{equation}
\iota_{{\rm j},\alpha}=\left[5\left(\langle~X^2_{{\rm j},\alpha}~\rangle-\langle~X_{{\rm j},\alpha}~\rangle^2\right)\right]^{\frac{1}{2}}
\label{eq:Moments}
\end{equation}

\noindent are utilised \citep{1994ApJ...433..757M}. Using Equations (\ref{eq:Moments}), we define the aspect ratio of the filament along $j=2,3$ as 

\begin{equation}
\xi_{{\rm j},\alpha}=\frac{\iota_{{\rm j},\alpha}}{\iota_{1,\alpha}}.
\label{eq:AspectRatio}
\end{equation}

In a similar way, the corresponding dispersion of the $j$-component of the velocity and the transverse velocity read

\begin{equation}
\delta_{{\rm v}_{{\rm j},\alpha}}=\left(\langle~v^2_{{\rm j},\alpha}~\rangle-\langle~v_{{\rm j},\alpha}~\rangle^2\right)^{\frac{1}{2}},
\label{eq:rmsVelocityComponent}
\end{equation}

\noindent and

\begin{equation}
\delta_{{\rm v}_{\alpha}}\equiv|\bm{\delta_{{\rm v}_{\alpha}}}|=\sqrt{\sum_{j=1,3}\delta_{{\rm v}_{{\rm j},\alpha}}^2},
\label{eq:rmsVelocity}
\end{equation}

\noindent respectively. Note that the cloud acceleration can be studied by analysing the behaviour of $\langle~v_{2,\alpha}~\rangle$. We also measure the degree of mixing between cloud and wind gas by using a mixing fraction expressed as a percentage

\begin{equation}
f_{{\rm mix}_{\alpha}}=\frac{\int \rho C^{\asterisk}_{\alpha}dV}{M_{{\rm cl},0}}\times 100\%,
\label{eq:MixingFraction}
\end{equation}

\noindent where the numerator is the mass of mixed gas, with $0.1\leq C^{\asterisk}_{\alpha}\leq 0.9$ tracking material in mixed cells, and $M_{{\rm cl},0}$ represents the mass of the cloud at time $t/t_{\rm cc}=0$. Additionally, the flux of mass through two-dimensional surfaces transverse to the $X_2$ axis is calculated from

\begin{equation}
{F_{\alpha}}=|\bm{{F_{\alpha}}}(X_{\rm cut})|=\left|\int \rho C_{\alpha}(\bm{v\cdot{\rm \hat{e}_{\rm 2}}}) dS~\bm{{\rm \hat{e}_{\rm 2}}}\right|,
\end{equation}

\noindent where $X_{\rm cut}$ defines the location at which we place the reference surface (e.g., at the rear side of the simulation domain), and $dS$ is a differential element of that surface. The surface elements are squares in our case as we are using equidistant uniform grids without mesh refinement. To maintain the scalability of our results, we report the mass fluxes normalised with respect to the initial flux of wind mass through the same reference surface defined by $X_{\rm cut}$, namely $F_{\rm wind,0}=|\bm{{F_{\rm wind,0}}}(X_{\rm cut})|=\left|\int \rho_{\rm w} (\bm{v_{\rm w}\cdot{\rm \hat{e}_{\rm 2}}}) dS~\bm{{\rm \hat{e}_{\rm 2}}}\right|$.\par

Another set of diagnostic quantities include those related to the energetics involved in the formation of magnetotails. The enhancement of kinetic energy in cloud (filament) material is proportional to the mass-weighted velocity of the structure, so its behaviour can be studied by analysing the evolution of $\langle~v_{{\rm j},\alpha}~\rangle$. On the other hand, the variation of the magnetic energy contained in filament material at a specific time, $t$, can be studied with
\begin{equation}
\Delta E_{M_{\alpha}}=\frac{~E_{M_{\alpha}}~-~E_{M_{\alpha,0}}~}{~E_{M_{\alpha,0}}~},
\label{eq:MEenhancement}
\end{equation}

\noindent where $~E_{M_{\alpha}}~=\int \frac{1}{2} |\bm{B}|^2C_{\alpha}dV$ is the total magnetic energy in cloud (filament) material, and $~E_{M_{\alpha,0}}~$ is the initial magnetic energy in the cloud.

\subsection{Dynamical time-scales}
\label{subsec:DynamicalTime-Scales}
 Three important dynamical time-scales in our simulations include:

\noindent a) The cloud-crushing time (as defined in \citealt{1996ApJ...473..365J})

\begin{equation}
t_{\rm cc}=\frac{2r_{\rm c}}{v_{\rm s}}=\left(\frac{\rho_{\rm c}}{\rho_{\rm w}}\right)^{\frac{1}{2}}\frac{2r_{\rm c}}{{\cal M_{\rm w}} c_{\rm w}}=\chi^{\frac{1}{2}}\frac{2r_{\rm c}}{{\cal M_{\rm w}} c_{\rm w}};
\label{eq:CloudCrushing}
\end{equation}

\noindent where $v_{\rm s}={\cal M_{\rm w}} c_{\rm w}\chi^{-\frac{1}{2}}$ is the speed of the internal shock transmitted to the cloud by the wind after the initial collision. Hereafter, times reported in this paper are normalised with respect to the cloud-crushing time in order to maintain scalability.

\noindent b) The simulation time, which in our case is

\begin{equation}
t_{\rm sim}=1.225\:t_{\rm cc}.
\label{eq:SimulationTime}
\end{equation}

\noindent c) The wind-passage time (as defined in \citealt{2005ApJ...619..327F})

\begin{equation}
t_{\rm wp}=\frac{2r_{\rm c}}{v_{\rm w}}=\frac{1}{\chi^{\frac{1}{2}}}\:t_{\rm cc}=0.032\:t_{\rm cc};
\label{eq:WindPassage}
\end{equation}

In reference to time resolution, our simulations are sensitive to changes occurring on time-scales of the order of $2.5\times 10^{-5}\:t_{\rm cc}$, owing to the small time steps involved, and output files are written at intervals of $\Delta t=8.2\times 10^{-3}\:t_{\rm cc}$ to ensure that sequential snapshots adequately capture the evolution of the filaments' morphology.\par

\section{Cloud Disruption}
\label{sec:CloudDisruption}
A thorough description of the results obtained in our simulations is presented in this and subsequent sections. In Section \ref{sec:CloudDisruption} we summarise the processes leading to the disruption of clouds when they are swept up by supersonic winds. In Section \ref{sec:FilamentFormation}, we describe the mechanisms involved in the formation of filamentary structures and how they evolve over time in a purely hydrodynamic case; in two models with different magnetic field geometries, namely aligned with and transverse to the wind velocity; and in the more general case in which the field is oblique (i.e., it has both aligned and transverse components). We also discuss the effects of varying the initial field strength as defined by the plasma beta ($\beta$) and of softening the equation of state by changing the adiabatic index ($\gamma$). Finally, in Section \ref{sec:Entrainment} we discuss the entrainment of clouds and filaments in global winds.\par

The processes leading to the break-up of clouds and the formation of filaments in wind-cloud interactions are intimately related. The relationship is, in fact, a causal one in which a filament forms as a result of the steady disruption of the cloud. Thus, to study the formation, structure, and evolution of filamentary structures, we first need to understand the mechanisms responsible for the destruction of the cloud. Even though distinct sets of initial conditions can result in morphologically different filamentary structures as we show in Section \ref{sec:FilamentFormation}, cloud disruption can be considered as a universal four-stage process regardless of the initial conditions. As we explain below, the main aspects of the evolution remain the same in models with distinct initial configurations with differences arising solely due to time lags in the emergence of fluid instabilities and turbulence. A summary of the processes leading to the disruption of a wind-swept cloud is below:

\begin{figure}
\begin{center}
  \begin{tabular}{c}
  \resizebox{80mm}{!}{\includegraphics{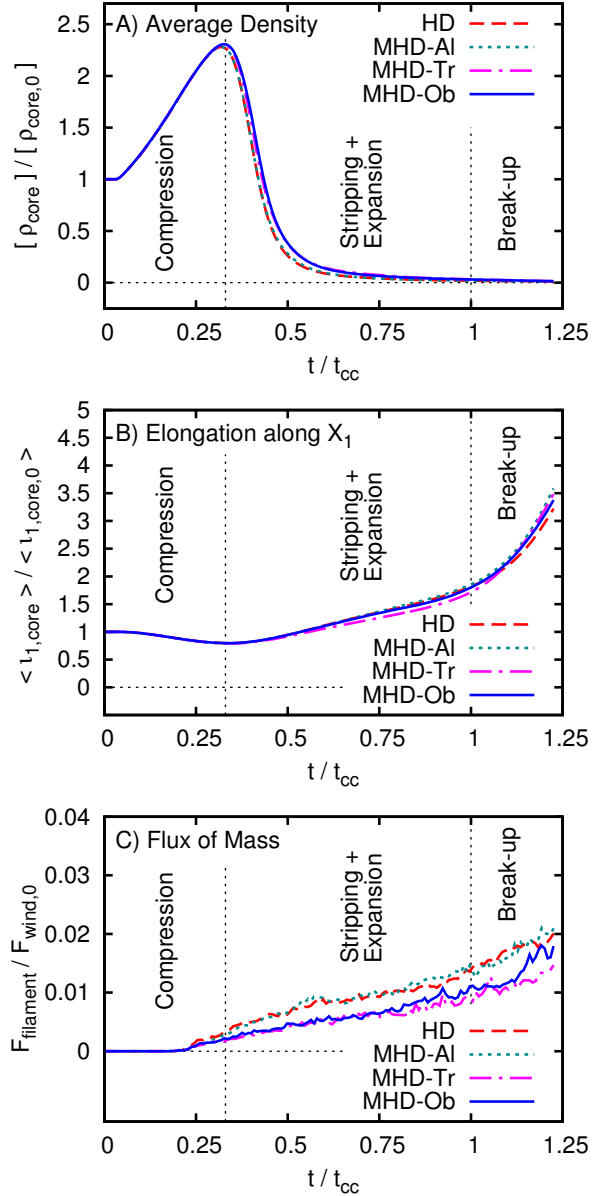}} \\
  \end{tabular}
  \caption{Panel A shows the time evolution of the average core density in models HD (dashed line), MHD-Al (dotted line), MHD-Tr (dash-dotted line), and MHD-Ob (solid line). Panel B indicates the evolution of the elongation of the cloud core in the $X_1$ direction: compression, stripping, expansion, and break-up phases are identified. Panel C shows the mass flux of stripped material flowing through the back surface of the simulation domain as a function of time. Note that the mass flux has been normalised with respect to that of the wind flowing through the same surface at the beginning of the computation (see Section \ref{sec:CloudDisruption}).}
  \label{Figure3}
\end{center}
\end{figure}

\begin{enumerate}
  \item \textbf{\textit{Compression phase:}} At the earliest stage, wind material commences interacting with the front surface of the cloud and produces two effects: a) shock waves are triggered in both media: one shock wave is reflected back into the upstream wind medium forming a high-pressure, bow shock, while an internal shock is transmitted into the cloud; and b) the wind ram pressure starts to compress the cloud gas in all directions, increasing the core density to twice its original value and reducing the lateral size of the core by $\sim25\,\%$ (see Panels A and B of Figure \ref{Figure3}, respectively). The shock transmitted to the cloud travels through its environment at a speed: $v_{s}\simeq{\cal M_{\rm w}}c_{\rm w}\chi^{-\frac{1}{2}}=0.126\:c_{\rm w}=0.032\:v_{\rm w}$, and arrives at its rear surface in a time of approximately $t/t_{\rm cc}=1.0$. The compression phase lasts until $\sim t/t_{\rm cc}=0.3$ and is common to all models as can be seen in Figure \ref{Figure3}.
  \item \textbf{\textit{Stripping phase:}} Meanwhile, wind material starts to flow downstream and wraps around the cloud converging behind it in $t/t_{\rm wp}=1.0$ (see Equation \ref{eq:WindPassage}). The convergence of flow on the axis at the rear of the cloud drives a transient biconical shock into the ambient gas. The coupling region in the biconical structure, formed by low-density gas, moves upstream (i.e., against the flow) towards the rear surface of the cloud and contributes to its flattening. Concurrently, the wind material moving downstream also interacts with the outer layers of the cloud and instigates stripping of its gas (see the flux of cloud/filament mass in Panel C of Figure \ref{Figure3}). Stripping occurs primarily due to the onset of the Kelvin-Helmholtz (hereafter KH) instability at the wind-cloud interface. As a result, cloud and wind material begin to mix downstream and the low-density gas in the envelope of the cloud is steadily removed and funnelled into the flow. The stripping phase occurs at all times, but it is more dynamically important from $\sim t/t_{\rm cc}=0.2$ until $t/t_{\rm cc}=0.5$. Different initial configurations can change the growth time of the KH instability at shear layers and speed up or slow down the stripping process. Note, for example, that models HD and MHD-Al exhibit higher mass fluxes than models MHD-Tr and MHD-Ob throughout most of the evolution. In Section \ref{sec:FilamentFormation} we describe how stripping leads to the formation of filamentary tails behind the cloud and how the structure of these tails changes when the initial magnetic field changes.
  \item \textbf{\textit{Expansion phase:}} The shock transmitted into the cloud travels through it, transporting energy with it. Without an efficient mechanism to remove the extra energy from the system, this is added in full to the internal energy of the gas, $\epsilon$. The resultant changes in thermal pressure then lead to adiabatic heating, the temperature rises, and the cloud expands (note e.g., how the elongation along the $X_1$ direction starts to increase after $t/t_{\rm cc}=0.3$ in Panel B of Figure \ref{Figure3}). Cloud material becomes more vulnerable to stripping caused by the wind ram pressure as the effective cross section upon which the wind exerts its force increases and this accelerates gas mixing. The arrival of the internal shock at the back surface of the cloud also allows denser gas to flow downstream and occupy low-pressure regions previously created by rarefaction waves in the aforementioned biconical structure. The expansion phase lasts from $t/t_{\rm cc}=0.5$ to $t/t_{\rm cc}=1.0$ and is qualitatively similar in all models regardless of the initial conditions. We note, however, that the degree of compression and expansion of cloud material is connected to the equation of state assumed for the gas, so if a softer polytropic index is used (i.e., $\gamma=1.1$), compression can be largely enhanced and expansion delayed (see Section \ref{subsubsec:Isothermal} in which we describe the evolution of a quasi-isothermal model in detail).
   \item \textbf{\textit{Break-up phase:}} The net effect of the drag force exerted by the wind on the cloud is to accelerate it. As material is removed from the cloud, this acceleration increases and the associated Rayleigh-Taylor (hereafter RT) instability develops more quickly. At about $t/t_{\rm cc}=1.0$, the cloud has been accelerated to about $0.60\:c_{\rm w}=0.09\:v_{\rm w}$ (see Section \ref{subsubsec:Terminalvelocity} for further details). This situation combined with an expanded cross sectional area favours the growth of more disruptive (long-wavelength) RT instability modes, which disrupt the cloud and break it up into smaller cloudlets (note how the lateral size of the core in the $X_1$ direction grows faster after $t/t_{\rm cc}=1.0$ in Panel B of Figure \ref{Figure3}). These cloudlets are further accelerated and should eventually acquire the full wind speed, if not destroyed by instabilities beforehand. However, we do not follow the evolution of these cloudlets beyond $t/t_{\rm cc}=1.2$, so that further investigation of this late-stage, co-moving phase is warranted. Even though long-wavelength RT perturbations are ultimately responsible for the destruction of the cloud in all cases, the break-up process can be sped up or slowed down depending on the initial configuration of the magnetic field. In Section \ref{subsubsec:Instabillities}, we provide a full description of the development of instabilities under different ambient conditions and their effect on the morphology of the resulting filaments.
\end{enumerate}

Previous simulations of shock-cloud interactions showed that the destruction of clouds can occur in several cloud-crushing times. In contrast, our models show that clouds, with the above density contrast, are disrupted in a single cloud-crushing time as defined by Equation (\ref{eq:CloudCrushing}). This result can be attributed to: 1) the employment of more realistic three-dimensional clouds with large density contrasts and gentle smoothing profiles; and 2) the fact that the clouds in our models are interacting with supersonic flows at all times. This is in agreement to what was found by \cite{2000ApJ...543..775G} despite the different initial conditions used in their work. In addition, we note that these phases overlap with each other (e.g., stripping occurs at all times), so the above description indicates the dominant effects at specific time intervals during the evolution. Previous authors described a similar four-phase process when studying shocked clouds (\citealt{1994ApJ...420..213K,2009ApJ...703..330C}). The evolution of shock-cloud systems was divided into a shock transmission phase followed by shock compression, cloud expansion, and cloud destruction phases. Since we are investigating wind-cloud interactions with high density contrasts, stripping is an important mechanism to support filamentary structures over extended periods of time. Thus, we believe that our division above is more relevant for the study of filament formation. Besides, the initial shock compression phase is triggered by the impact of the incident wind on the cloud nose, so that both can be seen as constituents of the same evolutionary stage.\par

We also note that different magnetic field strengths and orientations can lead to specific quantitative changes. In particular, the presence of transverse components in the initial magnetic field can change the dynamics of small-scale flows, i.e., they can enhance or suppress fluid instabilities at shear layers depending on how the geometry and strength of the field evolve (see e.g., \citealt{1991ApJ...376L..21C,1996ApJ...460..777F}). Before studying the effects of fluid instabilities in more detail, however, we first concentrate on how filamentary structures form in these interactions.

\section{Filament formation and evolution}
\label{sec:FilamentFormation}

\begin{figure*}
\begin{center}
  \begin{tabular}{c c c c}
    \resizebox{40mm}{!}{\includegraphics{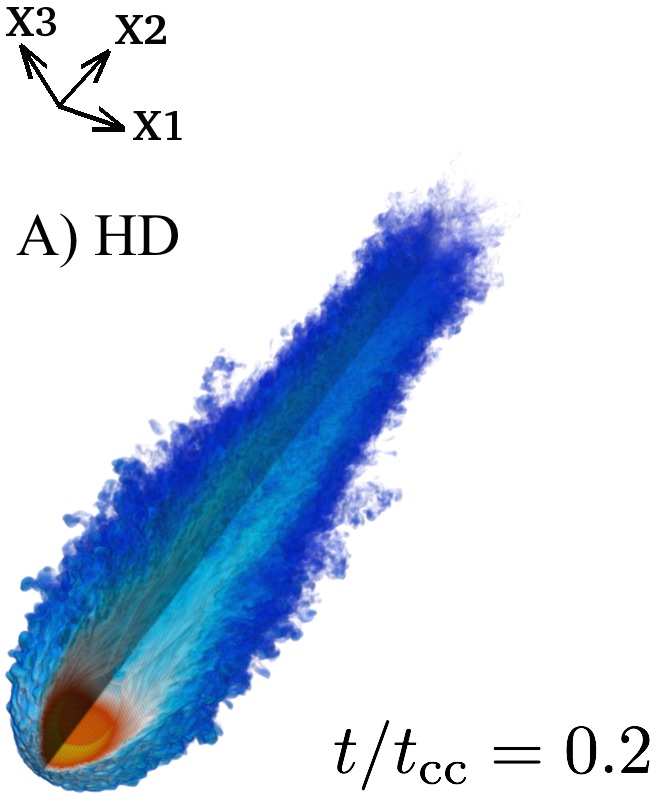}} & \resizebox{40mm}{!}{\includegraphics{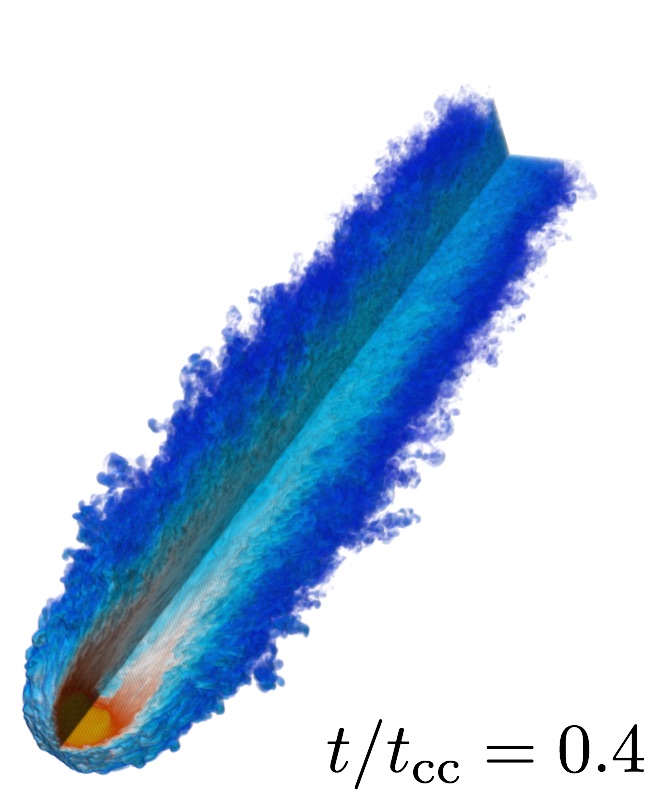}} & \resizebox{40mm}{!}{\includegraphics{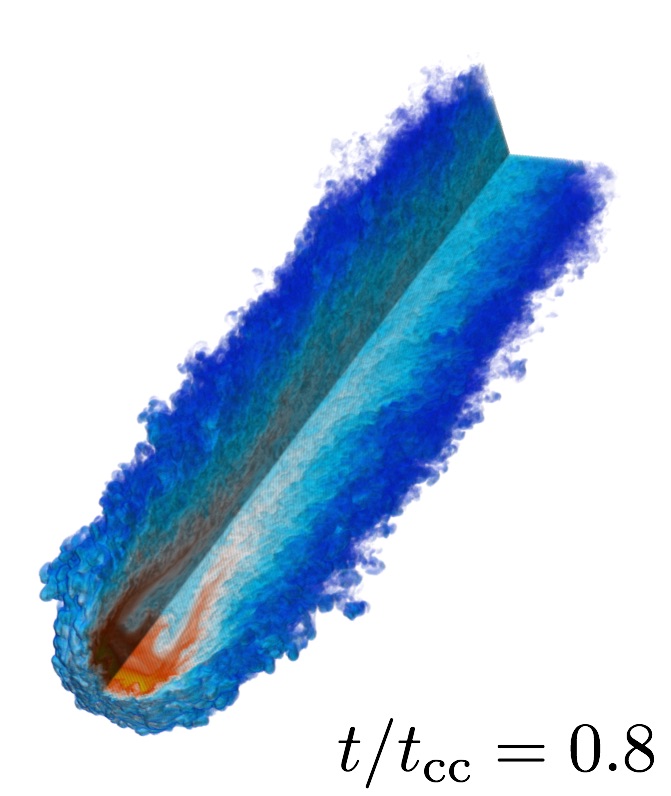}} & \resizebox{40mm}{!}{\includegraphics{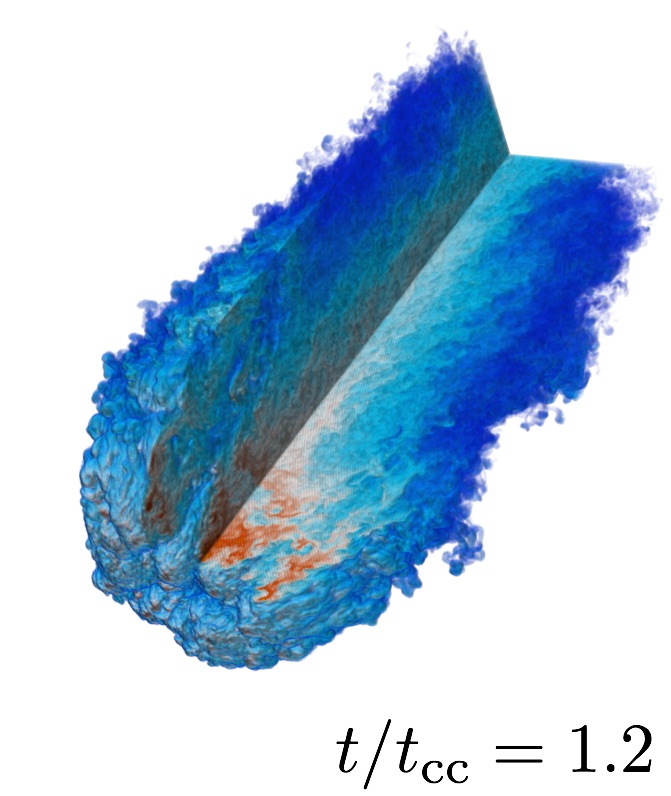}}\\
    \resizebox{40mm}{!}{\includegraphics{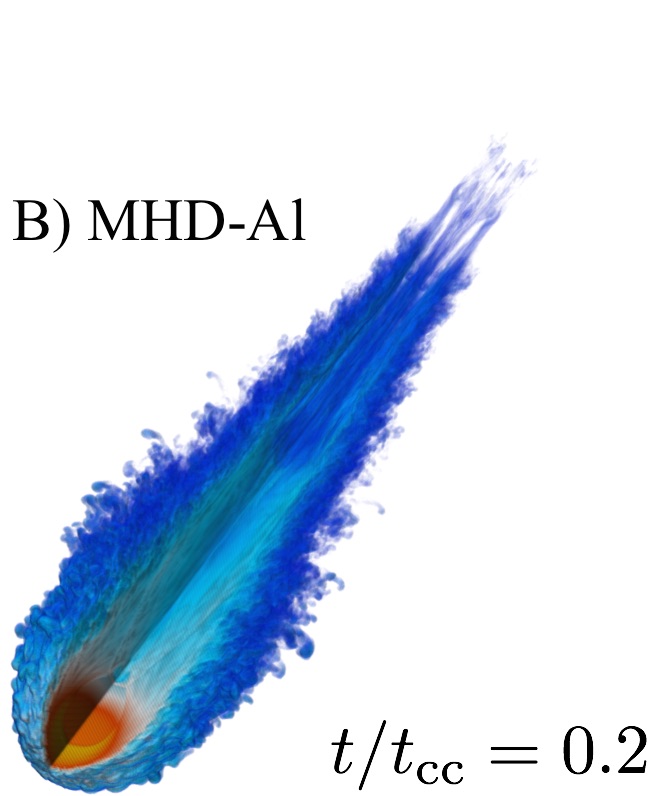}} & \resizebox{40mm}{!}{\includegraphics{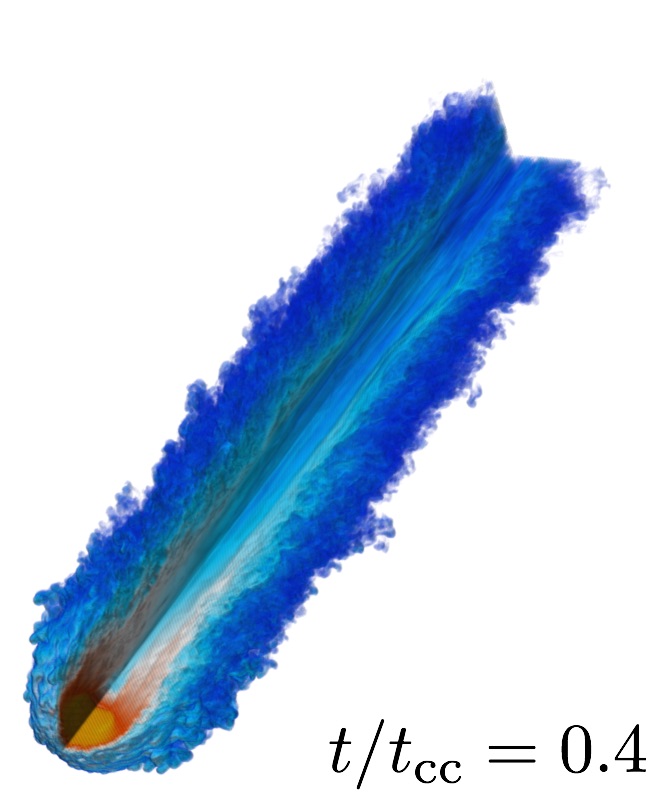}} & \resizebox{40mm}{!}{\includegraphics{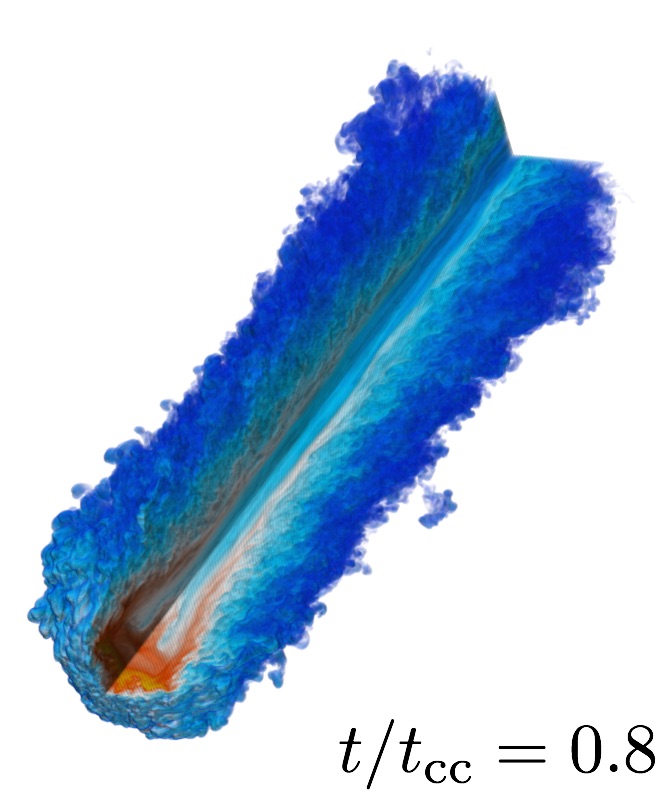}} & \resizebox{40mm}{!}{\includegraphics{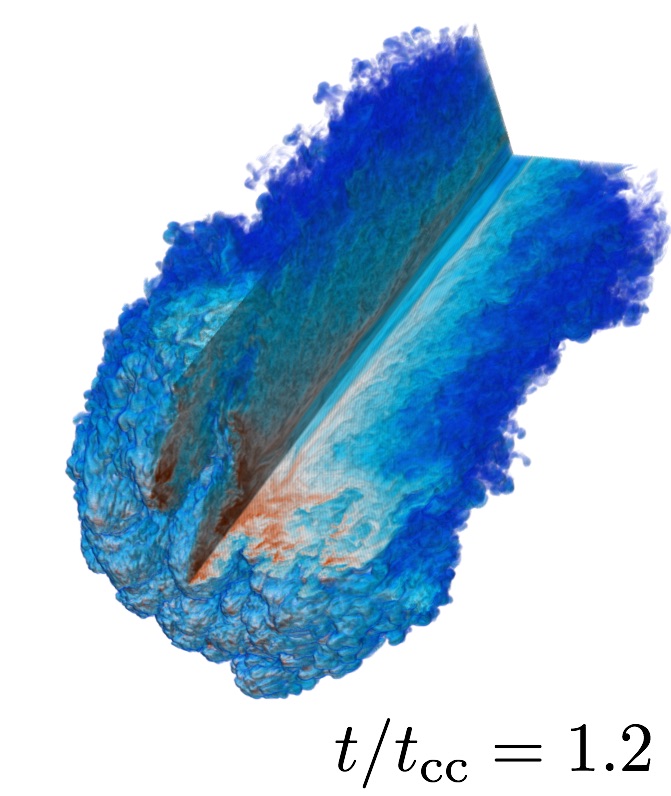}}\\
    \resizebox{40mm}{!}{\includegraphics{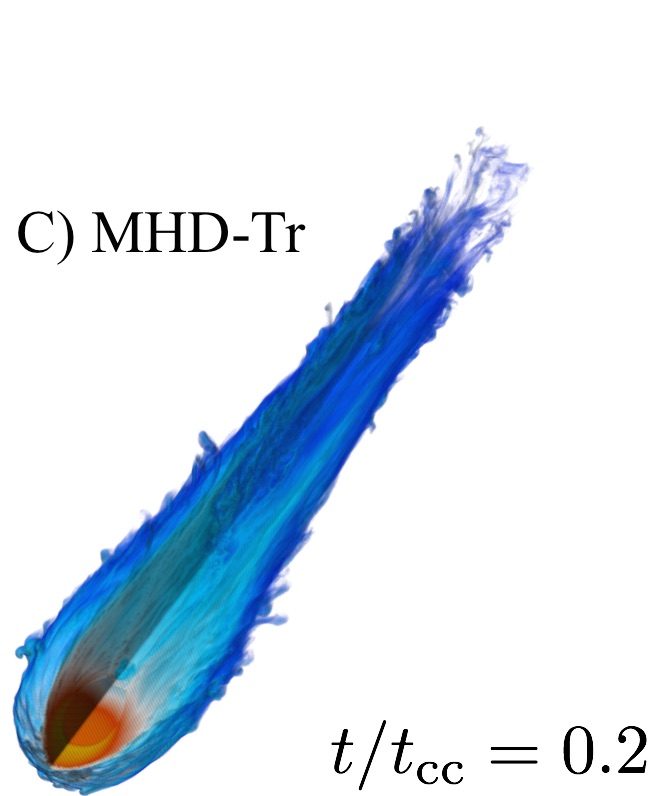}} & \resizebox{40mm}{!}{\includegraphics{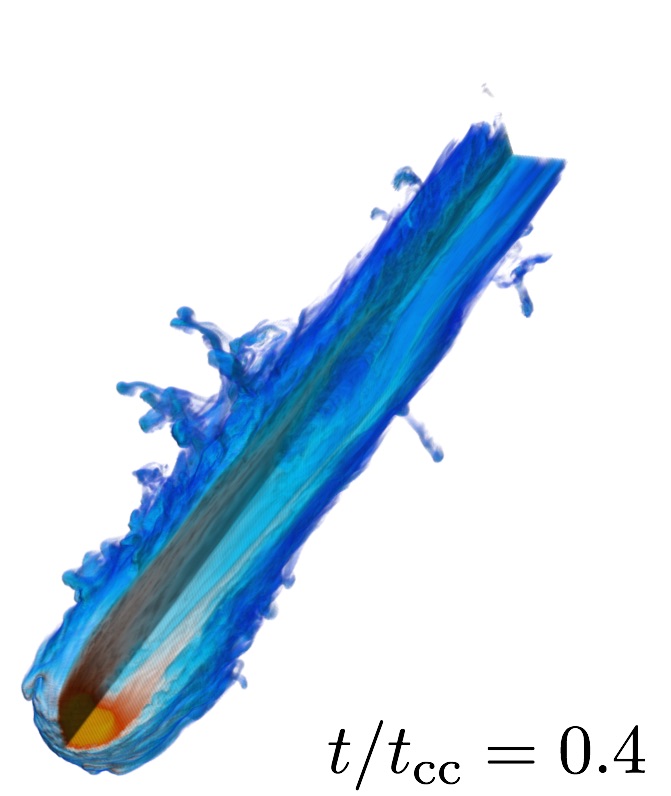}} & \resizebox{40mm}{!}{\includegraphics{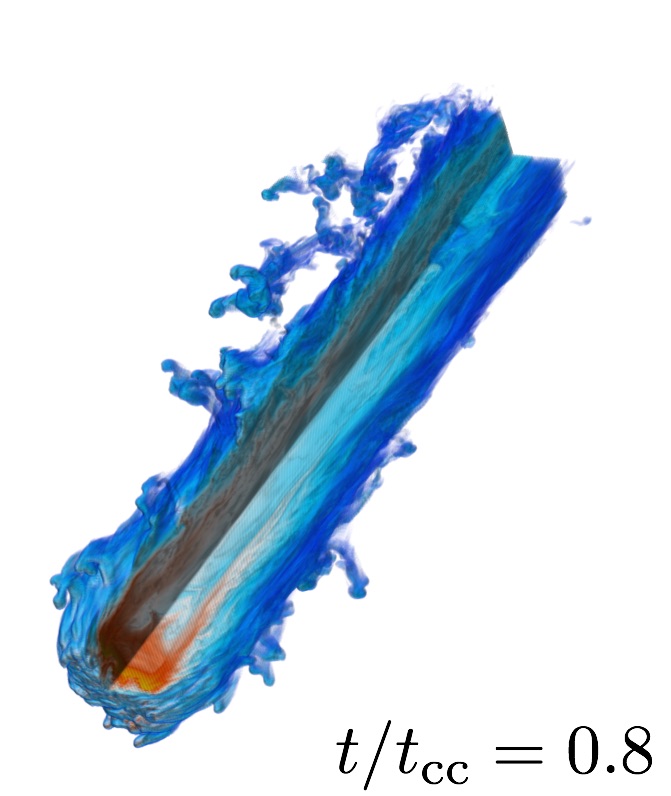}} & \resizebox{40mm}{!}{\includegraphics{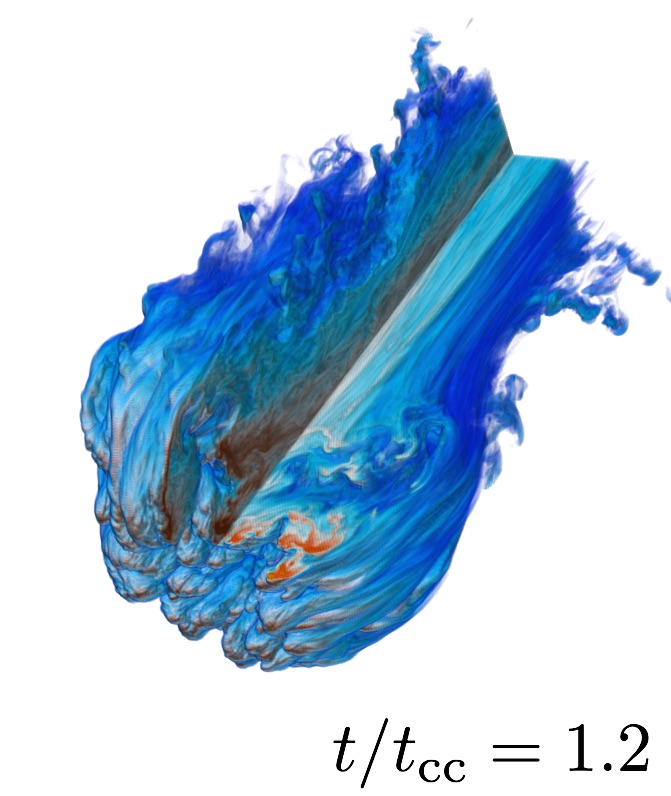}}\\
    \resizebox{40mm}{!}{\includegraphics{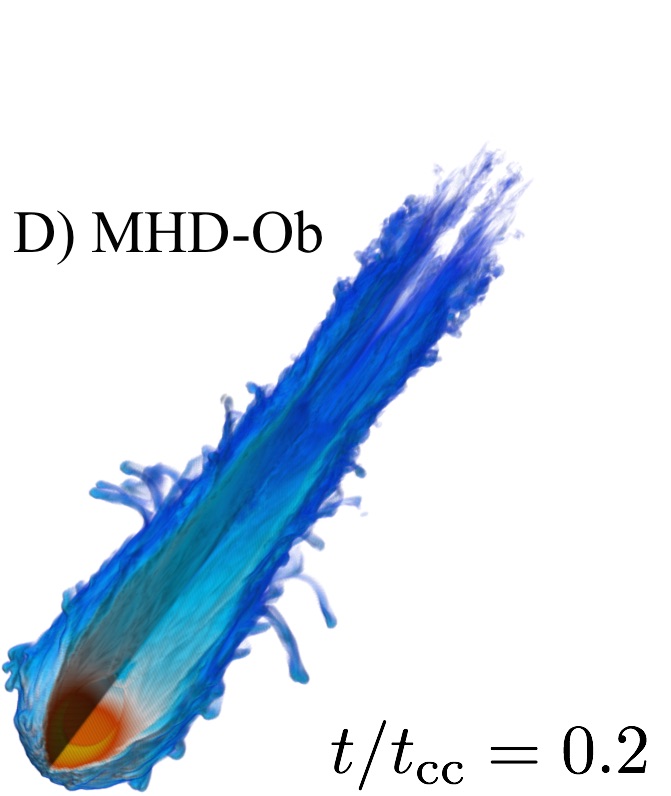}} & \resizebox{40mm}{!}{\includegraphics{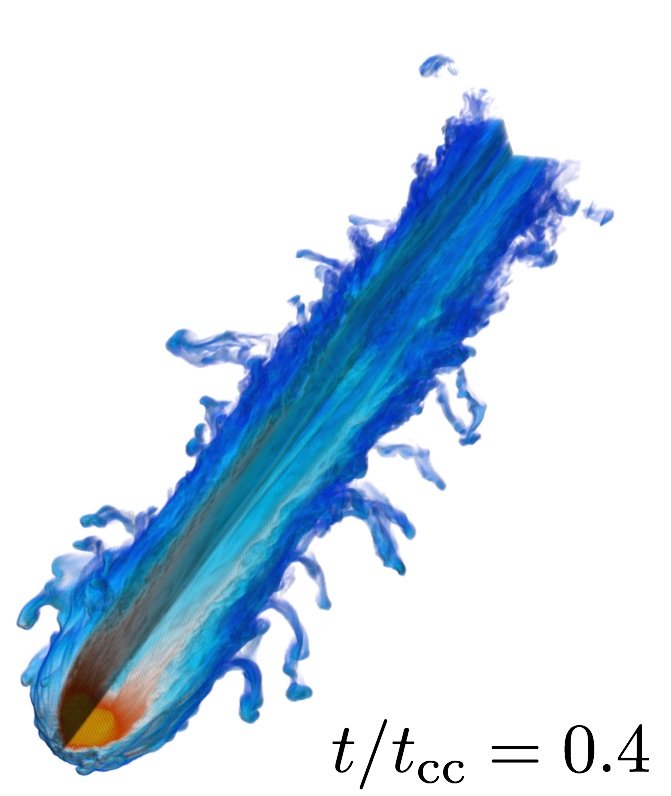}} & \resizebox{40mm}{!}{\includegraphics{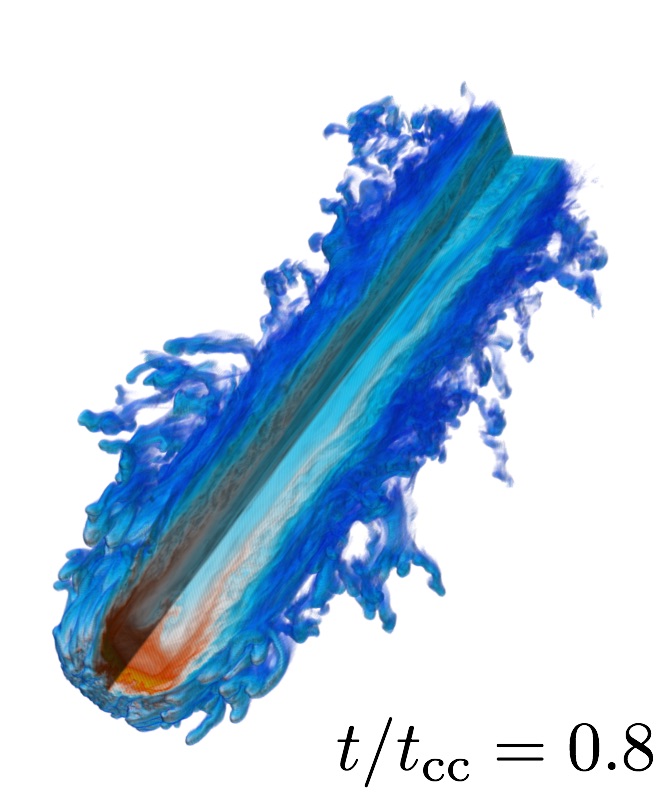}} & \resizebox{40mm}{!}{\includegraphics{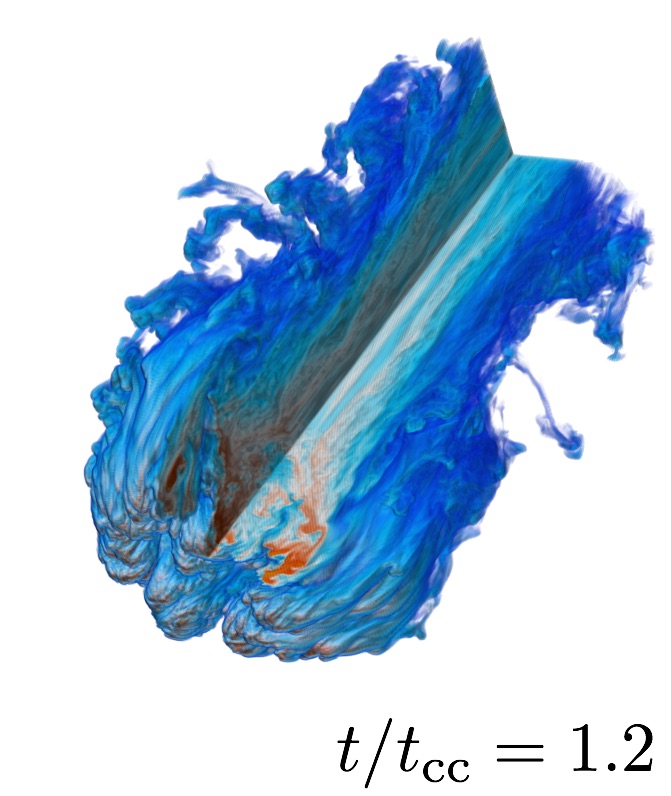}}\Dstrut \\
    \multicolumn{4}{c}{\resizebox{!}{8mm}{\includegraphics{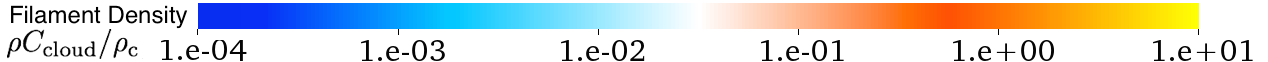}}}\\
  \end{tabular}
  \caption{3D volume renderings of the logarithm of the mass density in filaments normalised with respect to the initial cloud density, $\rho_{\rm c}$, at four different times: $t/t_{\rm cc}=0.2$, $t/t_{\rm cc}=0.4$, $t/t_{\rm cc}=0.8$, and $t/t_{\rm cc}=1.2$. Panel A shows the evolution in a purely hydrodynamic case, whilst the next three Panels: B, C, and D show the evolution of MHD wind-cloud systems with the magnetic field aligned, transverse, and oblique to the wind direction, respectively. Note that a quadrant has been clipped from the renderings to show the interior of the tails. Small-scale vorticity, gas mixing, and lateral expansion are more significant in models in which the initial magnetic field does not have transverse components (see Section \ref{sec:FilamentFormation} for further details). Magnetic field components transverse to the streaming direction suppress the KH instability and confine the gas (that has been stripped from the cloud) in narrow tails.} 
  \label{Figure4}
\end{center}
\end{figure*}

\begin{figure*}
\begin{center}
  \begin{tabular}{c c c c c c c}
       \multicolumn{7}{c}{\resizebox{!}{4.3mm}{\includegraphics{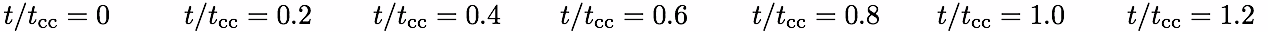}}}\Tstrut\Bstrut\\
      \resizebox{21mm}{!}{\includegraphics{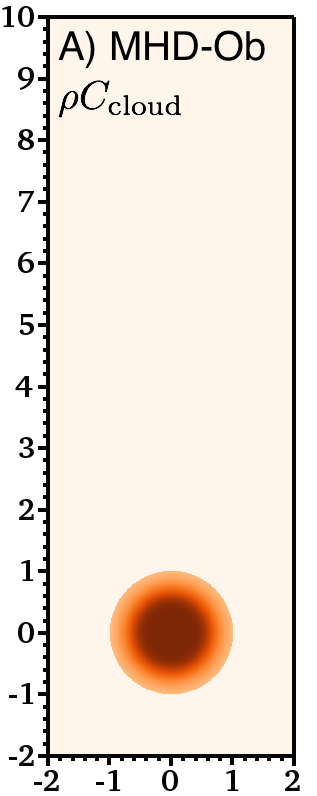}} & \resizebox{21mm}{!}{\includegraphics{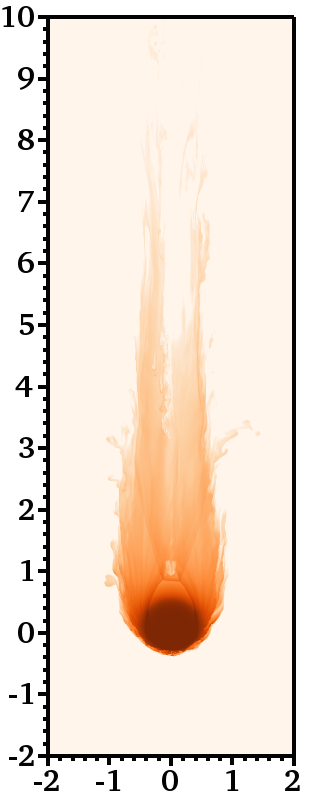}} & \resizebox{21mm}{!}{\includegraphics{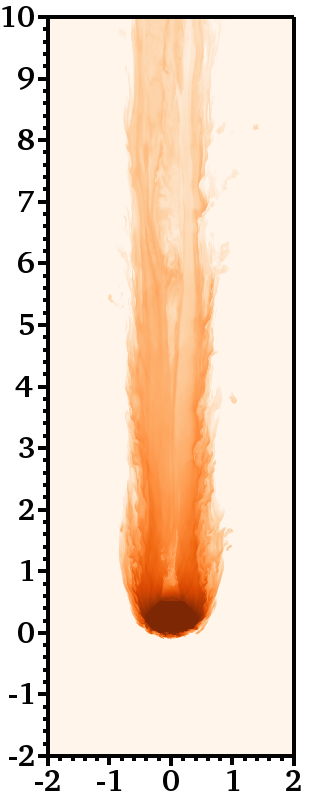}} & \resizebox{21mm}{!}{\includegraphics{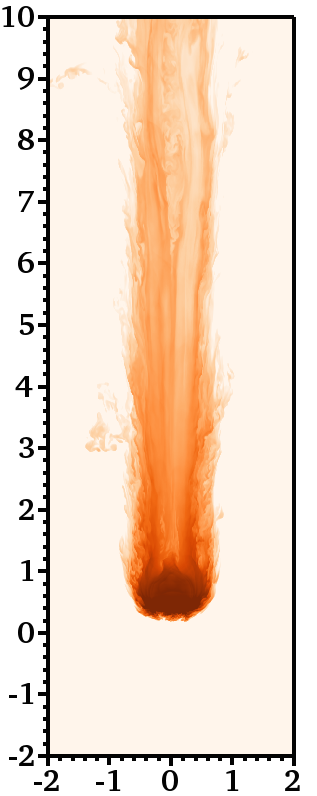}} & \resizebox{21mm}{!}{\includegraphics{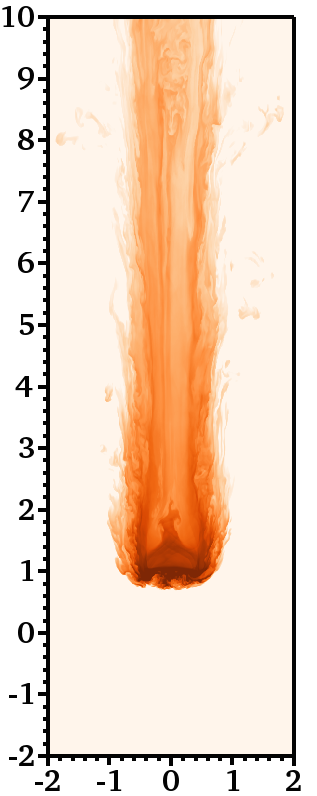}} & \resizebox{21mm}{!}{\includegraphics{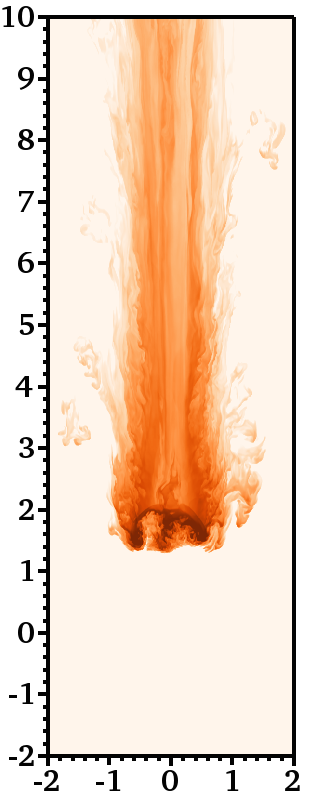}} & \resizebox{21mm}{!}{\includegraphics{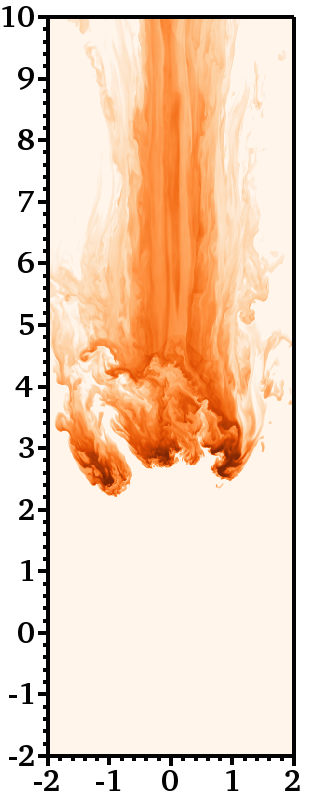}}\\           
      \resizebox{21mm}{!}{\includegraphics{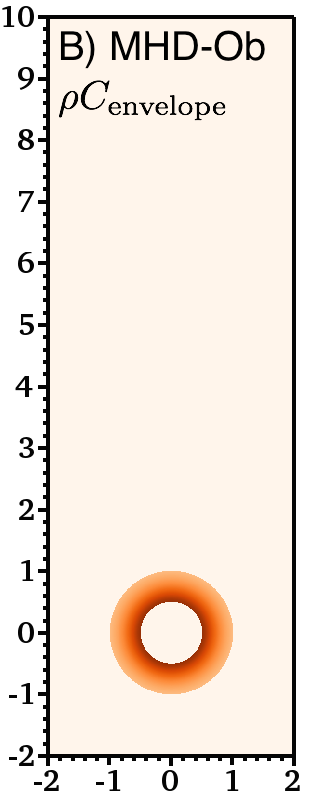}} & \resizebox{21mm}{!}{\includegraphics{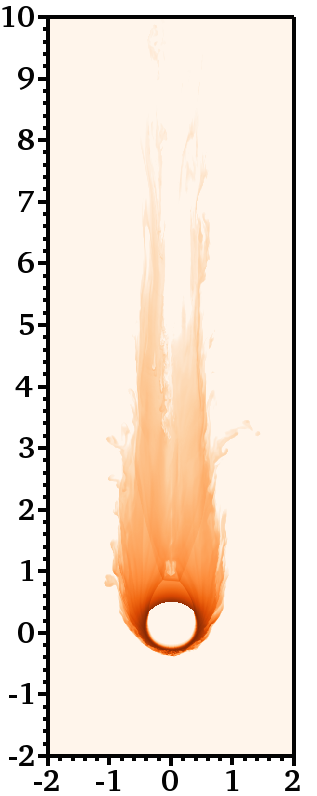}} & \resizebox{21mm}{!}{\includegraphics{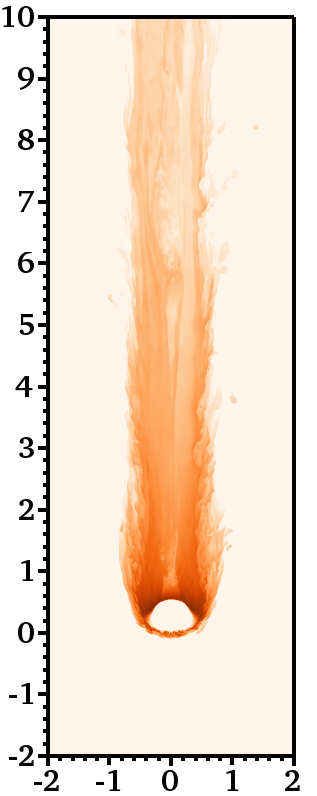}} & \resizebox{21mm}{!}{\includegraphics{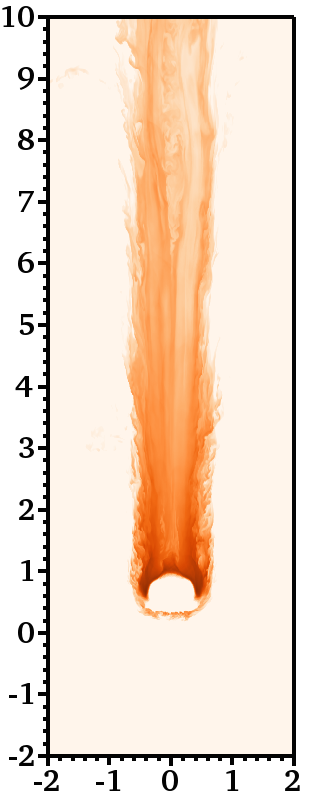}} & \resizebox{21mm}{!}{\includegraphics{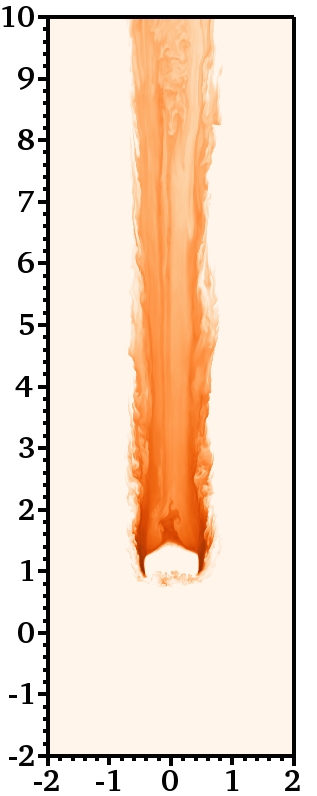}} & \resizebox{21mm}{!}{\includegraphics{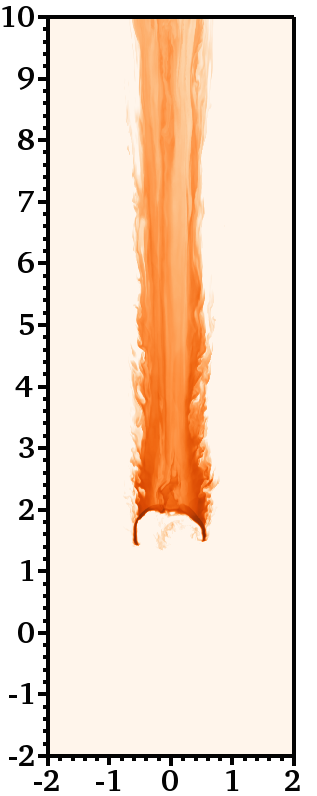}} & \resizebox{21mm}{!}{\includegraphics{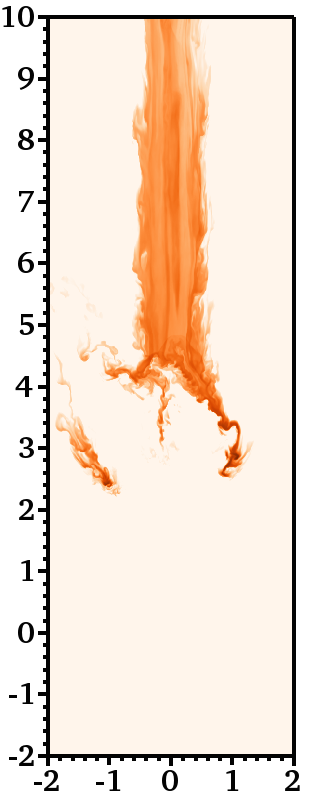}}\\       
      \resizebox{21mm}{!}{\includegraphics{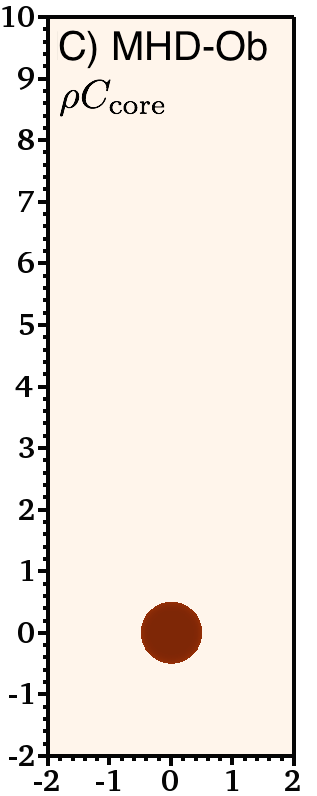}} & \resizebox{21mm}{!}{\includegraphics{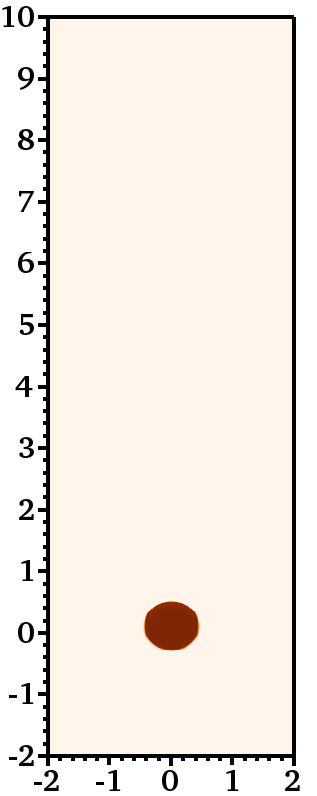}} & \resizebox{21mm}{!}{\includegraphics{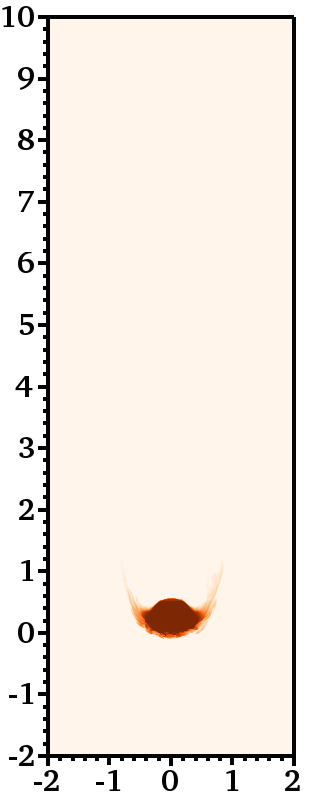}} & \resizebox{21mm}{!}{\includegraphics{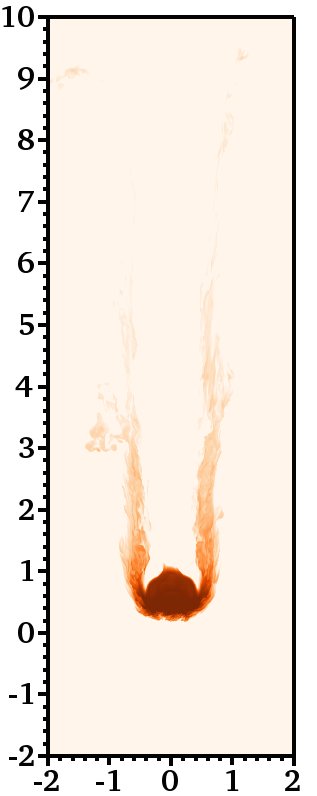}} & \resizebox{21mm}{!}{\includegraphics{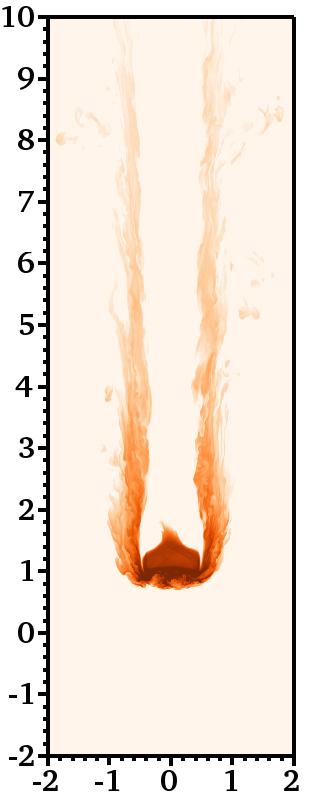}} & \resizebox{21mm}{!}{\includegraphics{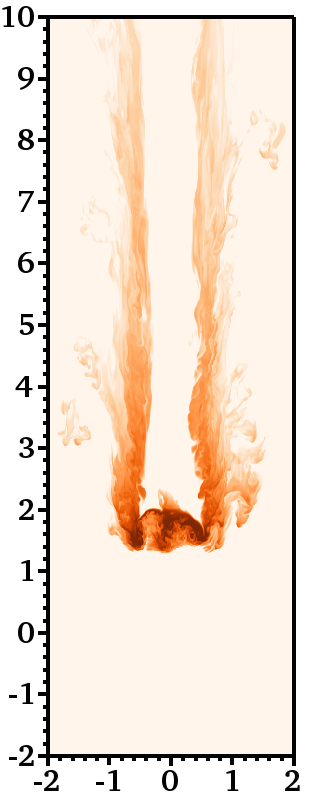}} & \resizebox{21mm}{!}{\includegraphics{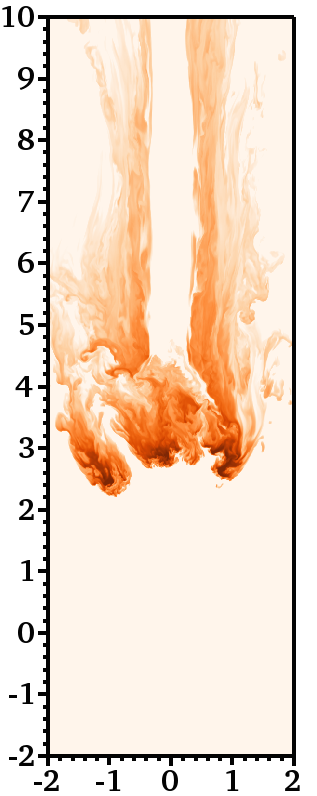}}\\
  \resizebox{!}{9mm}{\includegraphics{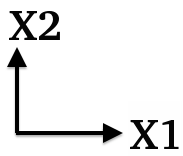}}& \multicolumn{6}{c}{\resizebox{!}{9mm}{\includegraphics{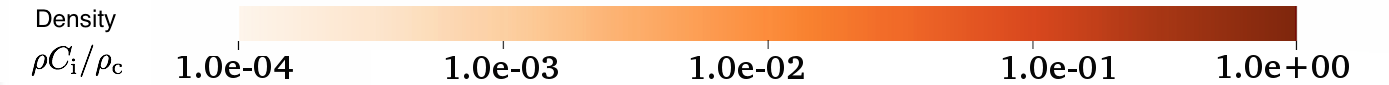}}}\Tstrut\\
  \end{tabular}
  \caption{2D slices at $X_3=0$ showing the evolution of the logarithm of the mass density in cloud/filament (Panel A), envelope/tail (Panel B), and core/footpoint (Panel C) material, normalised with respect to the initial cloud density, in model MHD-Ob at seven different times: $t/t_{\rm cc}=0$, $t/t_{\rm cc}=0.2$, $t/t_{\rm cc}=0.4$, $t/t_{\rm cc}=0.6$, $t/t_{\rm cc}=0.8$, $t/t_{\rm cc}=1.0$, and $t/t_{\rm cc}=1.2$. The time sequence shows that gas originally in the envelope of the cloud is transported downstream and deposited at the rear of the cloud to form the tail of the filament, while gas originally in the core of the cloud acts as the footpoint and late-stage outer layer of the filamentary structure (see Section \ref{sec:FilamentFormation} for further details). A similar behaviour is observed in the other models reported in this paper, i.e., in models HD, MHD-Al, MHD-Tr, MHD-Ob-S, and MHD-Ob-I.} 
  \label{Figure5}
\end{center}
\end{figure*}

In this section we address the principal aspects of the formation and evolution of filaments associated with wind-cloud systems. First, we study how the inclusion of magnetic fields affects the filament motion and its morphology as it travels through the ambient medium, and second, we analyse how magnetic fields inside and around the filament respond to that motion and change of shape. We provide a detailed description of the structure and magnetic configuration of both filaments and winds for different initial field orientations. Figure \ref{Figure4} shows the evolution of the logarithmic filament density in four different models, HD, MHD-Al, MHD-Tr, and MHD-Ob, at four different times, namely $t/t_{\rm cc}=0.2$, $t/t_{\rm cc}=0.4$, $t/t_{\rm cc}=0.8$, and $t/t_{\rm cc}=1.2$. Note that the density has been multiplied by the tracer $C_{\rm cloud}$, so that only filament gas can be seen in the images. In addition, a quarter of the volume in the rendering images has been clipped in order to show the internal structure in greater detail. A qualitative examination of Figures \ref{Figure4} and \ref{Figure5} reveals that the overall evolution of filaments associated with wind-cloud interactions comprises four stages:

\begin{figure*}
\begin{center}
  \begin{tabular}{c c}
        \textbf{Filament Tail (Cloud Envelope)} & \textbf{Filament Footpoint (Cloud Core)}\\ 
      \resizebox{80mm}{!}{\includegraphics{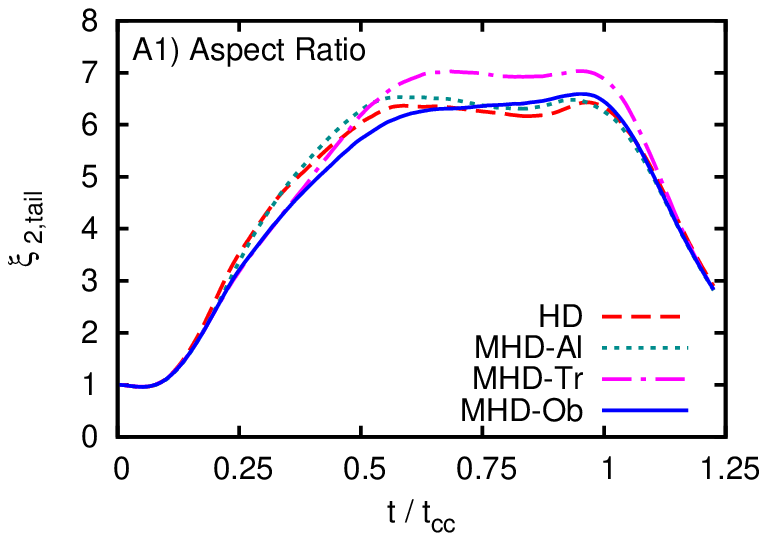}} & \resizebox{80mm}{!}{\includegraphics{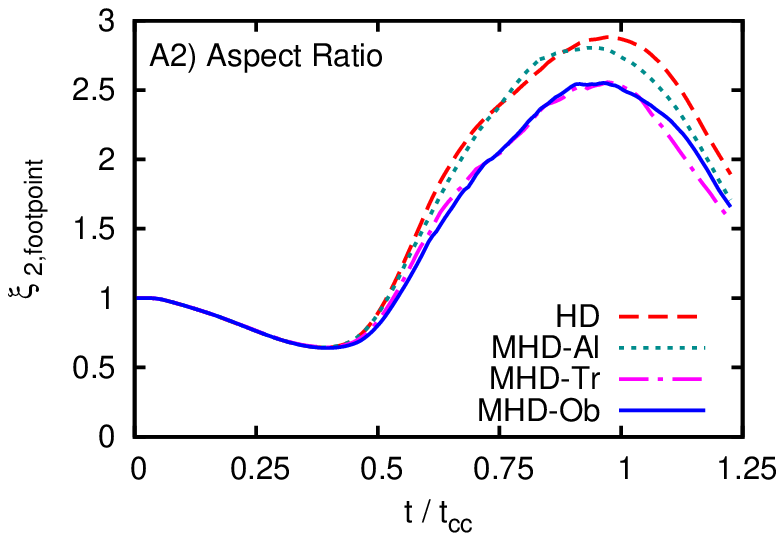}}\\       
      \resizebox{80mm}{!}{\includegraphics{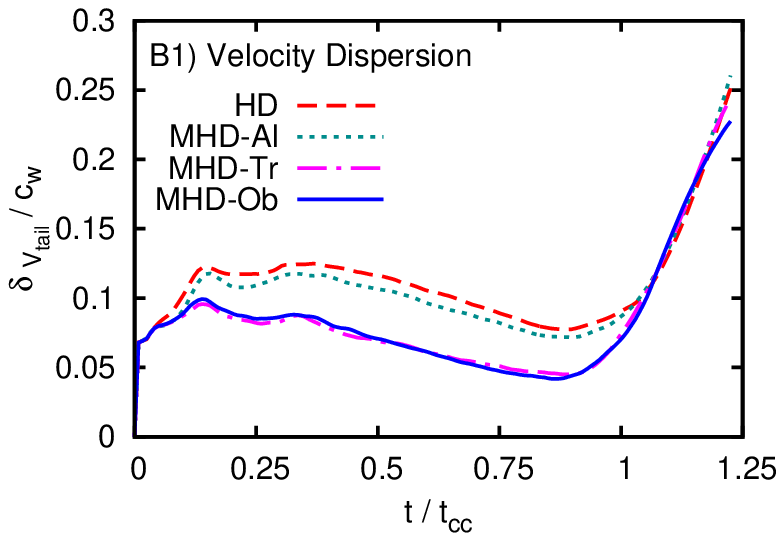}} & \resizebox{80mm}{!}{\includegraphics{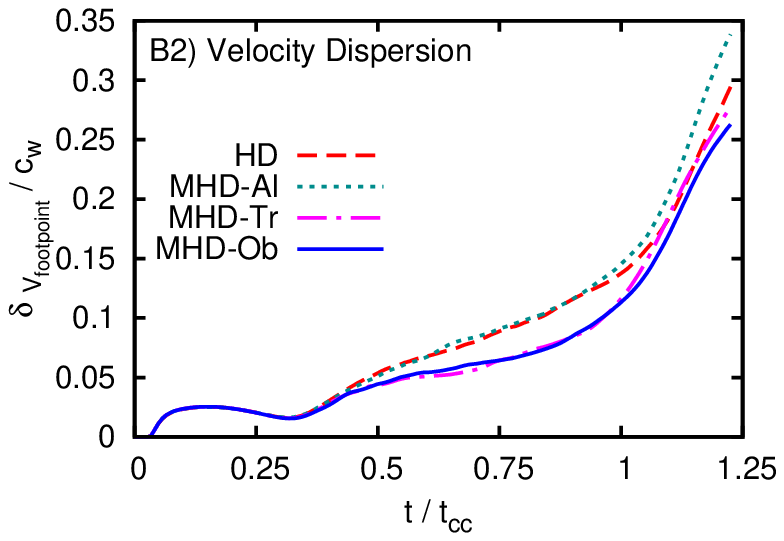}}\\
      \resizebox{80mm}{!}{\includegraphics{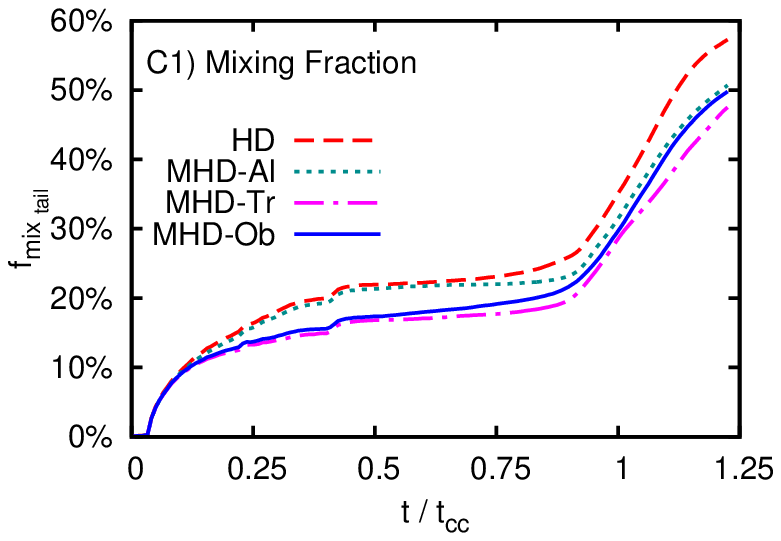}} & \resizebox{80mm}{!}{\includegraphics{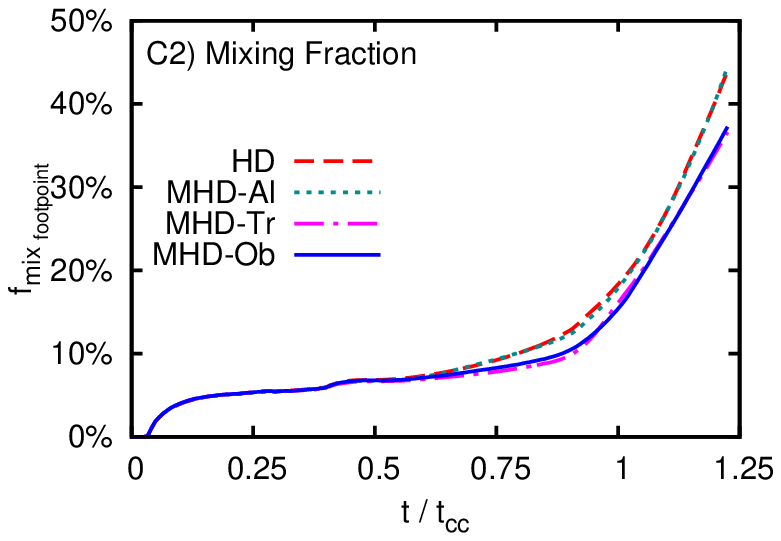}}\Tstrut\\
  \end{tabular}
  \caption{Time evolution of the filament tail (left column) and filament footpoint (right column) of three diagnostics: aspect ratio (Panels A1 and A2), transverse velocity dispersion (Panels B1 and B2), and mixing fraction (Panels C1 and C2) in models HD (dashed line), MHD-Al (dotted line), MHD-Tr (dash-dotted line), and MHD-Ob (solid line). Due to our finite simulation domain, the numerical quantities given for the aspect ratios in Panels A1 and A2 should be considered as lower limits of the actual diagnostics (see Section \ref{sec:FilamentFormation} and Appendix \ref{subsec:Appendix} for further details).} 
  \label{Figure6}
\end{center}
\end{figure*}

\begin{enumerate}
  \item \textbf{\textit{Tail formation phase:}} Filaments start to form during the stripping phase of the cloud disruption process. As the cloud is enveloped by the wind, instabilities remove material from its surface layers and the wind carries this material downstream (see Panels A and B of Figure \ref{Figure5}). The advection of envelope material follows pressure gradients, i.e., the material is deposited at the rear of the cloud at locations where the thermal pressure is low (at the beginning of the interaction, gas at the rear of the cloud is evacuated by the initial motion of the wind, leaving behind regions of relatively low gas pressure). As soon as a filament forms, we see that it is constituted by two substructures: a) a diffuse elongated tail and b) a dense footpoint, analogous to morphologies observed in cometary tails embedded in the Solar wind. As shown in Panel B of Figure \ref{Figure5}, the tail is formed by a mix of wind material and low-density material from the cloud envelope, the latter being the dominant component. The filament footpoint, on the other hand, is mainly composed of material originally in the cloud core. As shown in Panel C of Figure \ref{Figure5}, the core primarily serves as a footpoint for the newly formed filament, but it also acts as an outer layer of tail material at late times in the evolution ($t/t_{\rm cc}\gtrsim0.6$). In all our simulations, identifiable tails have fully formed by $t/t_{\rm cc}=0.2$ and they remain stable until the filaments' footpoints are broken up by disruptive instabilities. Panels A1 and A2 of Figure \ref{Figure6} show that the aspect ratios of tail and footpoint material, respectively, do not depend upon the model, i.e., similar filament elongations are expected in adiabatic simulations regardless of the initial conditions. Since envelope and core material start to leave the simulation domain at approximately $t/t_{\rm cc}=0.2$ and $t/t_{\rm cc}=0.6$, respectively, the reader should consider the numbers in these panels as lower limits after these times. In fact, a comparison of model MHD-Ob with an equivalent model in a larger domain shows that aspect ratios of $\xi_{2,\rm tail}\gtrsim12$ and $\xi_{2,\rm footpoint}\gtrsim 4$ should be contemplated (see Appendix \ref{subsec:Appendix} for further details).
  \item \textbf{\textit{Tail erosion phase:}} As can be seen in the renderings in Figure \ref{Figure4}, not only the cloud surface, but also the outermost layers of the tail are affected by dynamical instabilities. Once the tail of the filament has formed, it remains as a coherent elongated structure for most of the evolution. A shear layer emerges at the interface between the tail material and the surrounding wind. The wind velocity is approximately tangential to this boundary layer rendering the tail prone to the effects of KH instability perturbations. The degree of turbulence and the intensity of vortices in and around the tails are regulated by the KH instability, which in turn depends upon the initial magnetic field orientation as we show in Sections \ref{subsubsec:FilamentsHD}, \ref{subsec:MHD-Al}, and \ref{subsec:MHD-Tr}. Panels B1 and C1 of Figure \ref{Figure6} clearly show this dependence as models with transverse magnetic field components develop less turbulence than their counterparts. The transverse velocity dispersions in the filamentary tails evolve similarly in all simulations until $t/t_{\rm cc}\sim 0.1$, but then diverge for models with and without transverse magnetic field components, with the filaments in the latter models being more turbulent. For example, at $t/t_{\rm cc}=0.5$ the transverse velocity dispersion in the tails is $40\,\%-50\,\%$ higher in models HD and MHD-Al than in models MHD-Ob and MHD-Tr. The ratio of mixed gas to initially unmixed gas in the filamentary tails displays a similar behaviour. It rises more rapidly for models without transverse magnetic fields reaching values $5\,\%-6\,\%$ higher than those in the other pair of models. A similar trend is seen in Panels B2 and C2 corresponding to footpoint material, but the values of transverse velocity dispersions and mixing fractions are lower and the effect is delayed. The mixing fraction, for example, only increases to values higher than $10\,\%$ after the break-up time.
  \item \textbf{\textit{Footpoint dispersion phase:}} The next stage in the lifetime of a filament commences when its footpoint is dispersed by the combined effect of KH and (more importantly) RT instabilities (see the rightmost renderings of Panels A$-$D of Figure \ref{Figure4}). Panels B and C of Figure \ref{Figure5} show that the tail is attached to the original cloud and survives as a result of the support provided by the cloud core and the continuous supply of material from its envelope. When the cloud commences its expansion phase ($t/t_{\rm cc}\gtrsim0.6$), the gas in both the tail and the footpoint also expands laterally with it and the morphology of the filament changes. The roles of the envelope and the core in the cloud are inverted in the filament after this time, with low-density tail material being wrapped by dense material originally located in the footpoint. At $\sim t/t_{\rm cc}=1.0$, the structure of the filament starts to lose coherence (note e.g., how the tail and footpoint aspect ratios decrease after this time) as a result of the expanded cross-sectional area (see Panels A1 and A2 of Figure \ref{Figure6}). In association with this, both the transverse velocity dispersion and the amount of mixed gas, rapidly grow after $t/t_{\rm cc}=1.0$. Panels B1 and B2 of Figure \ref{Figure6} show that the velocity dispersions are three times higher in both the tail and footpoint at $t/t_{\rm cc}=1.2$ when compared to values before the break-up. A similar increase is seen in the values of the mixing fractions in both components (see Panels C1 and C2 of Figure \ref{Figure6}). As shown in the following sections, the cloud acceleration and the associated RT bubbles formed at the leading edge of the cloud are ultimately responsible for the break-up and dispersion of the footpoint. After the footpoint of the filament is destroyed, the tail of the filament is immersed in a highly turbulent environment and is consequently more susceptible to disruptive perturbations.
  \item \textbf{\textit{Filament free floating:}} Although some of the coherence of the filamentary structure is lost after the footpoint is dispersed, our simulations show that more diffuse tails and smaller filaments survive for longer periods of time (see  the rightmost 2D slices of Panels B and C of Figure \ref{Figure5}). We find that these structures linger either attached to cloudlets or as entrained structures moving freely in the flowing wind. In the latter scenario, the tails disconnect from the footpoints by $t/t_{\rm cc}=1.2$, suggesting that both filamentary tails and cloudlets could potentially be observed as independent structures at late-stages of wind-cloud interactions in the ISM. The tail disconnection phenomenon has been reported in both observations and simulations of cometary tails in the Solar system (see e.g., \citealt{1978ApJ...223..655N,2000Icar..148...52B,2007ApJ...668L..79V}), and of the Earth's magnetosphere (see e.g., \citealt{2012JGRA..117.6224B}). The size of our current simulation domains are, unfortunately, not sufficiently large to follow the evolution of these structures, so that further work along this line is warranted.
\end{enumerate} 

\subsection{Dynamical Instabilities}
\label{subsubsec:Instabillities}
Dynamical instabilities arise naturally in wind-cloud interactions and they not only deform the cloud but also alter the morphology of the associated filaments. Previous studies showed that four instabilities can have significant effects on the formation and evolution of wind-swept clouds. These are the KH, RT, Richtmyer-Meshkov (hereafter RM), and tearing-mode (hereafter TM) instabilities. The KH instability in our simulations results from shearing motions occurring at the boundary layer separating filament and ambient gas (see the 3D study of the KH instability by \citealt*{2000ApJ...545..475R}). The sinuosity observed in the lateral boundaries of the filamentary structures in the panels of Figure \ref{Figure4} is caused by the KH instability. The growth time-scale of the KH perturbations is given by 

\begin{dmath}
\frac{t_{\rm KH}}{t_{\rm cc}}\simeq\left[\frac{\rho'_{\rm c}\rho'_{\rm w}k_{\rm KH}^2}{(\rho'_{\rm c}+\rho'_{\rm w})^2}(v'_{\rm w}-v'_{\rm c})^2-\frac{2B^{'2}k_{\rm KH}^2}{(\rho'_{\rm c}+\rho'_{\rm w})}\right]^{-\frac{1}{2}}\frac{{\cal M}c_{\rm w}}{2r_{\rm c}\chi^{\frac{1}{2}}},
\label{KHtime}
\end{dmath}

\noindent where the primed quantities represent the values of the physical variables at the location of shear layers, and $k_{\rm KH}=\frac{2\pi}{\lambda_{\rm KH}}$ is the wavenumber of the KH perturbations \citep{1961hhs..book.....C}. In addition, the RT instability arises when the initially perturbed interface between the cloud and wind is allowed to grow under the influence of the wind-driven acceleration of dense gas (see \citealt{2007ApJ...671.1726S}). The growth time-scale of the RT perturbations is given by

\begin{dmath}
\frac{t_{\rm RT}}{t_{\rm cc}}\simeq\left[\left(\frac{\rho'_{\rm c}-\rho'_{\rm w}}{\rho'_{\rm c}+\rho'_{\rm w}}\right)ak_{\rm RT}-\frac{2B^{'2}k_{\rm RT}^2}{(\rho'_{\rm c}+\rho'_{\rm w})}\right]^{-\frac{1}{2}}\frac{{\cal M}c_{\rm w}}{2r_{\rm c}\chi^{\frac{1}{2}}},
\label{RTtime}
\end{dmath}

\noindent where the primed quantities represent the values of the physical variables at the leading edge of the cloud, $a$ is the local, effective acceleration of dense gas, and $k_{\rm RT}=\frac{2\pi}{\lambda_{\rm RT}}$ is the wavenumber of the RT perturbations \citep{1961hhs..book.....C}. Equations (\ref{KHtime}) and (\ref{RTtime}) correspond to analyses of the instabilities in the incompressible regime. Therefore, the values provided by them should be considered as indicative numbers for the growth time-scales of the KH and RT instabilities in the compressible case. Table \ref{Table3} provides reference time-scales for the growth of KH and RT instabilities, estimated from Equations (\ref{KHtime}) and (\ref{RTtime}) using simulation results as input quantities. The RM instability grows at the beginning of the interaction as a result of the impulsive acceleration produced by the refraction of the initial shock wave into the cloud (see \citealt{2012ApJ...758..126S}; \citealt*{PhysRevLett.111.205001} for recent studies). Bubbles and spikes are characteristic of both the RT and RM instabilities (see \citealt{2011NIMPA.653....2K}), however, the exponential growth rate of the RT modes makes the linearly-growing RM instability only important at the very early stages of the evolution. In MHD models, a fourth instability emerges, namely the TM instability, which grows when oppositely directed magnetic field lines are pushed together, leading to magnetic reconnection (see \citealt{1979cmft.book.....P}). The resulting morphology and endurance of filaments are determined by the growth rates of these instabilities, which in turn heavily depend upon whether or not the medium is magnetised and how the field is oriented when present. It is, therefore, convenient to describe some details of the evolution in each model independently. 

\subsection{Filaments in hydrodynamic models}
\label{subsubsec:FilamentsHD}
We commence our analysis with the purely hydrodynamic model, HD. As mentioned above, erosion of filament gas primarily occurs due to the emergence of the KH instability at the ambient-filament boundary layers, but the time-scales over which the KH instability grows depend upon the kinetic and magnetic conditions in those layers. If magnetic fields are absent or are dynamically unimportant (as in model HD), the growth time of a KH perturbation with wavenumber, $k_{\rm KH}$, is solely determined by the density contrast between both media, $\chi$, and the relative velocity at the boundary layer, $(v'_{\rm w}-v'_{\rm c})$ (i.e., by the first term in Equation \ref{KHtime}). Thus, higher relative velocities accelerate the KH growth, while higher density contrasts retard it. As the cloud in our models is initially at rest and the denser regions of the cloud only interact with the wind at later stages, short-wavelength KH instability modes emerge early in the evolution. The growth rate of these modes is fast at the beginning, but slows down as the simulation progresses (see reference time-scales in Table \ref{Table3}). As can be seen in Panel A of Figure \ref{Figure4}, the filament density in model HD presents a tower-like structure that remains unchanged for most of the evolution. The interior is highly turbulent with mass-weighted velocity dispersions in the transverse direction reaching $\sim 0.1$ of the wind speed in both the tail and its footpoint at $t/t_{\rm cc}=1.0$ (see Panels B1 and B2 of Figure \ref{Figure6}).\par

The HD filament has a density gradient dropping off from the $X_2$ axis outwards, except for the region immediately adjacent to the rear side of the cloud in which rarefaction effects vacate the gas and form a low-pressure cavity. During the cloud expansion phase described in the previous section (for $0.5<t/t_{\rm cc}<1.0$), the cloud core becomes W-shaped when dense material is sucked from the cloud to occupy the cavity (see the third evolutionary stage of Panel A of Figure \ref{Figure4}). The wavy structure of the filamentary tail remains stable and coherent during the stripping and expansion phases of cloud evolution. However, stability is lost when a combination of highly-disruptive KH and RT modes emerge and break-up the footpoint. In the absence of magnetic fields the growth time of the RT instability is determined by the effective acceleration of dense gas, $a$, and the RT instability mode wavenumber, $k_{\rm RT}$. By $t/t_{\rm cc}\sim 1.0$, i.e., towards the end of the expansion phase, the enlarged cross section of the filament footpoint caused by the internal heating of cloud gas, and the emergence of long-wavelength RT modes create low-density bubbles at the front of the cloud. These bubbles penetrate the denser layers of the cloud quite rapidly, break up the cloud into at least three cloudlets (see the fourth evolutionary stage of Panel A of Figure \ref{Figure4}) and disrupt the filament while doing so. Although we did not follow the evolution of these structures beyond $t/t_{\rm cc}=1.2$, we expect the remaining cloudlets to also expand and mix further with ambient gas.\par

\begin{table}\centering
\caption{Column 1 indicates the model. Columns 2 and 3 report the time-scales for the growth of the KH and RT perturbations, respectively. The time-scales are estimated semi-analytically, assuming $\lambda_{\rm KH}=\lambda_{\rm RT}=1.0\:r_{\rm c}$ in Equations (\ref{KHtime}) and (\ref{RTtime}).}
\begin{tabular}{c c c c c c c}
\hline
\textbf{(1)} & \textbf{(2)} & \textbf{(3)}\Tstrut\\
\textbf{Model} & $t_{\rm KH}/t_{\rm cc}$ & $t_{\rm RT}/t_{\rm cc}$\Bstrut \\ \hline
HD & $0.03$ & $0.18$\Tstrut \\
MHD-Al & $0.03$ & $0.19$\\
MHD-Tr & $0.11$ & $0.15$\\
MHD-Ob & $0.10$ & $0.15$\\
MHD-Ob-S & $0.41$ & $0.14$\\
MHD-Ob-I & $0.90$ & $0.25$\Bstrut\\\hline
\end{tabular}
\label{Table3}
\end{table}

\begin{figure*}
\begin{center}
  \begin{tabular}{c c c c}
    \resizebox{40mm}{!}{\includegraphics{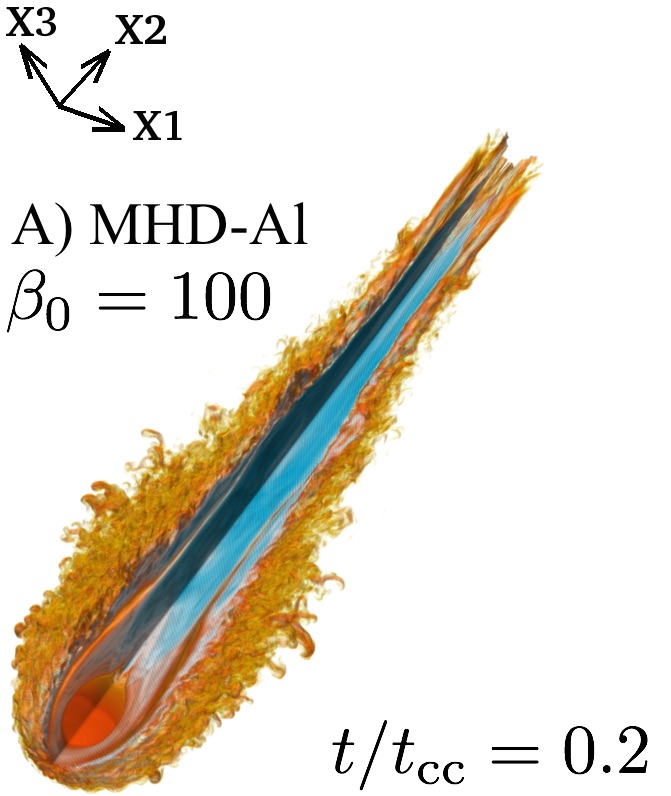}} & \resizebox{40mm}{!}{\includegraphics{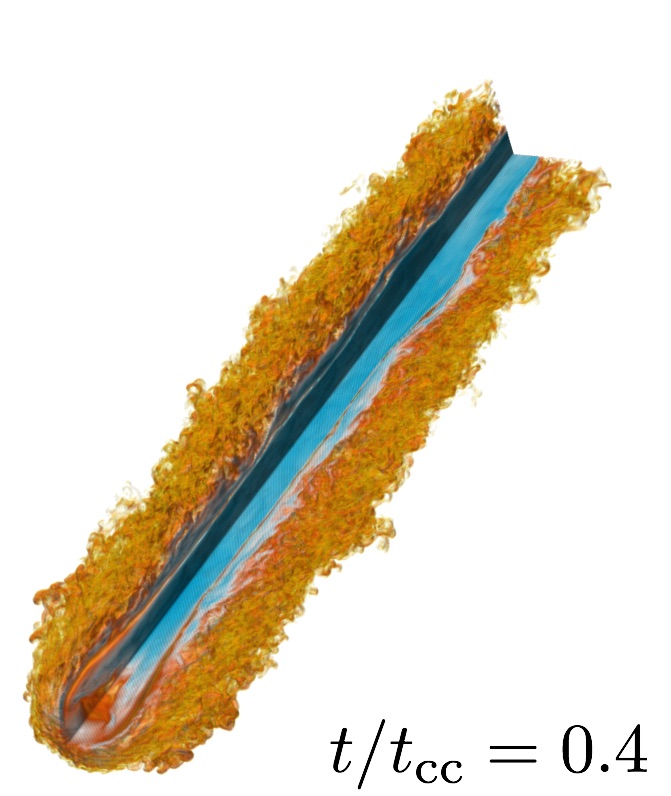}} & \resizebox{40mm}{!}{\includegraphics{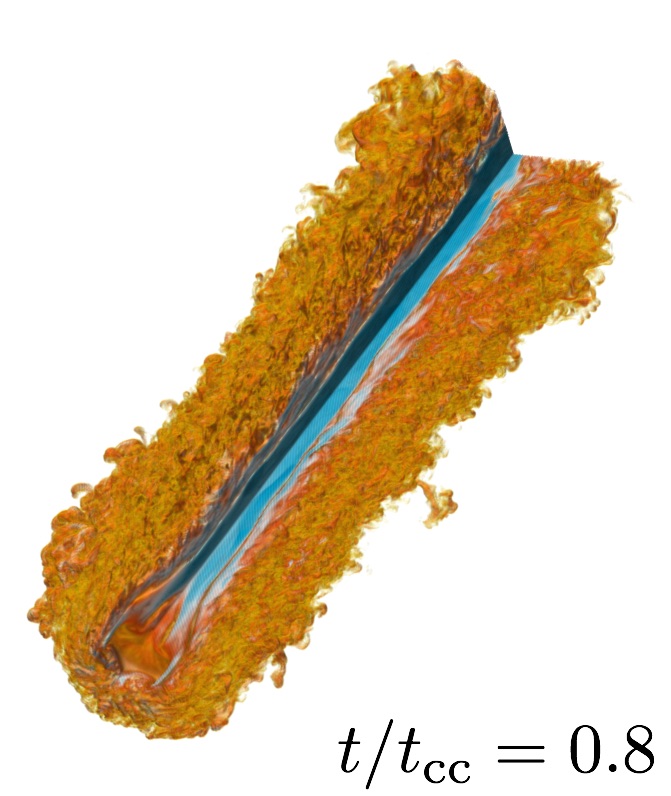}} & \resizebox{40mm}{!}{\includegraphics{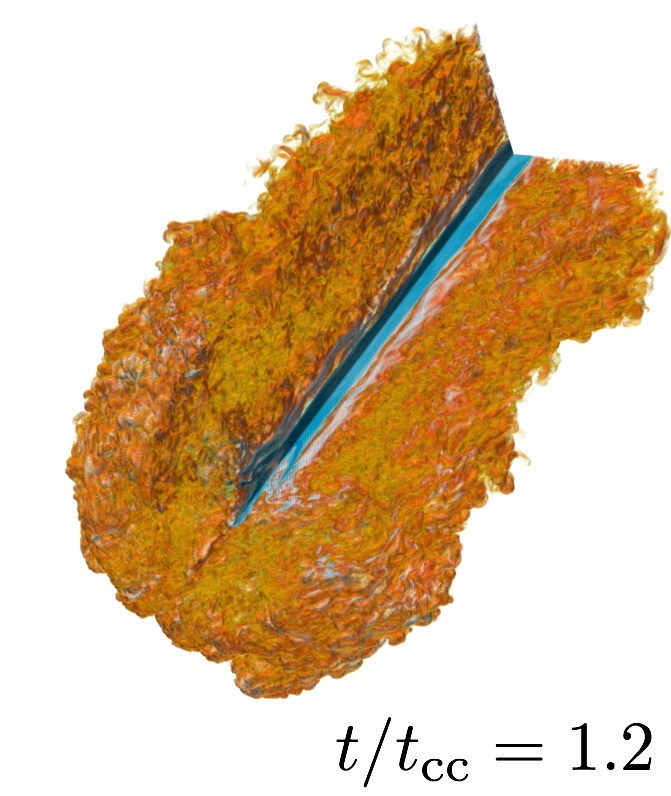}}\\
    \resizebox{40mm}{!}{\includegraphics{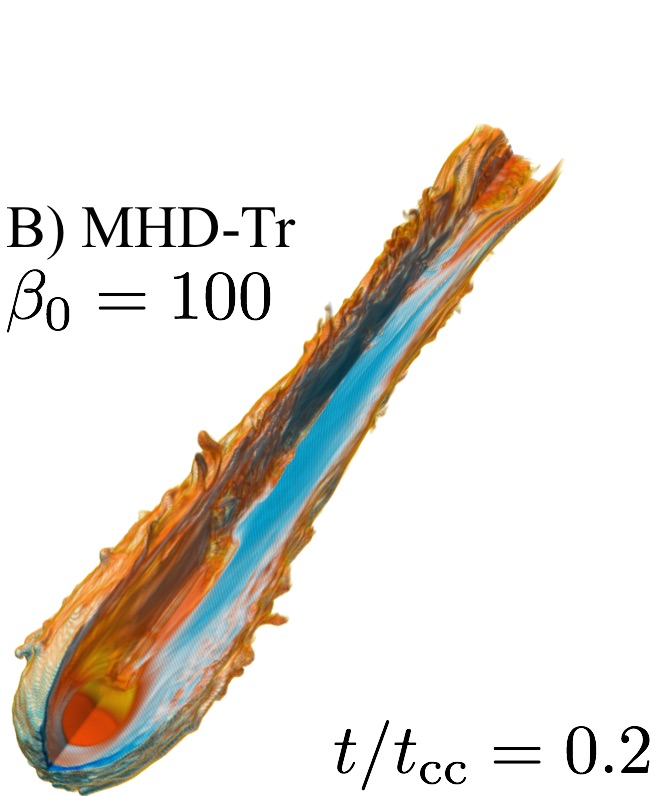}} & \resizebox{40mm}{!}{\includegraphics{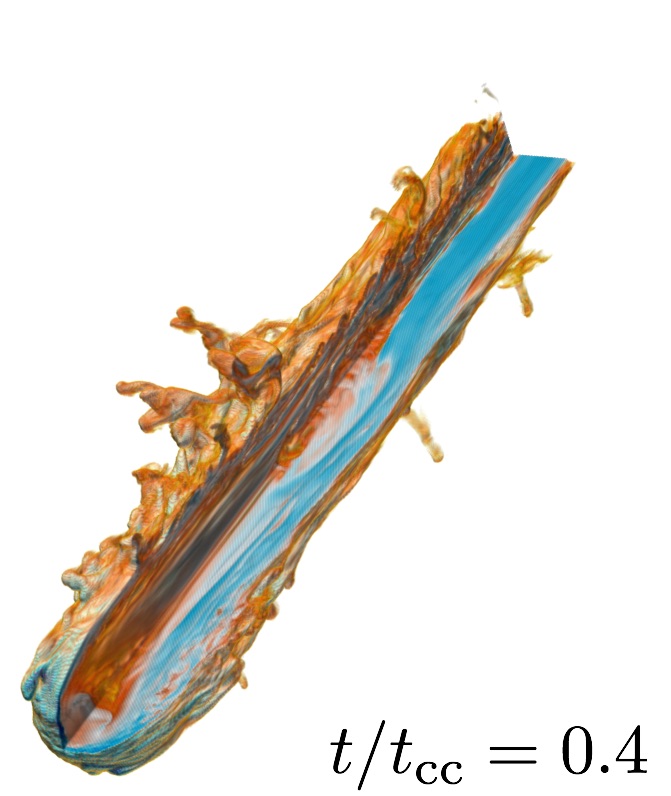}} & \resizebox{40mm}{!}{\includegraphics{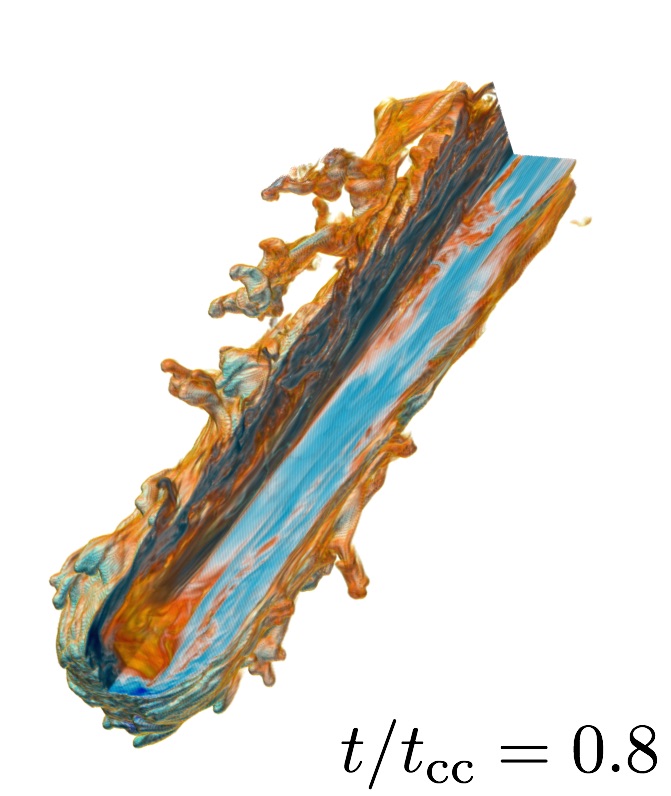}} & \resizebox{40mm}{!}{\includegraphics{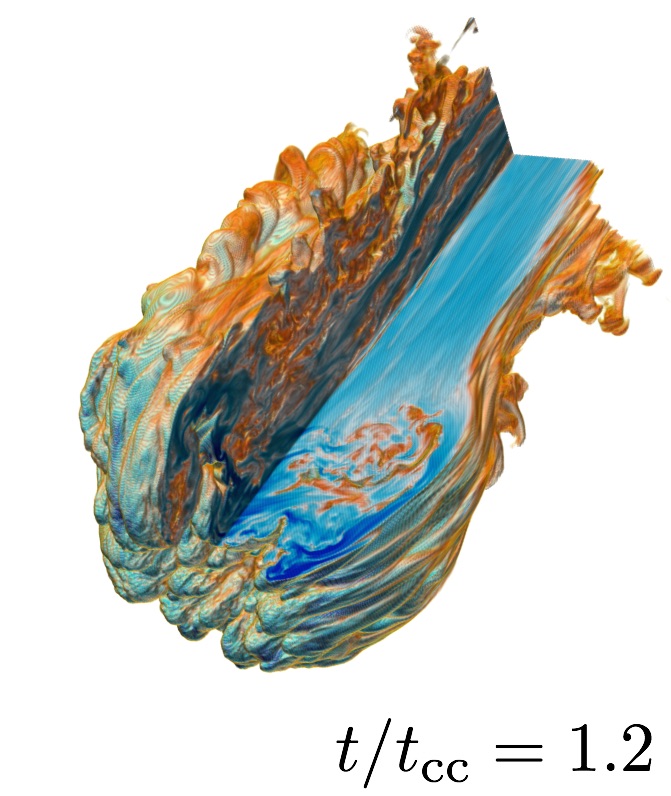}}\\
    \resizebox{40mm}{!}{\includegraphics{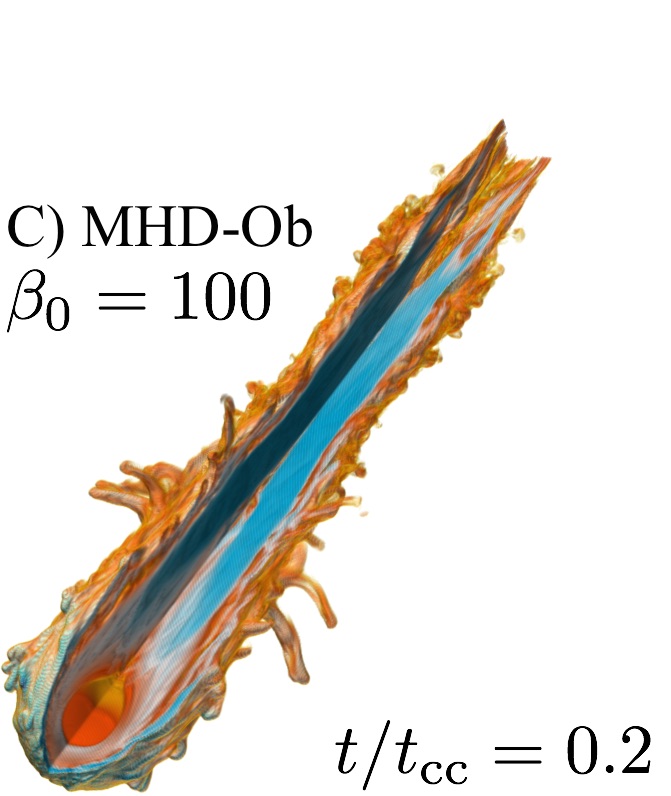}} & \resizebox{40mm}{!}{\includegraphics{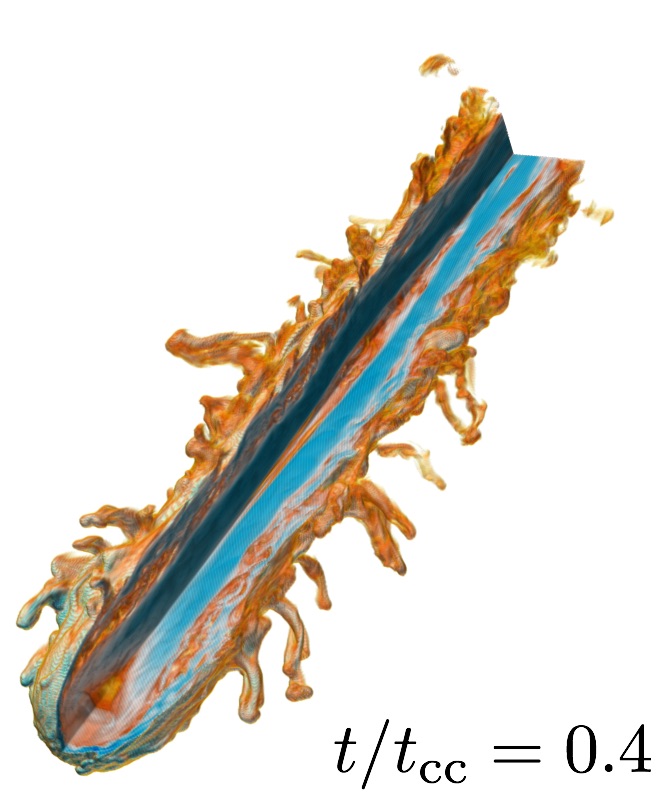}} & \resizebox{40mm}{!}{\includegraphics{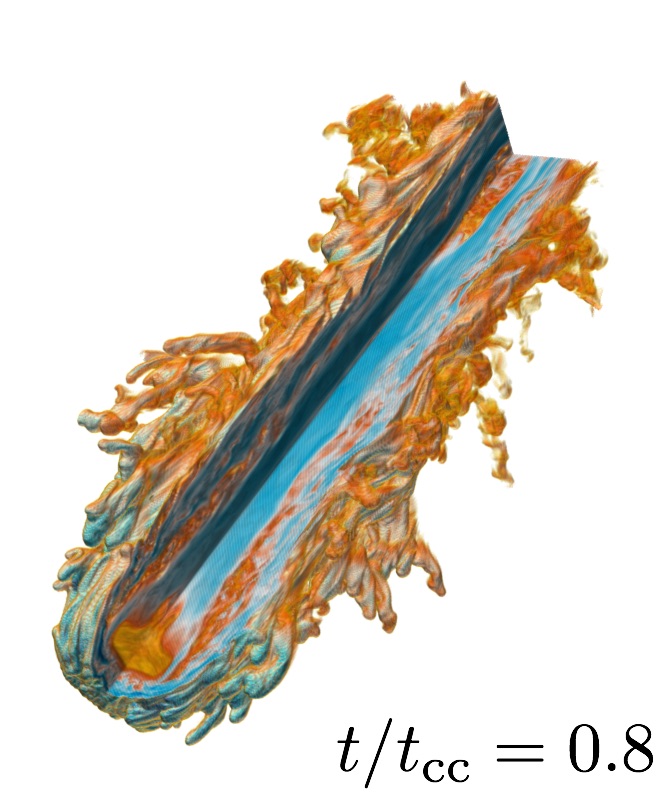}} & \resizebox{40mm}{!}{\includegraphics{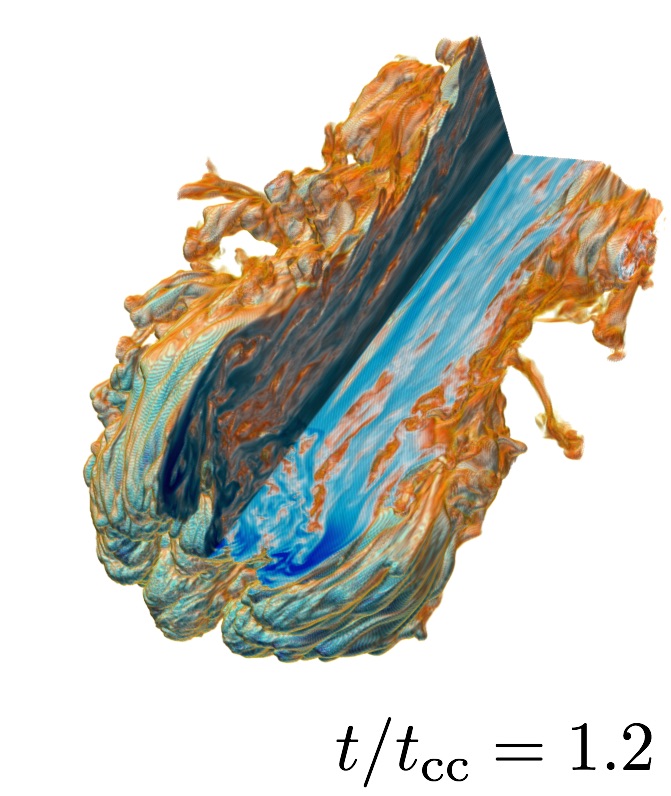}}\\
    \resizebox{40mm}{!}{\includegraphics{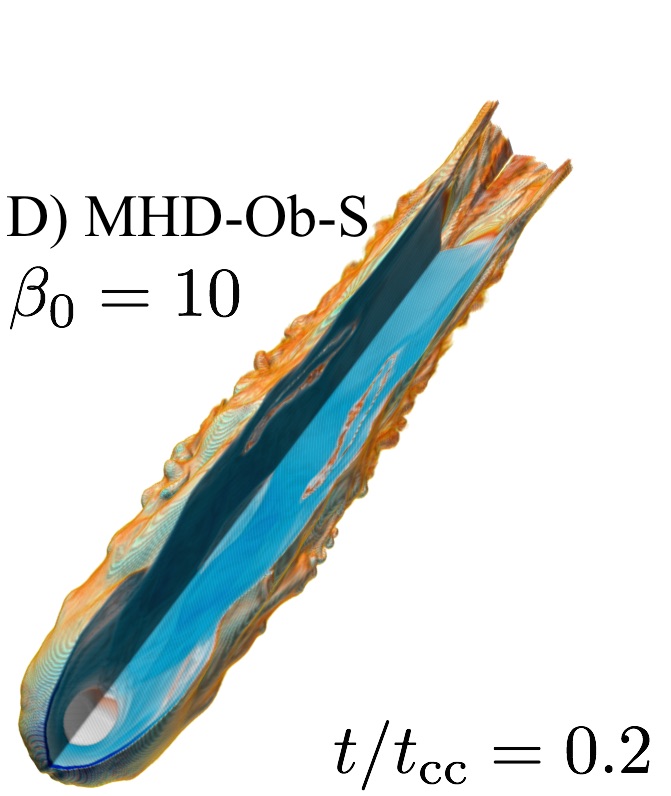}} & \resizebox{40mm}{!}{\includegraphics{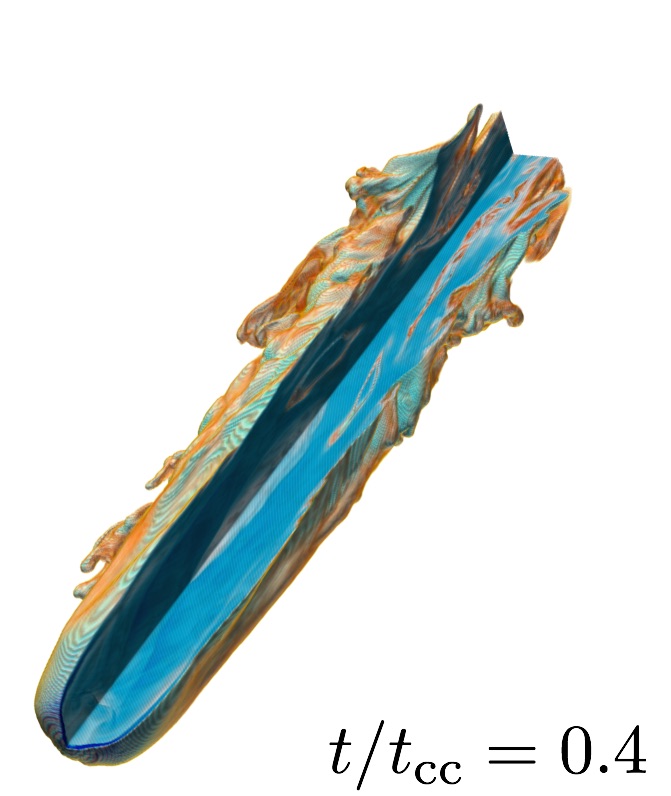}} & \resizebox{40mm}{!}{\includegraphics{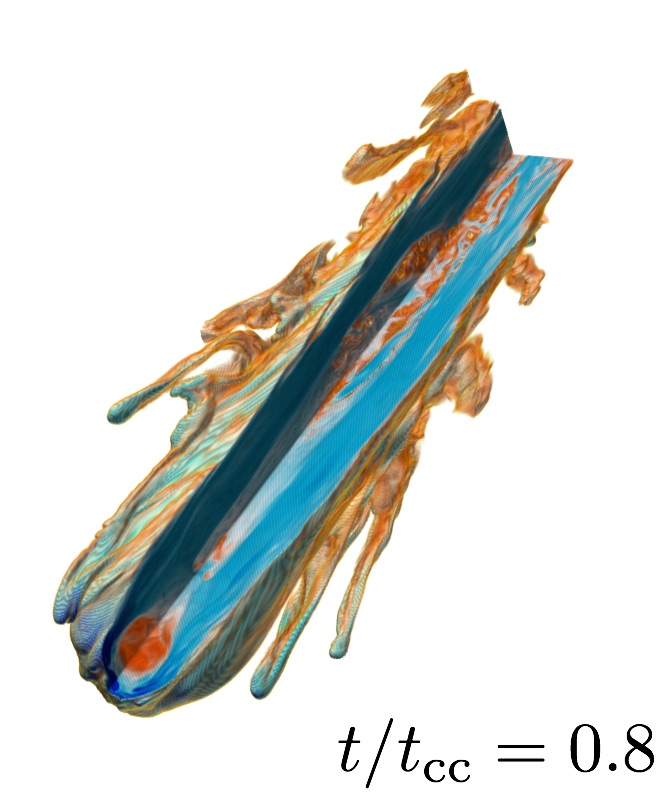}} & \resizebox{40mm}{!}{\includegraphics{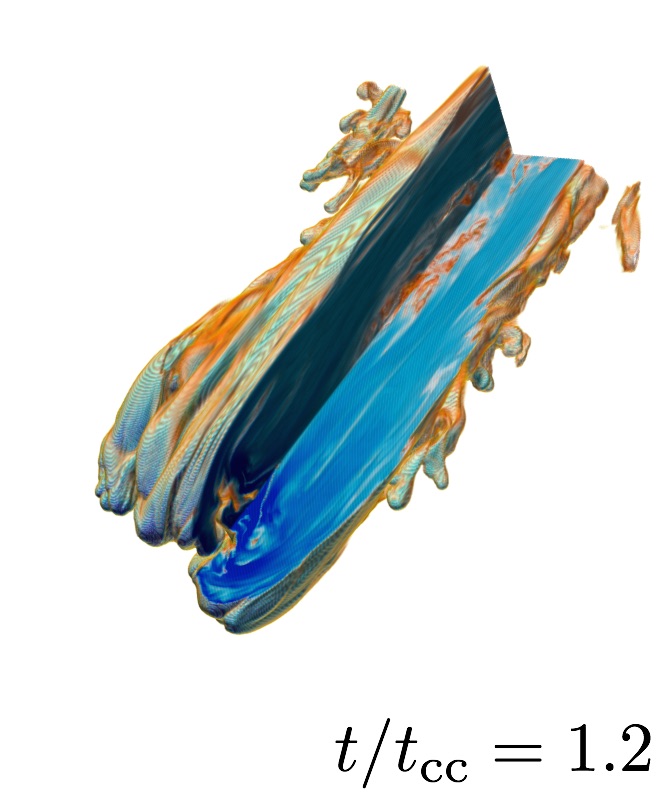}}\Dstrut\\
 \multicolumn{4}{c}{\resizebox{!}{8mm}{\includegraphics{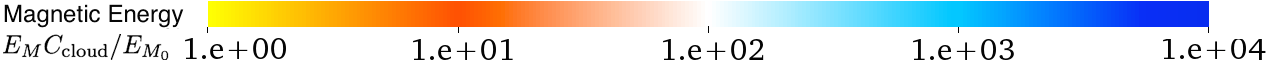}}}\\    
  \end{tabular}
  \caption{3D volume renderings of the logarithm of the magnetic energy density in filaments, normalised with respect to the initial magnetic energy density in the cloud, at four different times: $t/t_{\rm cc}=0.2$, $t/t_{\rm cc}=0.4$, $t/t_{\rm cc}=0.8$, and $t/t_{\rm cc}=1.2$. Panels A, B, and C show the evolution of wind-cloud systems with the magnetic field aligned, transverse, and oblique to the wind direction, respectively. Panel D shows the evolution with a slightly stronger initial oblique field (i.e., $\beta=10$). Note that a quadrant has been clipped from the renderings to show the interior of the tails. Magnetic field components aligned with the direction of the wind favour the formation of strongly-magnetised, rope-like structures in the tails. Magnetic filed components transverse to the streaming direction, on the other hand, form reconnection-prone current sheets. Increasing the strength of the initial magnetic field leads to further collimation of the gas in the tail, and accelerates the emergence of disruptive RT modes and subsequent break-up of the cloud core.} 
  \label{Figure7}
\end{center}
\end{figure*}

\subsection{The role of magnetic field components aligned with the flow}
\label{subsec:MHD-Al}
Magnetic fields are known to provide additional stability to ISM clouds in some circumstances (see \citealt{2011ApJ...730...40P,2012ApJ...761..156F,2015MNRAS.450.4035F} for recent discussions on the role of magnetic fields in cloud stability, MHD turbulence in ISM clumps, and star formation). However, previous studies show that, for some field orientations, magnetic fields actually help to disrupt clouds (\citealt{1999ApJ...527L.113G,2000ApJ...543..775G}). We find concordance with previous works in our simulations as we explain below. Figure \ref{Figure7} shows the evolution of the magnetic energy in filament material in four different MHD models, namely: MHD-Al, MHD-Tr, MHD-Ob, and MHD-Ob-S at four different times: $t/t_{\rm cc}=0.2$, $t/t_{\rm cc}=0.4$, $t/t_{\rm cc}=0.8$, and $t/t_{\rm cc}=1.2$.\par

When magnetic fields are present in the system, the growth rates of both the KH and RT instabilities depend on the field strength and orientation. If the magnetic field is oriented transverse to the layer, it does not affect the growth of the KH instability, while a component in the direction of streaming does suppress it. If the magnetic field at shear layers is weak, however, suppression is minimal (setting $B\simeq0$ in Equation (\ref{KHtime}) leads to our previous expression for the HD model). Strong magnetic fields suppress this instability regardless of the perturbation wavelength considered. As the magnetic field strength at the shear layers separating filament and wind gas is weak, suppression of the KH instability is not significant in model MHD-Al. The time-scales for the growth of KH perturbations with wavelengths comparable to the cloud radius in this model is of the order of $t/t_{\rm cc}\sim0.03$ (see Table \ref{Table3}). As a result, the filamentary tail in model MHD-Al is more turbulent than its counterparts in models MHD-Tr and MHD-Ob, and its turbulent profile is comparable to that of the HD model (see the behaviour of the velocity dispersion and mixing fraction in Panels B1, B2, C1, and C2 of Figure \ref{Figure6}, for example).\par

Despite this similarity, the presence of the magnetic field does affect the internal structure of the resulting filament. In the MHD-Al scenario, the field lines located at the rear of the cloud are compressed by the in-situ convergence of oppositely-directed wind flows. When wind gas has fully enveloped the cloud, different fronts converge at the rear of the cloud and advect the field lines towards the $X_2$ axis. This mechanism creates a linear region of high magnetic pressure that resembles flux ropes in the Solar corona. The location and extension of the rope-like structure can be seen in Panel B of Figure \ref{Figure4} and Panel A of Figure \ref{Figure7} as a low-density, high-magnetic-energy region (similar structures were reported by \citealt{1994ApJ...433..757M} and \citealt{2008ApJ...680..336S} for different sets of initial conditions). As the simulation progresses, the rope is further confined by the turbulent pressure of the surrounding gas. The W-shaped core previously observed in the model HD is also visible here (see the third evolutionary stage in Panel B of Figure \ref{Figure4}), but it is less pronounced as dense gas entrainment in the tail is impeded by the high magnetic pressure at the rope's upstream end.\par

The reference growth time-scale of the RT instability in this case is also similar to that in model HD. The field lines at the front of the cloud are neither stretched nor compressed, so the magnetic pressure at this location is not high enough to alter the development of RT perturbations. The growth time of RT modes in this model is of the order of $t/t_{\rm cc}\sim0.19$ (see Table \ref{Table3}), yielding a similar value as for the HD model. Indeed, models HD and MHD-Al are equally dominated by short-wavelength vorticity. Swirling motions not only strip material from the sides of the cloud but also lead to the formation of reconnection-prone topologies in the filament body. This can be seen in Panels A1 and A2 of Figure \ref{Figure8} where the evolution of the average plasma beta indicates that the thermal pressure can be three and four orders of magnitude higher than its magnetic counterpart in tail and footpoint material, respectively. From $t/t_{\rm cc}\sim0.3$ onwards the steady annihilation of the internal magnetic field keeps the plasma beta roughly constant.\par

In addition, Panels B1 and B2 of Figure \ref{Figure8} reveal that the magnetic energy in the tail increases faster than in the footpoint, with the magnetic energy already being enhanced $60$ times by $t/t_{\rm cc}=0.25$. This behaviour can be attributed to line stretching occurring at the back of the cloud. In the core, on the other hand, the magnetic energy only starts to grow after the compression phase finalises, i.e., at $t/t_{\rm cc}=0.3$. As the core expands, the associated stretching at its sides then leads to field amplification, with the magnetic energy being $\sim 100$ times higher than the initial value (at $t/t_{\rm cc}=1.2$). Even though the average plasma beta in the filamentary tail and footpoint are higher than $100$, we note that the region where the flux rope is located inside the filament body does contain gas with plasma betas of the order of $\beta\sim10$ or even less throughout the simulation. Panel A of Figure \ref{Figure9} shows the morphology of the rope in the filamentary tail at different evolutionary stages. Note also that the turbulent pressure in the surroundings of this structure increases, making the rope thinner as time progresses. Panel A of Figure \ref{Figure9} also reveals that: a) the magnetic field vectors follow the direction of the flux rope forming a coherent linear structure along the direction of streaming (i.e., along the $X_2$ axis); and b) the filamentary tail survives the cloud break-up phase to become an independent structure seemingly detached from its footpoint.

\subsection{The role of magnetic field components transverse to the flow}
\label{subsec:MHD-Tr}
We now consider the evolution in the case where the initial magnetic field has a component perpendicular to the wind direction, such as in model MHD-Tr. The evolutionary stages in Panel B of Figure \ref{Figure7} show that when the initial magnetic field has a transverse component, the degree of stripping is lower than in models HD and MHD-Al. The shear layers separating wind and filament gas are less affected by vortical motions in this model than in its counterparts. In principle, the KH instability should not be affected by a magnetic field component transverse to the direction of streaming, such as in this case. However, the change in topology of the magnetic field around the filament results in the suppression of KH perturbations at boundary layers. Despite the fact that the field orientation is initially transverse, as time progresses, advection of the magnetic field lines at the leading edge of the cloud eventually aligns the field at the sides of the filament with the direction of streaming. The lines surround the cloud body without slipping through its sides (i.e., they become stretched), and KH instability modes are suppressed or retarded as a consequence.\par 

\begin{figure*}
\begin{center}
  \begin{tabular}{c c}
        \textbf{Filament Tail (Cloud Envelope)} & \textbf{ Filament Footpoint (Cloud Core)}\\ 
      \resizebox{80mm}{!}{\includegraphics{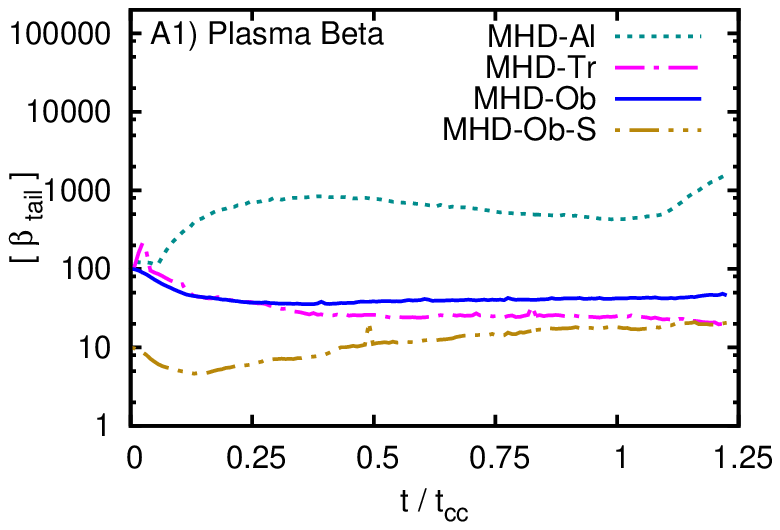}} & \resizebox{80mm}{!}{\includegraphics{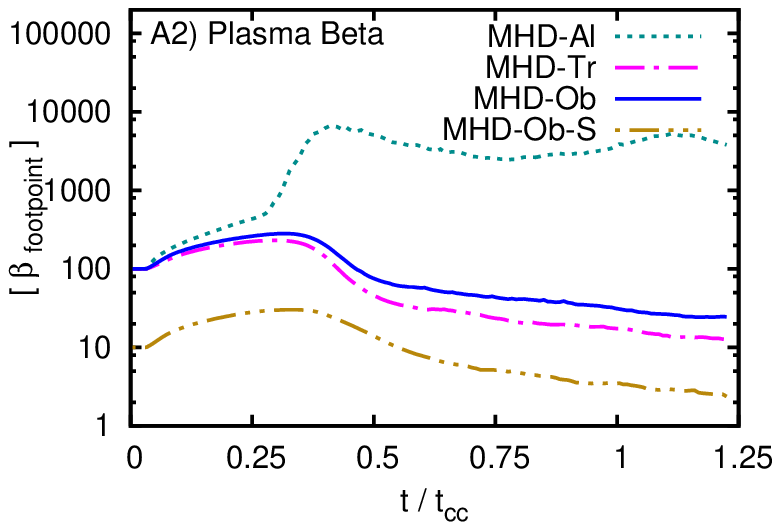}}\\
      \resizebox{80mm}{!}{\includegraphics{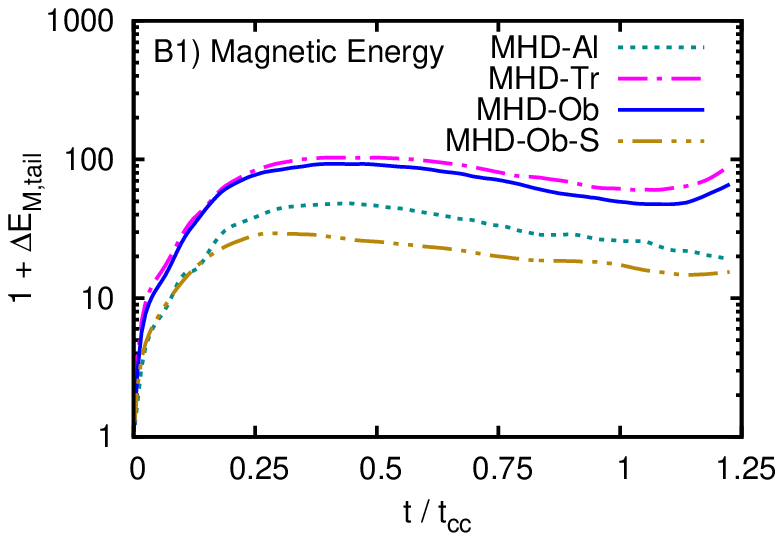}} & \resizebox{80mm}{!}{\includegraphics{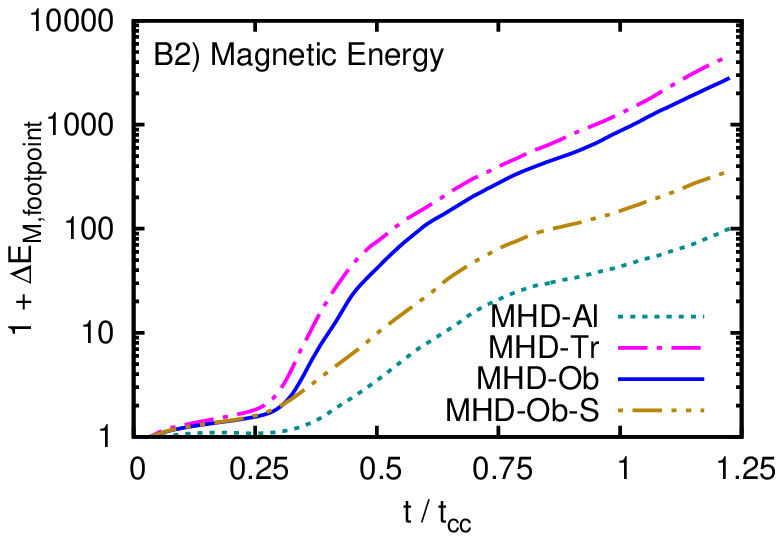}}\Tstrut\\
  \end{tabular}
  \caption{Time evolution of the plasma beta (Panels A1 and A2) and the magnetic energy enhancement (Panels B1 and B2) in the filaments' tails and footpoints in four different MHD models, including MHD-Al (dotted line), MHD-Tr (dash-dotted line), MHD-Ob with $\beta=100$ (solid line), and MHD-Ob-S with $\beta=10$ (double-dot-dashed line). The enhancement of magnetic energy is dominated by stretching of field lines in the perimeters of the filamentary tail. Folding of magnetic field components transverse to the direction of streaming leads to further amplification as evidenced in models MHD-Tr and MHD-Ob (see Sections \ref{subsec:MHD-Tr} and \ref{subsec:MHD-Ob}). Due to our finite simulation domain, the numerical quantities given for the plasma beta and the magnetic energy enhancement in tail material in Panels A1 and B1 should be considered as upper and lower limits, respectively, of the actual diagnostics after $t/t_{\rm cc}=0.25$ (see Appendix \ref{subsec:Appendix} for further details).}
  \label{Figure8}
\end{center}
\end{figure*}

The reference time-scale for the development of KH perturbations with wavelengths comparable to the cloud radius is an order of magnitude higher in model MHD-Tr than in models with null or aligned magnetic field components (see Table \ref{Table3}). Contrary to what we found in the previously analysed models, in this case the KH time-scale is regulated by both terms in Equation (\ref{KHtime}), owing to the magnetic pressure in shear layers becoming comparable to the wind ram pressure. It is, therefore, expected that the strong magnetic tension of field lines parallel to the direction of the wind suppresses the KH instability at these boundary layers. This effect radically changes the degree of turbulence in the filaments and has important implications for the generation of vortical motions in the wind. In model MHD-Tr, small-scale swirling vortices are damped as a result of field tension enhancement at the boundary between wind and filament gas (a magnetic shield forms around the filament). Hence, the downstream flow becomes more laminar, i.e., it is dominated by large-scale vortices. On the other hand, the formation of a magnetic shield around the filamentary tail effectively decreases the amount of stripping at small scales, and the tail becomes less turbulent than in models HD and MHD-Al. Panels B1 and B2 of Figure \ref{Figure6} reveal that the transverse velocity dispersions in both tail and footpoint material are $\sim 25\,\%$ lower in model MHD-Tr than in models HD and MHD-Al (at $t/t_{\rm cc}=1.0$). As a result, magnetic field annihilation triggered by turbulent motions in the filament body is prevented in the evolution of this model or is unimportant to the dynamics, if present.\par

\begin{figure*}
\begin{center}
  \begin{tabular}{c c c c c c c}
       \multicolumn{7}{c}{\resizebox{!}{4.1mm}{\includegraphics{bar-times.png}}}\Tstrut\Bstrut\\    
      \resizebox{21mm}{!}{\includegraphics{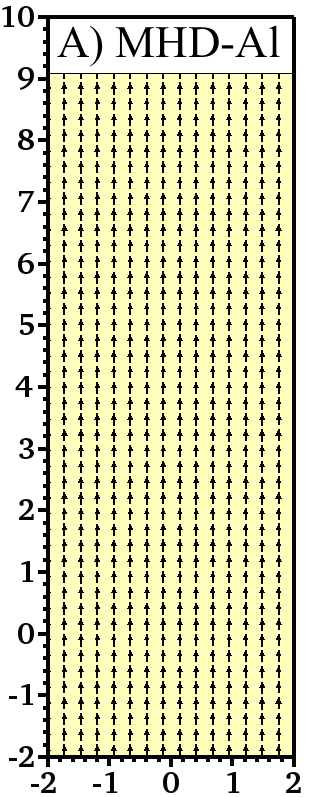}} & \resizebox{21mm}{!}{\includegraphics{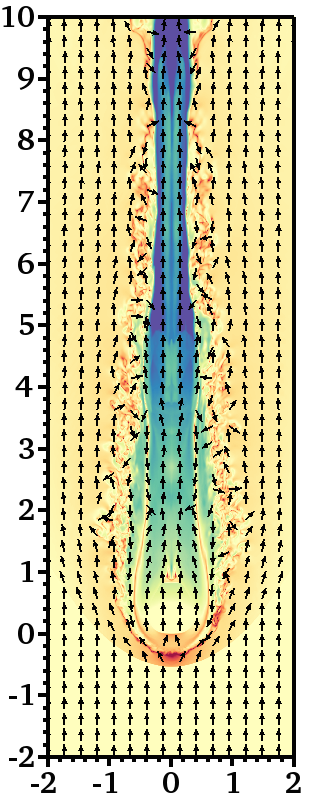}} & \resizebox{21mm}{!}{\includegraphics{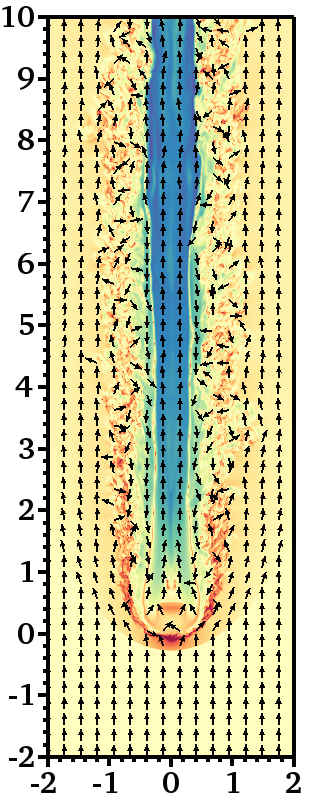}} & \resizebox{21mm}{!}{\includegraphics{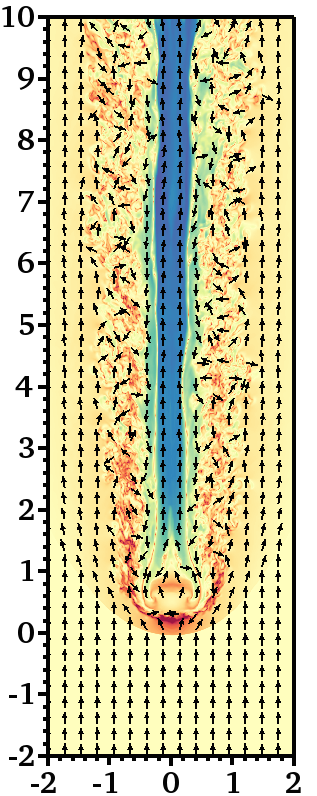}} & \resizebox{21mm}{!}{\includegraphics{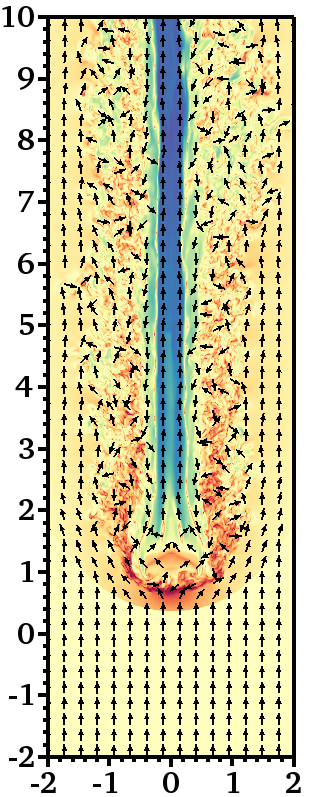}} & \resizebox{21mm}{!}{\includegraphics{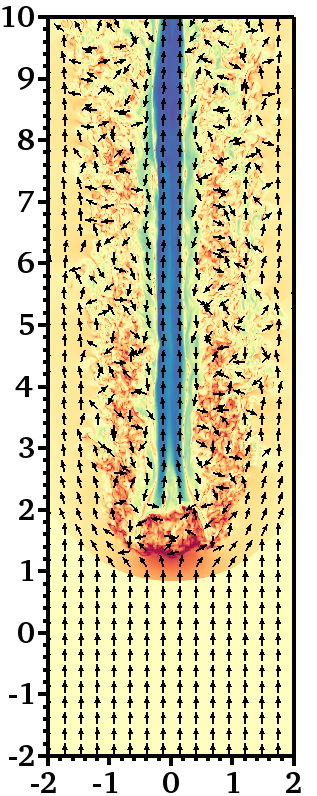}} & \resizebox{21mm}{!}{\includegraphics{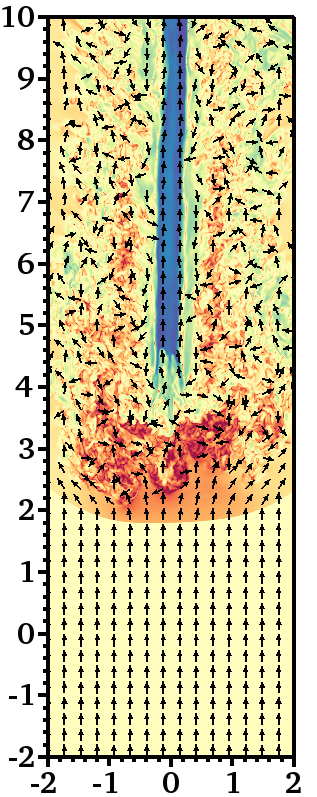}}\\
      \resizebox{21mm}{!}{\includegraphics{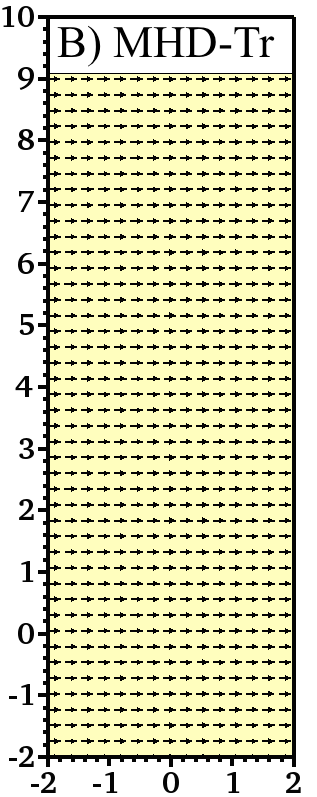}} & \resizebox{21mm}{!}{\includegraphics{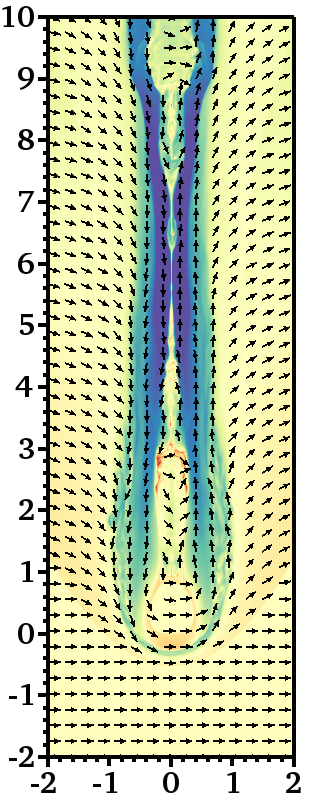}} & \resizebox{21mm}{!}{\includegraphics{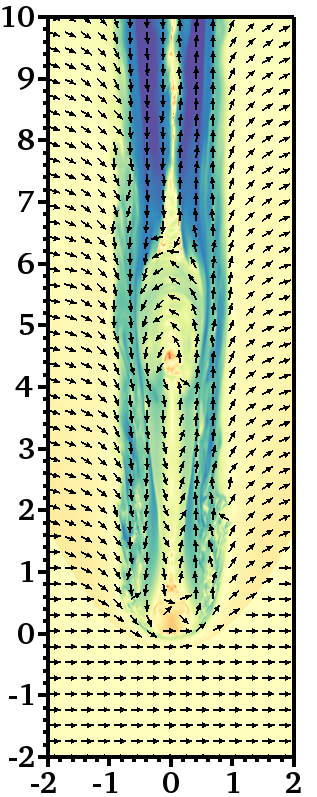}} & \resizebox{21mm}{!}{\includegraphics{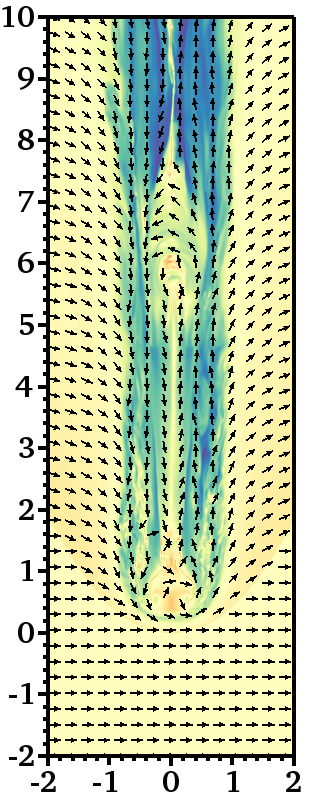}} & \resizebox{21mm}{!}{\includegraphics{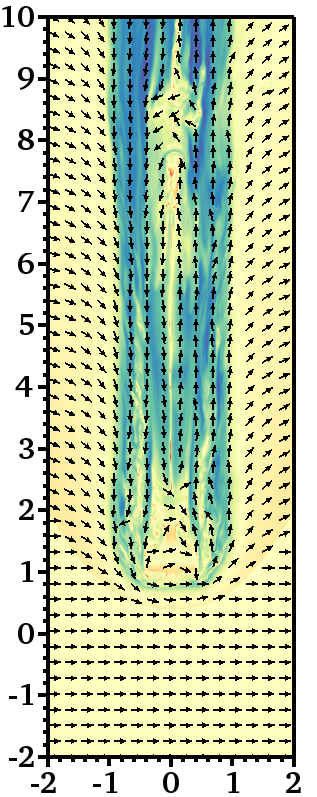}} & \resizebox{21mm}{!}{\includegraphics{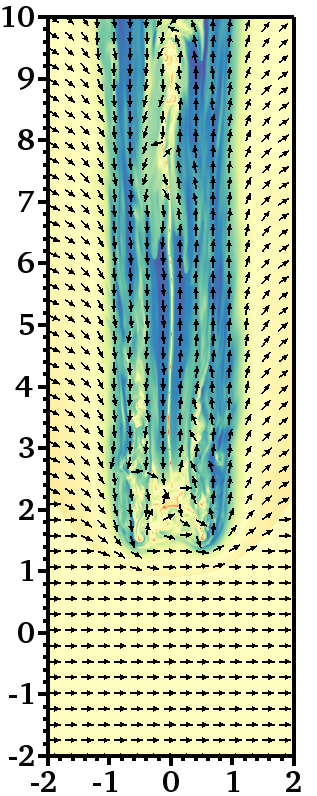}} & \resizebox{21mm}{!}{\includegraphics{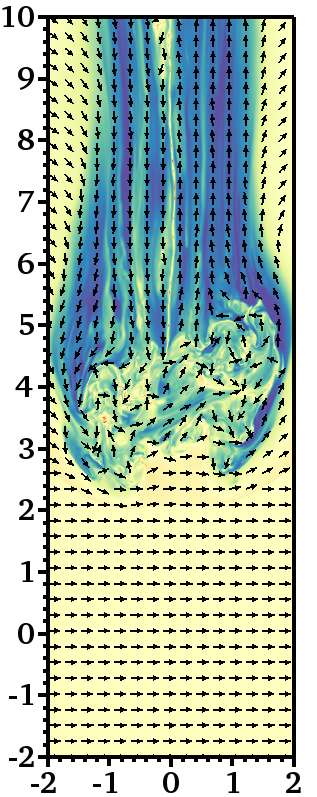}}\\
      \resizebox{21mm}{!}{\includegraphics{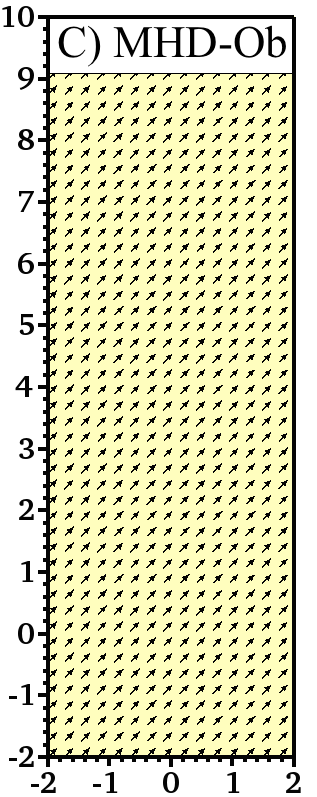}} & \resizebox{21mm}{!}{\includegraphics{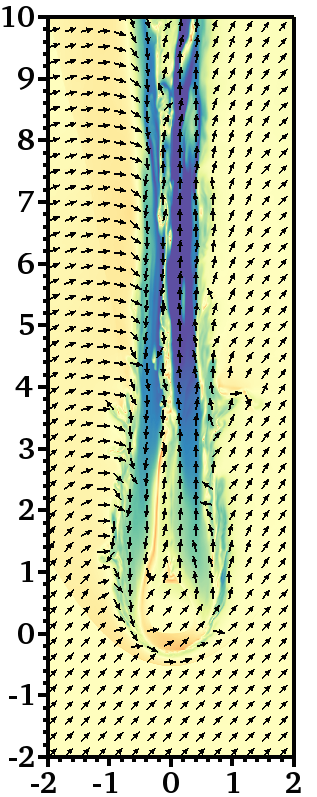}} & \resizebox{21mm}{!}{\includegraphics{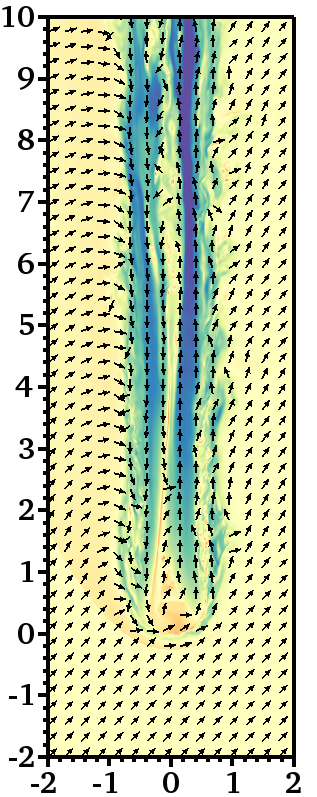}} & \resizebox{21mm}{!}{\includegraphics{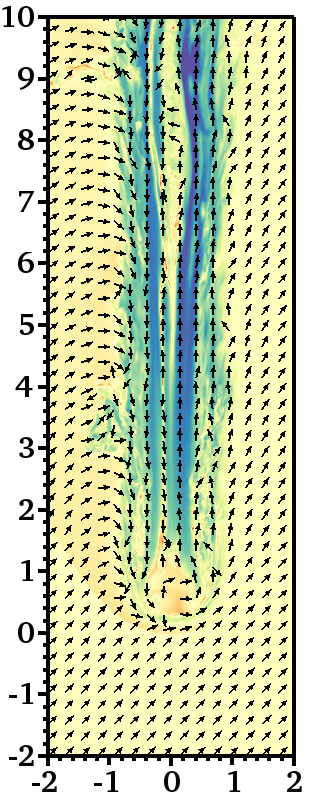}} & \resizebox{21mm}{!}{\includegraphics{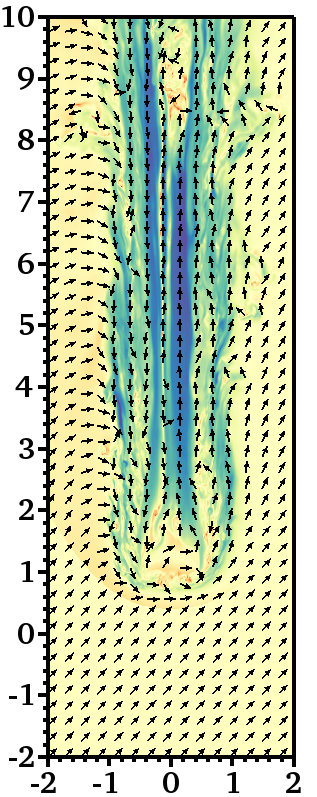}} & \resizebox{21mm}{!}{\includegraphics{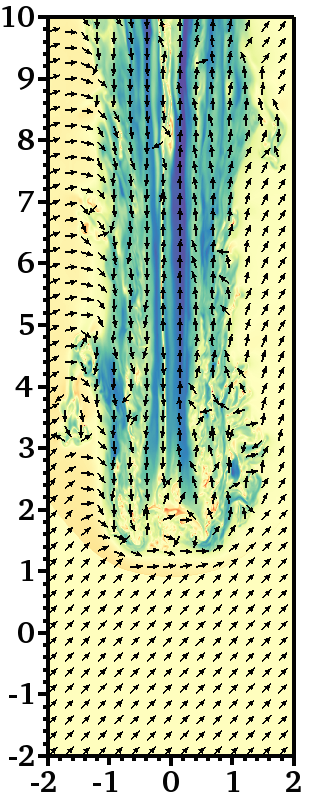}} & \resizebox{21mm}{!}{\includegraphics{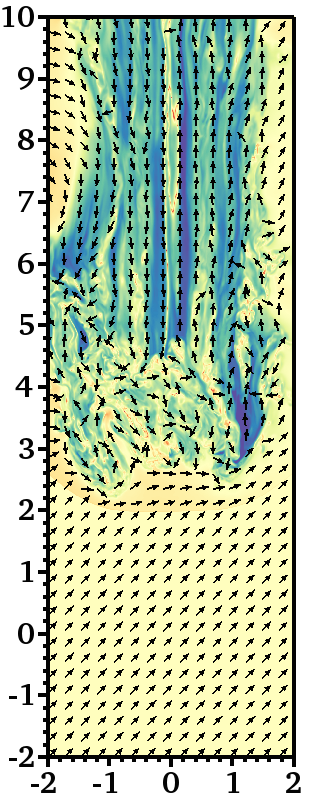}}\\
  \resizebox{!}{9mm}{\includegraphics{Axes.png}}& \multicolumn{6}{c}{\resizebox{!}{6.8mm}{\includegraphics{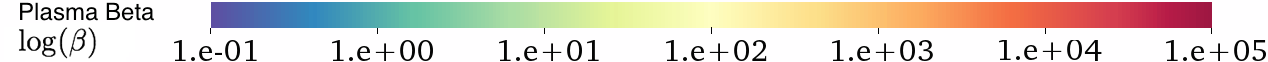}}}\Tstrut\\
  \end{tabular}
  \caption{2D slices at $X_3=0$ showing the evolution of plasma beta at $7$ different times: $t/t_{\rm cc}=0$, $t/t_{\rm cc}=0.2$, $t/t_{\rm cc}=0.4$, $t/t_{\rm cc}=0.6$, $t/t_{\rm cc}=0.8$, $t/t_{\rm cc}=1.0$, and $t/t_{\rm cc}=1.2$, for three different scenarios: MHD-Al (Panel A), MHD-Tr (Panel B) and MHD-Ob (Panel C). A rope-like structure is identified in the evolution of model MHD-Al in which the initial magnetic field is aligned with the direction of streaming (see Section \ref{subsec:MHD-Al}). Current sheets arise in models MHD-Tr and MHD-Ob in which the initial magnetic field has a component perpendicular to the wind direction (see Sections \ref{subsec:MHD-Tr} and \ref{subsec:MHD-Ob}). The plots showing the final stages of the evolution (i.e., $t/t_{\rm cc}\geq1.0$) indicate that small, magnetised filamentary tails survived the footpoint dispersion phase. The arrows indicate the direction and orientation of magnetic fields.} 
  \label{Figure9}
\end{center}
\end{figure*}

In addition, the RT instability is also important in the evolution of the filamentary structure. As explained above, the growth of RT bubbles at the leading edge of the cloud is, in fact, responsible for the break-up of the cloud. The compression of field lines at the front of the cloud substantially enhances the field strength at this location so that both terms in Equation (\ref{RTtime}) become important. It follows that the post-initial-shock ram pressure of the wind and the local magnetic pressure become comparable in magnitude and contribute equally to the acceleration of the cloud in models with transverse magnetic field components. This can be seen in Panels B1 and B2 of Figure \ref{Figure8} where the evolution of the magnetic energy in the tail and footpoint, respectively, indicates that amplification of the field is initiated as soon as the wind impacts the cloud. By $t/t_{\rm cc}=0.25$, the magnetic field strength has been amplified by a factor of $100$ in the tail. In the core, the amplification is delayed and only reaches $100$ of the initial value at $t/t_{\rm cc}=0.5$. Panel B of Figure \ref{Figure9} shows that values around and below $\beta=10$ are present at the leading edge of the cloud and along the filamentary structure, respectively.\par

The convergence of shock waves behind the cloud combined with the folding of field lines around the filament leads to the formation of a current sheet along the $X_2$ axis behind the cloud (similar sheet-like structures were reported by \citealt{2000ApJ...543..775G}). Low-density gas is trapped between the folding lines and starts to form vortices with moderate plasma betas ($\beta=10-100$) at the rear of the cloud. These vortices grow over time and slip between the field lines through the newly-formed sheet, eventually to be expelled from the region. Magnetic reconnection occurs at the boundary layer between the upstream-oriented field on the left and the downstream-oriented field on the right as a result of the growth of the TM instability. Field annihilation due to reconnection becomes dynamically significant after $t/t_{\rm cc}=0.5$ and prevents the magnetic tension becoming extremely large at this location. Panels A1 and A2 of Figure \ref{Figure8} shows how, after the onset of the TM instability, the plasma beta in the tail and footpoint remains nearly constant until $t/t_{\rm cc}=1.2$. Following the break-up of the filament footpoint, some of the coherence of the current sheet in the tail is lost. Then, turbulent motions disperse the magnetic structure to form a collection of highly-magnetised, small-scale filaments. The magnetic energy is further enhanced during the sub-filamentation process (after $t/t_{\rm cc}=1.0$) as revealed in Panels B1 and B2 of Figure \ref{Figure8}). Despite this, the original current sheet does survive the destruction of the footpoint as shown in Panel B of Figure \ref{Figure9}, with its upstream end being the location at which the tails splits into smaller rope-like filaments. Limitations in the size of our simulation domain, however, do not allow us to study the evolution of these substructures beyond $t/t_{\rm cc}=1.2$, indicating that further numerical work is needed.

\subsection{MHD filaments: Oblique field case}
\label{subsec:MHD-Ob}
In this section we consider the more general situation in which the magnetic field has components both aligned and transverse to the wind velocity (i.e., models MHD-Ob, MHD-Ob-S, and MHD-Ob-I). \cite{1996ApJ...473..365J} showed that a quasi-transverse behaviour is to be expected whenever the angle between the wind velocity and the magnetic field satisfies: $\theta\geq {\rm arctan}\left(\frac{1}{\cal M}\right)\simeq14^{\circ}$. Since the inclination angle between the magnetic field and the $X_1$-$X_3$ plane in our simulations does comply with this condition, we expect the evolution of model MHD-Ob to be similar to that of model MHD-Tr. A quick look at Panel C of Figure \ref{Figure7} indicates that this is, in fact, the case. We notice, however, that both the tail and the footpoint in model MHD-Ob are slightly more deformed and turbulent throughout the simulation. Global quantities, such as the transverse velocity dispersion (Panels B1 and B2 of Figure \ref{Figure6}) and the mixing fraction (Panels C1 and C2 of Figure \ref{Figure6}), indicate this behaviour as well. The curves corresponding to the plasma beta in both the filament tail and its footpoint are also similar in models MHD-Ob and MHD-Tr at the beginning of the evolution (see Panels A1 and A2 of Figure \ref{Figure8}, respectively). However, after $t/t_{\rm cc}=0.3$, folding and stretching of the stronger transverse component of the initial magnetic field in model MHD-Tr make the average plasma beta in both the tail and the footpoint at least $20\,\%$ lower than in the MHD-Ob scenario (e.g., at $t/t_{\rm cc}$=1.0). These values of plasma beta suggest that a stronger field component perpendicular to the wind direction favours magnetic shielding. Panels B1 and B2 of Figure \ref{Figure8} show that the enhancement in magnetic field tension is effectively less pronounced in model MHD-Ob than in model MHD-Tr.\par

In addition, Panel C of Figure \ref{Figure9} shows an asymmetric distribution of the plasma beta values, being higher on the left side of the $X_2$ axis. On the right side of this axis, the plasma beta values are rather low at locations where the field is preferentially aligned with the flow. Compression of the aligned components of the magnetic field on this side of the $X_2$ axis is responsible for the asymmetric enhancement of magnetic energy. The direction and orientation of the magnetic field vectors revealed in Panel C of Figure \ref{Figure9} suggests that a current sheet is also formed in this model, but it lies on a plane perpendicular to the plane containing both the initial wind velocity and the initial $\bm B$ field. We also note that the sub-filamentation process, observed in model MHD-Tr, is also visible in model MHD-Ob. Towards the final stages of the evolution, a collection of small filamentary tails survive the break-up of the filament footpoint. These tails retain some linear coherence despite the strong vortical motions in the surrounding gas, but further numerical work is needed to study their longevity once they are entrained in the wind. In order to understand how the filament morphology is affected by variations in the field strength and in the equation of state, we report below the results from two additional models, namely MHD-Ob-S and MHD-Ob-I.

\begin{figure}
\begin{center}
  \begin{tabular}{c c}
    \resizebox{40mm}{!}{\includegraphics{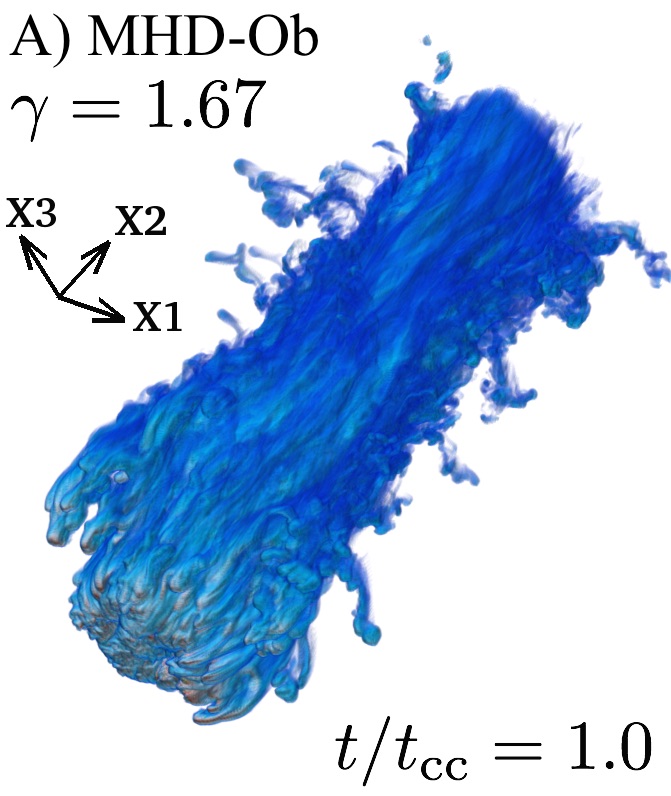}} & \resizebox{40mm}{!}{\includegraphics{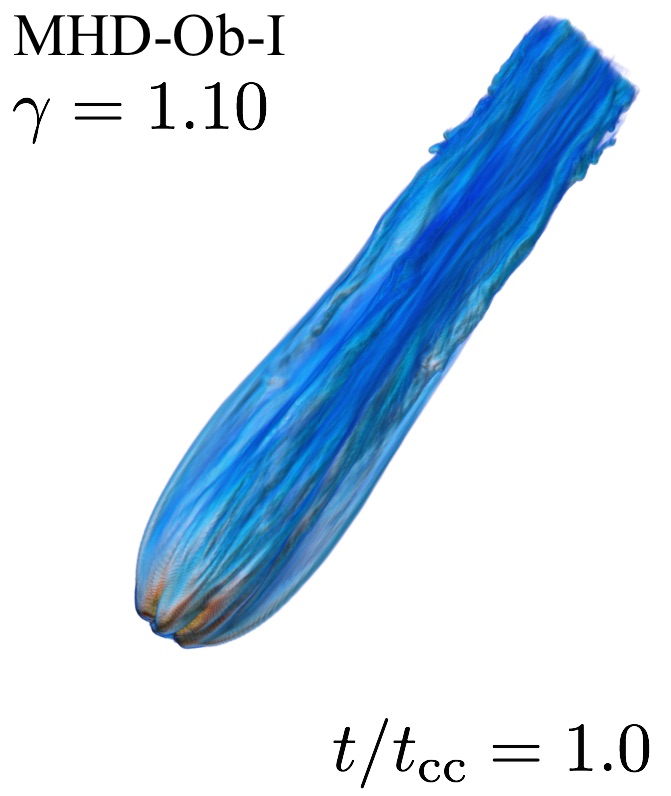}}\Bstrut \\
          \multicolumn{2}{c}{\resizebox{!}{6.5mm}{\includegraphics{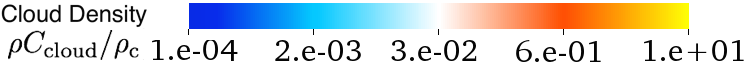}}}\Dstrut \\
      \resizebox{40mm}{!}{\includegraphics{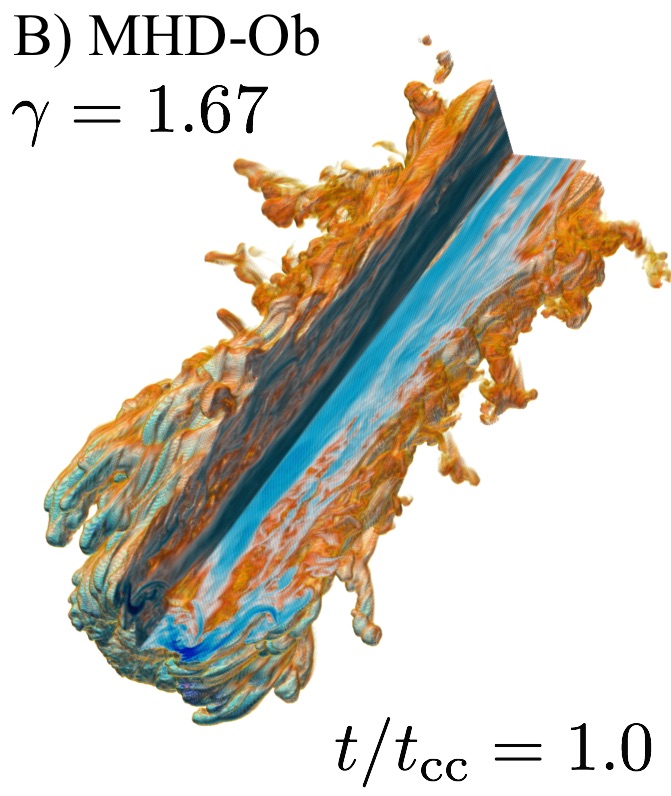}} & \resizebox{40mm}{!}{\includegraphics{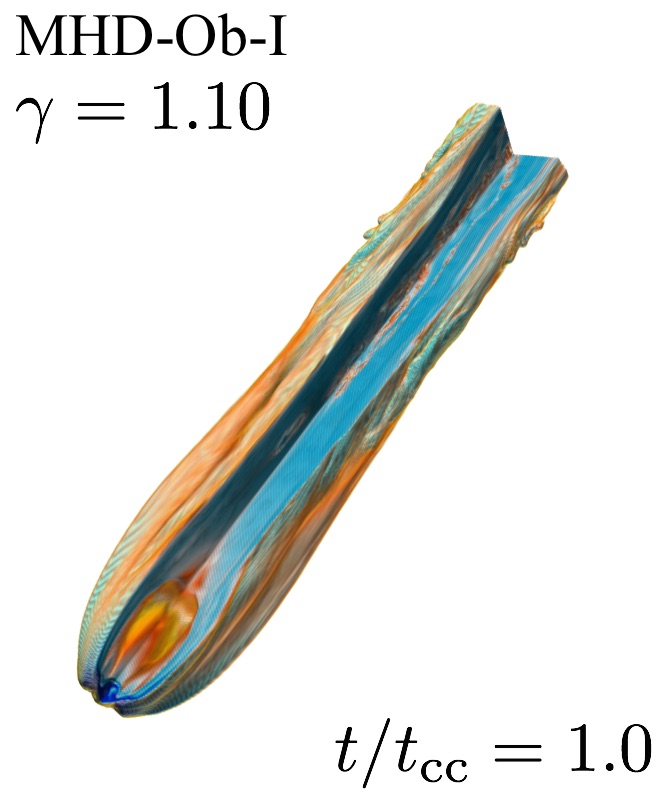}}\Bstrut \\
      \multicolumn{2}{c}{\resizebox{!}{6.5mm}{\includegraphics{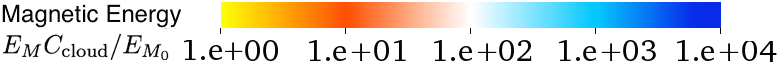}}}\\ 
  \end{tabular}
  \caption{Comparison of 3D volume renderings of the logarithmic mass density (Panel A) and magnetic energy (Panel B) in two different scenarios: adiabatic with $\gamma=1.67$ (left column) and quasi-isothermal with $\gamma=1.10$ (right column). All renderings correspond to snapshots of the simulations at $t/t_{\rm cc}=1.0$. Note that a quadrant has been clipped from the renderings in Panel B to show the interior of the filamentary tails. The inclusion of a softer, nearly-isothermal equation of state leads to the emergence of a more linear, less turbulent filament compared to adiabatic interactions (see Section \ref{subsubsec:Isothermal} for further details).} 
  \label{Figure10}
\end{center}
\end{figure}

\subsubsection{Dependence on the field strength}
\label{subsubsec:FieldStrength}
As mentioned above, several plasma instabilities can be dynamically important in wind-cloud interactions. Their influence on the evolution of these systems depends on how fast their perturbations grow and how disruptive these are. Varying the strength of the initial magnetic field naturally modifies the growth rate of the KH and RT instabilities and alters the morphology of the resulting filaments. Panels C and D of Figure \ref{Figure7} show the time evolution of the magnetic energy of filaments in two models with the same initial field orientation (i.e., oblique with respect to the wind direction), but different magnetic field strengths ($\beta=100$ in MHD-Ob and $\beta=10$ in MHD-Ob-S). We find that the KH instability is present in the model with a stronger field, but it occurs at larger scales and is less pronounced than it is in its weak-field counterpart. The broadening of the shear layer between wind and filament gas, in which line stretching occurs, leads to the formation of a stronger magnetic shield around the filament in model MHD-Ob-S. The high magnetic pressure in this region inhibits and delays the emergence of KH perturbations around the cloud (see the KH growth times reported in Table \ref{Table3}). The presence of mild deformations on the surface of the filamentary tail suggests that the Alfv\'{e}n speed locally exceeds the relative speed between wind and filament gas. The suppression of KH unstable modes also affects the flow around the filament, which becomes smoother and more laminar in model MHD-Ob-S with respect to previous scenarios.\par

\begin{figure*}
\begin{center}
  \begin{tabular}{c c}
        \textbf{Filament Tail (Cloud Envelope)} & \textbf{Filament Footpoint (Cloud Core)}\\ 
      \resizebox{80mm}{!}{\includegraphics{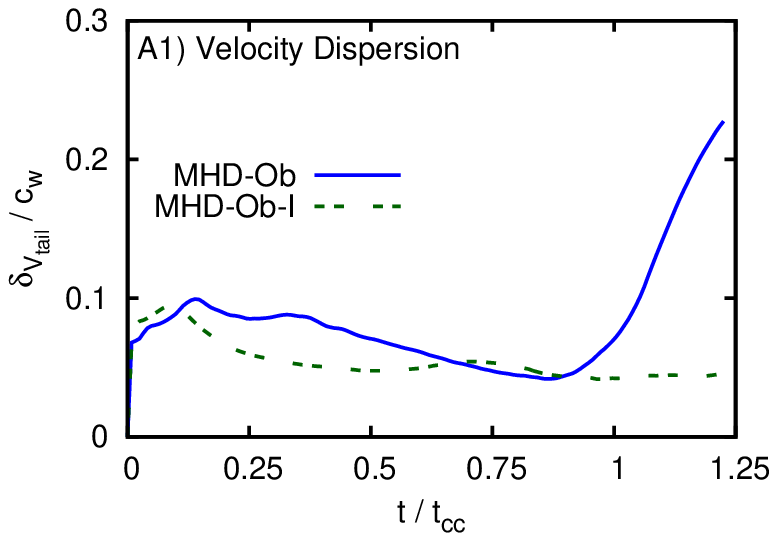}} & \resizebox{80mm}{!}{\includegraphics{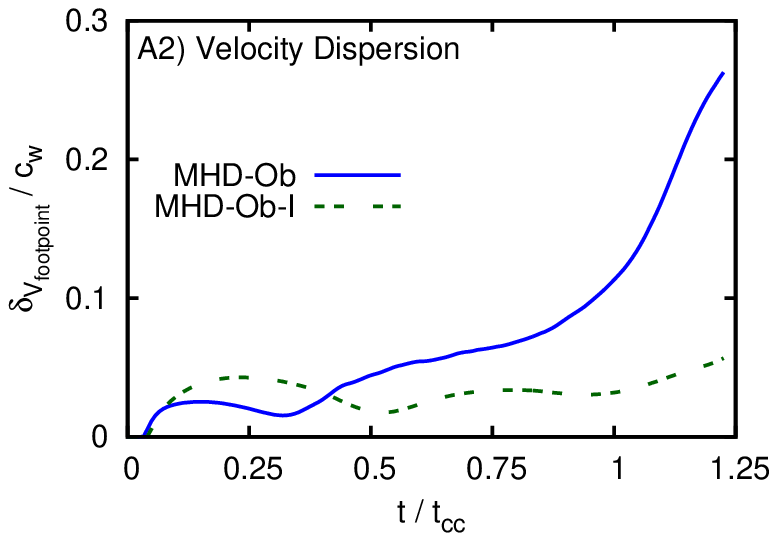}}\\       
      \resizebox{80mm}{!}{\includegraphics{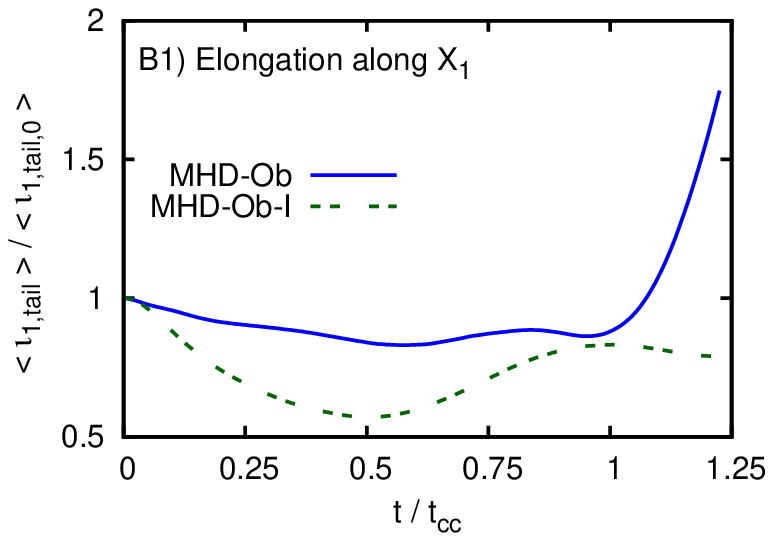}} & \resizebox{80mm}{!}{\includegraphics{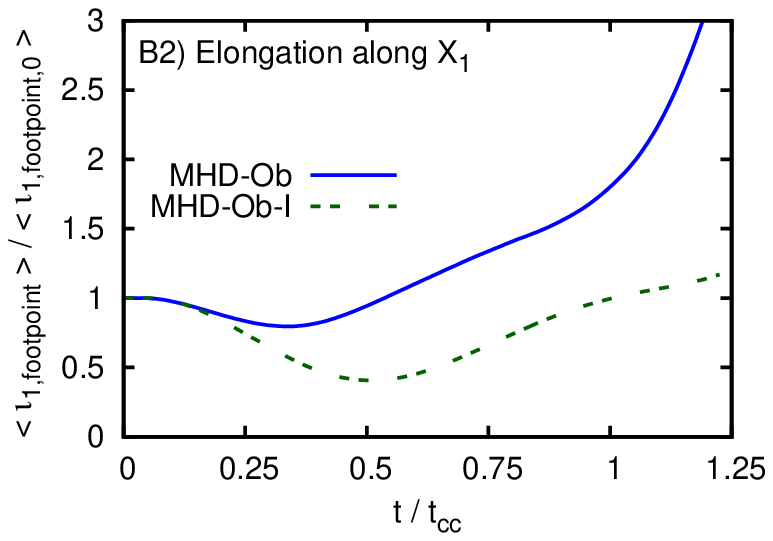}}\Tstrut\\
  \end{tabular}
  \caption{Comparison between the time evolution of the transverse velocity dispersions (Panels A1 and A2), and elongations along the $X_1$ direction (Panels B1 and B2) of filaments in the adiabatic model MHD-Ob (solid line), and the quasi-isothermal model MHD-Ob-I (double-dashed line). The filamentary tail is less turbulent in the latter as evidenced by its lower velocity dispersions with respect to the adiabatic case. The wind-swept cloud gas in the quasi-isothermal case is also confined in a more linear filament with lower transverse elongation than its adiabatic counterpart (see Section \ref{subsubsec:Isothermal} for a complete discussion).} 
  \label{Figure11}
\end{center}
\end{figure*}

On the other hand, the growth of the RT instability is also affected in the strong-field scenario. We find that the disruption of the filament footpoint is enhanced in model MHD-Ob-S as a result of a faster growth of RT unstable modes (see Table \ref{Table3}). This enhanced growth is driven by a stronger magnetic flux density at the leading edge of the cloud (magnetic bumper) as a result of the stretching of the field lines anchored in that region. This result is in agreement to what was found by \cite{1999ApJ...527L.113G,2000ApJ...543..775G}; and \cite{2008ApJ...680..336S} at lower resolution. The panels of Figure \ref{Figure8} also reveal that the overall process of field amplification is conducted in a similar fashion in models MHD-Ob and MHD-Ob-S. However, the stronger initial magnetic field prescribed for the latter results in larger magnetic pressures along the wind-filament boundary. Thus, the stronger magnetic shielding seen in model MHD-Ob-S secludes tail material earlier than in model MHD-Ob, enhancing its thermal pressure and saturating field amplification (see Panels A1 and B1 of Figure \ref{Figure8}). On the other hand, the early increase of magnetic energy observed in the tail is not seen in the footpoint, where the magnetic energy remains nearly constant until $\sim t/t_{\rm cc}=0.3$ (see Panels A2 and B2 of Figure \ref{Figure8}). The growth of magnetic field in the footpoint is slowed down when compared with that of MHD-Ob, and it only increases when the footpoint commences its expansion. From this time onwards, the magnetic field lines anchored at the leading edge of the cloud core are steadily stretched by the wind, which enhances its magnetic energy by an order of magnitude by $t/t_{\rm cc}=0.5$. Even though this magnetic energy enhancement is $10$ times lower in model MHD-Ob-S when compared to that of model MHD-Ob, it is significant enough to maintain the average plasma beta in the footpoint low ($\beta\lesssim10$) during the remainder of the evolution.\par

\subsubsection{Dependence on the equation of state}
\label{subsubsec:Isothermal}
Figure \ref{Figure10} shows how a softened equation of state (with $\gamma=1.1$) leads to the production of a much less turbulent tail. The renderings in Panel A compare the external morphology of the logarithmic filament density, while the renderings in Panel B present the internal magnetic configurations of models MHD-Ob and MHD-Ob-I. The snapshots correspond to $t/t_{\rm cc}=1.0$. As we approach the isothermal regime, the gas more efficiently converts the shock-driven heat into kinetic energy, thus significantly reducing the expansion of the cloud in the transverse direction. The compression at the front end and lateral sides of the cloud is also significantly higher and this translates into higher densities in the footpoint and a reduced bulk velocity (see Panel A2 of Figure \ref{FigureAA}). As the density contrast between wind and filament gas increases, the interface becomes stable to KH instability modes. The suppression occurs at long and short wavelengths, so the KH instability is less pronounced in this scenario, i.e., stripping occurs at a much slower rate than in adiabatic scenarios. Figure \ref{Figure11} shows how two parameters, namely the transverse velocity dispersion (Panels A1 and A2) and the lateral size (Panels B1 and B2) of the filaments' tails and footpoints, respectively, vary depending on the equation of state. We see that these parameters are significantly affected by $\gamma$. At $t/t_{\rm cc}=0.5$, for example, the velocity dispersion in both the tail and footpoint in model MHD-Ob-I is reduced by $\sim 20\,\%$ with respect to that in model MHD-Ob. Similarly, the elongation along the $X_1$ direction of both tail and footpoint material is reduced by a factor of $2$ at this time, indicating that the nearly isothermal filament is more collimated than its adiabatic counterparts.\par

The growth time-scale of disruptive RT modes is also delayed by the higher density contrasts and lower effective acceleration at the leading edge of the cloud. As a result, the filament footpoint is disrupted and dispersed in time-scales longer than the cloud-crushing time (defined in Section \ref{subsec:DynamicalTime-Scales} as a reference). Since we maintained the wind speed constant in all our simulations, the survival time of the filament in the quasi-isothermal scenario is effectively higher than the survival times of the adiabatic filaments. The fast growth of the transverse velocity dispersions observed in adiabatic models after the time of break-up ($t/t_{\rm cc}=1.0$) is not present in model MHD-Ob -I (see Panels A1 and A2 of Figure \ref{Figure11}). If radiative cooling were properly treated, a similar evolution would be expected, with gas in the cloud cooling down to form dense footpoints, which would, in turn, support the filamentary tails longer. Distinct cooling regimes would, however, produce different effects on the evolution of filaments as pointed out by \cite{2005ApJ...619..327F} and \cite{2013ApJ...774..133L}, so the longevity of the filament in our quasi-isothermal simulation is only indicative of what we should expect for radiative clouds. The increased density in the footpoint has other implications for the morphology of the filament. Since the lateral size of the filamentary structure is reduced in the nearly isothermal model, the aspect ratio of a highly-radiative filament is expected to be higher than that of its adiabatic counterparts. Our results in this section are in agreement with those reported by \cite{2008ApJ...674..157C,2009ApJ...703..330C}.

\section{Mass entrainment in winds}
\label{sec:Entrainment}
In this section we discuss the implications of our study to mass entrainment processes in global winds. Here we examine  the transport of clouds (filaments) by the supersonic wind in all the aforementioned models, namely HD, MHD-Al, MHD-Tr, MHD-Ob, MHD-Ob-S, and MHD-Ob-I.

\subsection{Displacement in the direction of streaming}
\label{subsubsec:FilamentDisplacement}
\begin{figure*}
\begin{center}
  \begin{tabular}{c c}
      \resizebox{80mm}{!}{\includegraphics{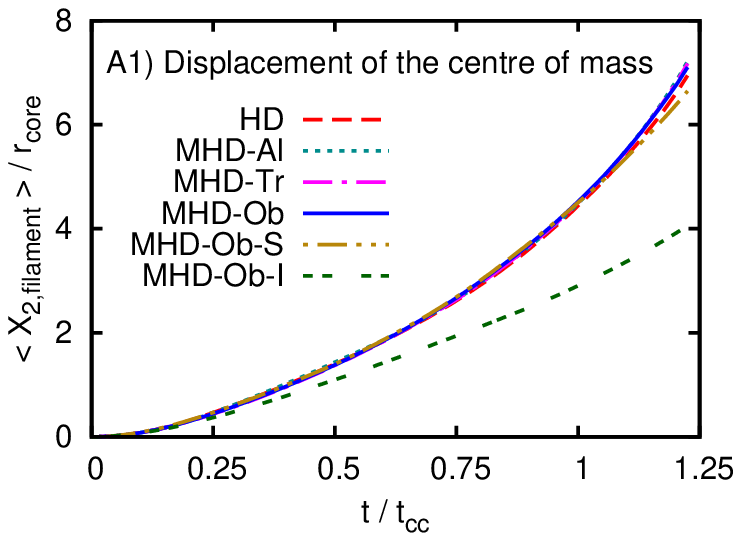}} & \resizebox{80mm}{!}{\includegraphics{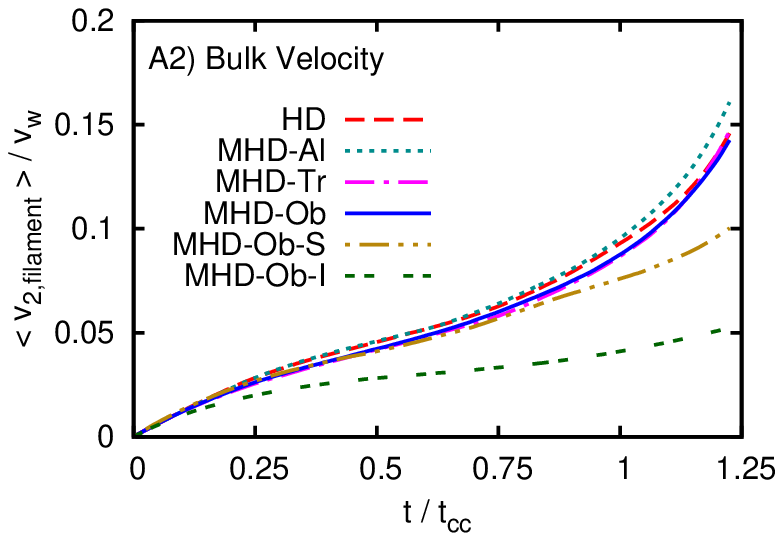}}\Tstrut\\
  \end{tabular}
  \caption{Displacement of the centre of mass in the direction of streaming (Panel A1) and bulk velocity of the filaments entrained in the wind (Panel A2) in six different models, namely: HD (dashed line), MHD-Al (dotted line), MHD-Tr (dash-dotted line), MHD-Ob with $\beta=100$ (solid line), MHD-Ob-S with $\beta=10$ (double-dot-dashed line), and MHD-Ob-I with $\gamma=1.1$ (double-dashed line).} 
  \label{FigureAA}
\end{center}
\end{figure*}

The longevity of a wind-swept cloud and its associated filamentary tail when immersed in a hot wind is determined by its ability to maintain coherence against the wind ram pressure and dynamical instabilities. Consequently, a natural question to be addressed in our study is how far they travel before they are dispersed in the wind. We investigate the displacement of the filamentary structure in the direction of streaming by examining the change in position of its centre of mass for $0\leq t/t_{\rm cc}\leq1.2$. Panel A1 of Figure \ref{FigureAA} provides the distances travelled by filamentary structures in models with different initial conditions, as measured by $\langle X_{2,\rm filament}\rangle$ normalised with respect to the initial radius of the cloud core. Our results show that the wind is capable of transporting dense material over distances equivalent to $3-5$ times the original size of the cloud core in the direction of streaming (measured at $t/t_{\rm cc}=1.0$, i.e., at the time when the footpoint is dispersed). Nevertheless, as mentioned above, smaller cloudlets and more distorted filamentary structures survive the disruptive effects of the wind ram pressure and dynamical instabilities, so these numbers are conservative first estimates for the distances that dense gas and filaments could effectively travel after the break-up.

\subsection{Velocity of entrained filaments}
\label{subsubsec:Terminalvelocity}
Another question to be discussed in the same context is what range of velocities dense gas and its associated filaments acquire until the time of break-up. We investigate this by analysing the evolution of the mass-weighted bulk velocity of the cloud gas in the direction of streaming (i.e., along the $X_2$ axis). Panel A2 of Figure \ref{FigureAA} provides the velocities of filamentary structures in models with different initial conditions, as measured by $\langle v_{2,\rm filament}\rangle$ normalised with respect to the wind speed, $v_{\rm w}$. Our results show that the velocity in adiabatic models is nearly unaffected by the presence of magnetic fields, with values around $\langle v_{2,\rm filament}\rangle/v_{\rm w}\simeq0.08-0.1$. In contrast, the velocity in a quasi-isothermal scenario is slower compared to its counterparts: $\langle v_{2,\rm filament}\rangle/v_{\rm w}\simeq 0.04$. Note, however, that after the break-up time, the bulk velocity of the remaining material starts to increase faster as described in Section \ref{sec:FilamentFormation}, suggesting that the surviving structures entrained in the wind could reach even higher velocities before fully evaporating and/or mixing with the hot gas.

\section{Limitations and Future Work}
\label{sec:FutureWork}
Despite the advances towards the understanding of filament formation in the interstellar medium that we have presented here, our current models have various limitations. For example, the simulations in this paper have only considered spherical clouds with smooth density profiles and magnetic fields distributed uniformly in the simulation domain. The ISM is, however, not uniform and homogenous. The velocity and density fields in ISM clouds and clumps, for example, are best described by turbulence (e.g., \citealt{1981MNRAS.194..809L}; \citealt*{1997MNRAS.288..145P,2004RvMP...76..125M,2009ApJ...692..364F}; \citealt{2012ApJ...761..156F}). Consequently, future work should consider more realistic scenarios, including clouds with self-contained, turbulent density, velocity, and magnetic fields. We shall study such scenarios in subsequent work (Paper II). On the other hand, specific applications of the results presented here should be discussed in the future as well. The inclusion of source terms, such as radiative cooling, thermal conduction, photo-evaporation, and gravity, in the MHD formulation is problem dependent, so that future work addressing applications will also need to incorporate additional physics, depending upon the context.

\section{Summary and concluding remarks}
\label{sec:Summary}
The lack of high-resolution, three-dimensional, MHD numerical studies on filamentary structures arising from wind-cloud interactions in the current literature motivated us to investigate such systems. We employ scale-free configurations and perform a parameter survey over three quantities: a) the magnetic field orientation (aligned, MHD-Al; transverse, MHD-Tr; and oblique, MHD-Ob, to the wind direction), b) the magnetic field strength (weak, MHD-Ob; and strong, MHD-Ob-S), and c) the polytropic index (adiabatic, MHD-Ob; and quasi-isothermal, MHD-Ob-I). We use 3D renderings and 2D slices in combination with volume- and mass-weighted-averaged quantities to follow the formation and evolution of filaments emerging from such interactions. Our initial conditions are selected so that they represent more realistic ISM conditions than previous 3D simulations, including high density contrasts between the wind and cloud of $\chi=10^3$, wind Mach numbers of the order of ${\cal M_{\rm w}}=4-4.9$, and initial plasma betas of $\beta=10-100$. The aim of this work is to determine how the presence of the magnetic field affects the morphology of the filamentary tail, and how the presence of the filament entrained into the wind reacts back on the magnetic field in the surrounding gas. The importance of ram pressure and dynamical instabilities in the formation, evolution, and longevity of the filaments is discussed for each model. The conclusions of our study are as follows:

\begin{enumerate}
 \item As in previous simulations, we find that the wind ram pressure combined with shocks and plasma instabilities arising at the wind-cloud interface are responsible for the break-up of clouds. The cloud disruption is a four-stage process that involves: 1) a compression phase in which the ram pressure of the wind generates reflected and refracted shocks in the ambient and cloud gas, respectively; 2) a stripping phase in which the KH instability deposits material originally in the cloud envelope on the symmetry axis behind the cloud; 3) an expansion phase triggered by shock-induced adiabatic heating; and 4) a break-up phase in which RT bubbles penetrate the cloud core and expand, destroying the cloud core and forming smaller cloudlets.
 \item Our results confirm that wind-swept clouds can lead to the formation of elongated, long-lived filamentary structures provided that the density contrast is reasonably large, i.e., $\chi\gtrsim10^3$. The evolution of filaments is comprised of four phases: 1) a tail formation phase in which the transport of gas, stripped from the sides of the cloud by the wind ram pressure and KH instabilities, forms a tail of gas behind the cloud; 2) a tail erosion phase in which KH instabilities acting on the boundary layers between filament and ambient gas shape the downstream gas producing tail morphologies specific to each model; 3) a footpoint dispersion phase in which the cloud core is disrupted by RT unstable modes and the filamentary tail either becomes distorted or breaks up into multiple strands; and 4) a free-floating phase in which smaller filaments are entrained in the wind and are transported downstream maintaining some coherence.
 \item Filaments consist of two main substructures, namely tails and footpoints. Our simulations show that gas from the cloud envelope is the main constituent of filamentary tails, while footpoints are predominantly formed by gas originally in the cores of clouds. When the core starts to expand, the roles are inverted as dense material, stripped from it, flows downstream and envelops the low-density tail. Filaments in models HD and MHD-Al are both dominated by small-scale KH perturbations, while medium-scale (large-scale) vorticity is favoured in models MHD-Tr and MHD-Ob (MHD-Ob-S and MHD-Ob-I). As a result: 1) Models HD and MHD-Al have higher mixing fractions and velocity dispersions than the models with magnetic field components transverse to the wind direction; 2) Owing to folding and stretching of field lines, the total magnetic energy contained in the filamentary tails formed in models MHD-Tr and MHD-Ob is twice as high as that in the other models; and 3) Either a stronger initial field or a higher mechanical compression leads gas in the tail of the filaments (produced in models MHD-Ob-S and MHD-Ob-I) to be confined in more elongated, less turbulent magnetotails than their counterparts.
 \item The shape and structure of the (magneto)tails ultimately depend on whether or not the medium is magnetised, and on what the initial topology of the magnetic field is, if present. Four different kinds of features are identified in filaments: highly-turbulent, tower-like tails arise in purely hydrodynamic models (HD); tails with strongly-magnetised flux ropes arise in models dominated by magnetic fields aligned with the flow (MHD-Al); tails with reconnection-prone current sheets emerge in models dominated by magnetic fields transverse to the flow (MHD-Tr and MHD-Ob); and highly-confined, tube-like tails emerge in models subjected to strong magnetic shielding (MHD-Ob-S and MHD-Ob-I). The morphology of the filaments remains coherent until the cloud is broken up (at $t/t_{\rm cc}=1.0$) by the combined effects of the wind ram pressure and dynamical instabilities. Movies with the full-time sequence of the snapshots in Figures \ref{Figure4}, \ref{Figure5}, \ref{Figure7}, and \ref{Figure9} are available online\footnote{\url{https://sites.google.com/site/wlady721/phd-research/wind-cloud-interactions}}.
 \item Our simulations show that dense gas in the cloud is effectively transported over distances equivalent to $3-5$ times the initial radius of the cloud core. The advected gas reaches velocities $\sim0.1$ of the wind speed by the time at which the cloud is broken up (or even higher speeds at late stages). Even though some of the coherence of filamentary tails is lost after their footpoints are dispersed, small cloudlets and highly distorted, magnetised filaments survive the break-up phase and become entrained in the wind. We find evidence for tail disconnection events occurring in all models, with filamentary tails detaching from dense gas as a result of magnetic reconnection occurring at the rear of the cloud. In model HD, the filament detaches from the footpoint and both structures evolve separately, while in models MHD-Tr and MHD-Ob the upstream end of the current sheet is dispersed, causing sub-filamentation and the appearance of strongly magnetised, small sinuous tails.
\end{enumerate}

Finally, the results presented here are scalable and relevant to understanding the formation of a variety of cosmic structures at both small or large scales. In particular, the discussion presented in Section \ref{sec:Entrainment} is relevant to understanding the transport of dense material from low to high latitudes in galactic outflows, such as the one observed in the Milky Way (see \citealt{2003ApJ...582..246B,2013Natur.493...66C,2013ApJ...770L...4M} for observations, and \citealt{2012MNRAS.423.3512C} for a theoretical overview). Our simulations suggest that dense clouds and their associated filamentary tails survive ablation and can effectively be advected by a global wind to potentially reach high latitudes.

\subsection*{Acknowledgements}
We thank the anonymous referee for providing insightful comments on the manuscript. This research was supported by the Research School of Astronomy and Astrophysics and by the NCI Facility at the Australian National University (grants~n72~and~ek9). WBB thanks Prof. Mark Morris for hosting him during a visit to UCLA and for insightful discussions on the scope and applications of the simulations presented here. In addition, WBB thanks the National Secretariat of Higher Education, Science, Technology, and Innovation of Ecuador, SENESCYT, for funding this project through a doctoral scholarship (CI:1711298438) under the ``Universidades de Excelencia" program. CF~acknowledges funding provided by the Australian Research Council's Discovery Projects (grants~DP130102078 and~DP150104329). WBB is the recipient of the Olin Eggen Scholarship at Mount Stromlo Observatory. RMC is the recipient of an Australian Research Council Future Fellowship (FT110100108). This work was further supported by resources provided by the Pawsey Supercomputing Centre with funding from the Australian Government and the Government of Western Australia.

\bibliography{paper1.bib}{}

\appendix{}
\section{Comparison with a large-domain simulation}
\label{subsec:Appendix}
In this Appendix we show the effects of our choice of simulation domain size on our diagnostics. For this purpose, we compare several measurements of two simulations with the same initial conditions and different domain sizes. The initial conditions of these simulations correspond to those of model MHD-Ob and both have been performed at resolutions of $64$ cells per cloud radius (i.e., $R_{64}$). The linear dimensions of the domain in model MHD-Ob(Large) is twice that of model MHD-Ob(Small), i.e., it covers a physical spatial range $-4r_{\rm c}\leq X_1\leq4r_{\rm c}$, $-2r_{\rm c}\leq X_2\leq22r_{\rm c}$, and $-4r_{\rm c}\leq X_3\leq 4r_{\rm c}$, where $r_{\rm c}$ is the radius of the cloud.\par

\begin{figure*}
\begin{center}
  \begin{tabular}{c c}
        \textbf{Filament Tail (Cloud Envelope)} & \textbf{Filament Footpoint (Cloud Core)}\\ 
      \resizebox{80mm}{!}{\includegraphics{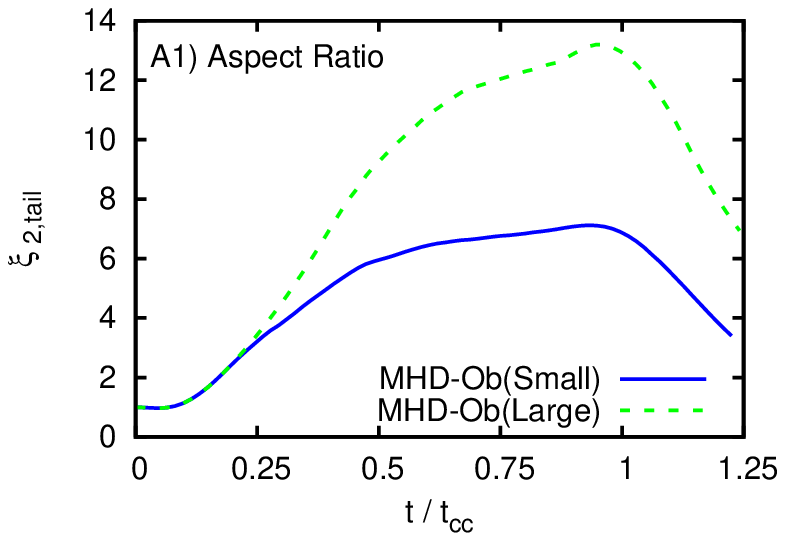}} & \resizebox{80mm}{!}{\includegraphics{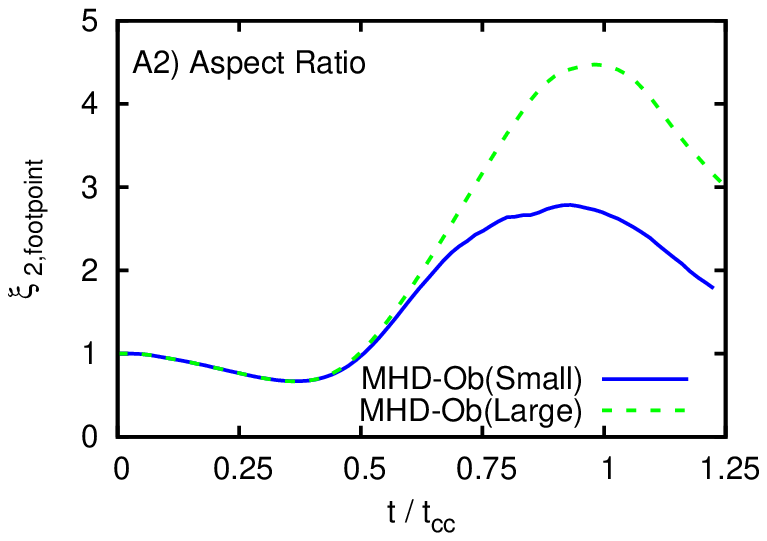}}\\       
      \resizebox{80mm}{!}{\includegraphics{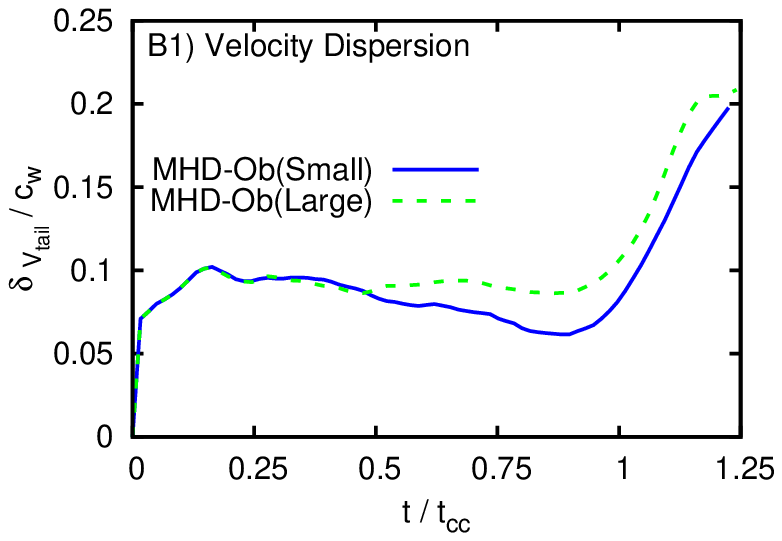}} & \resizebox{80mm}{!}{\includegraphics{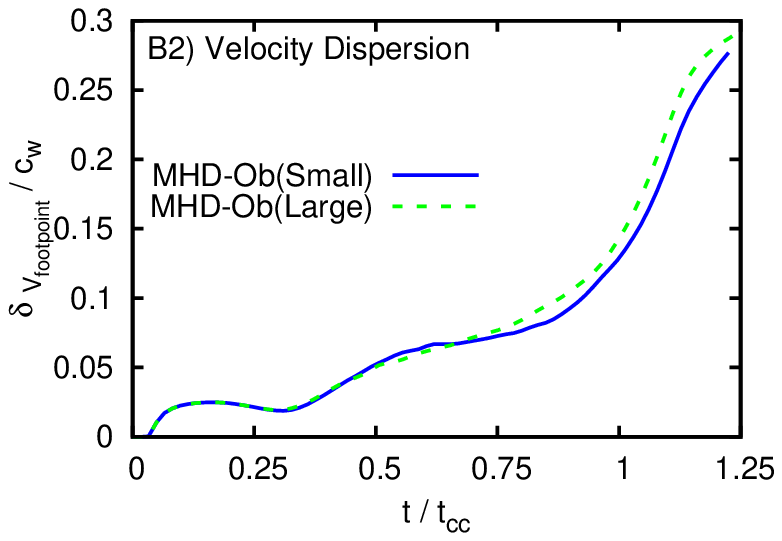}}\\
      \resizebox{80mm}{!}{\includegraphics{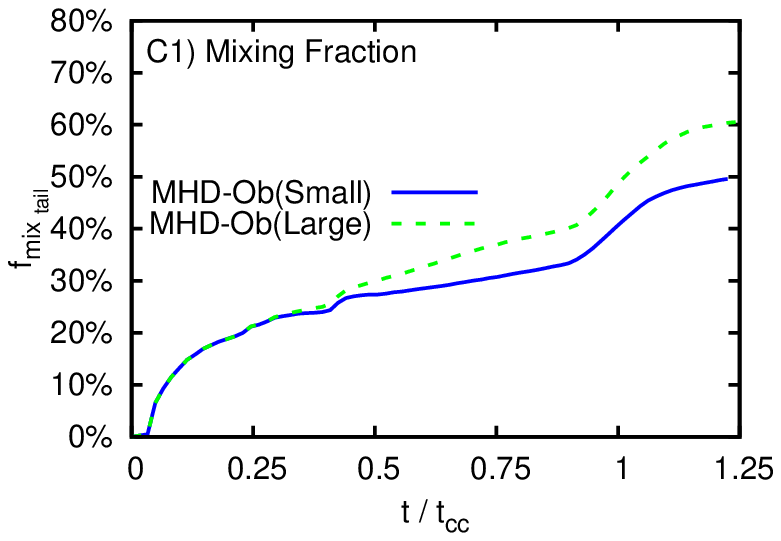}} & \resizebox{80mm}{!}{\includegraphics{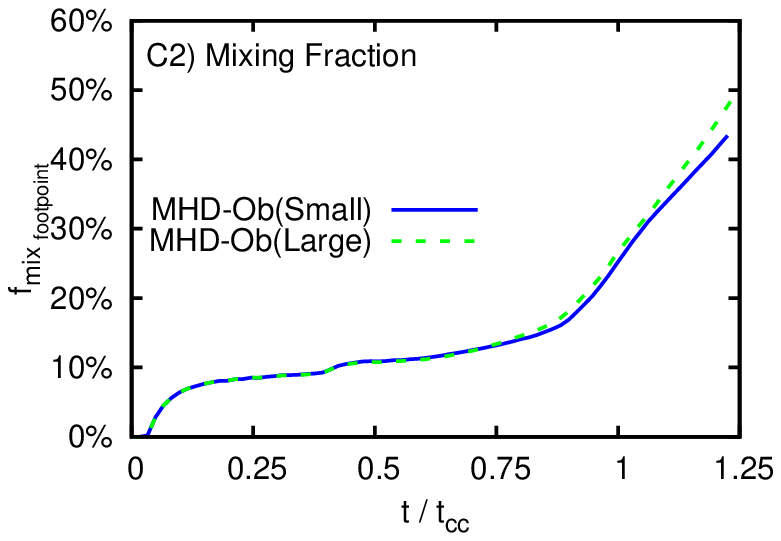}}\\
      \resizebox{80mm}{!}{\includegraphics{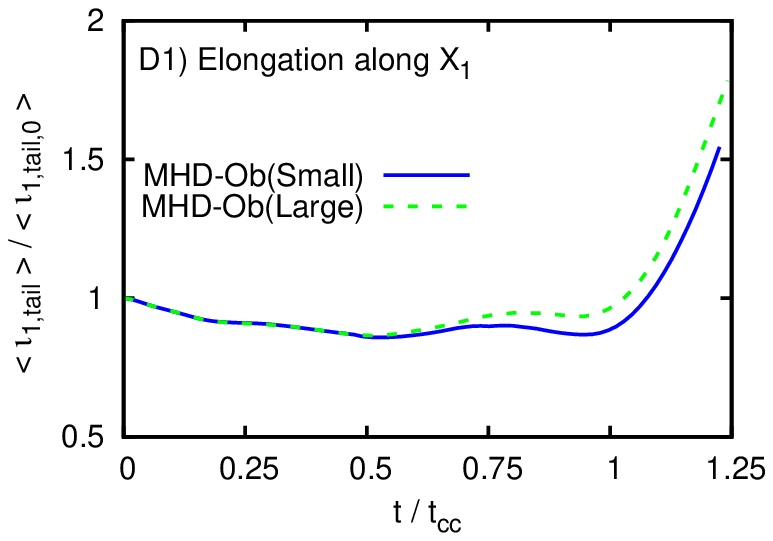}} & \resizebox{80mm}{!}{\includegraphics{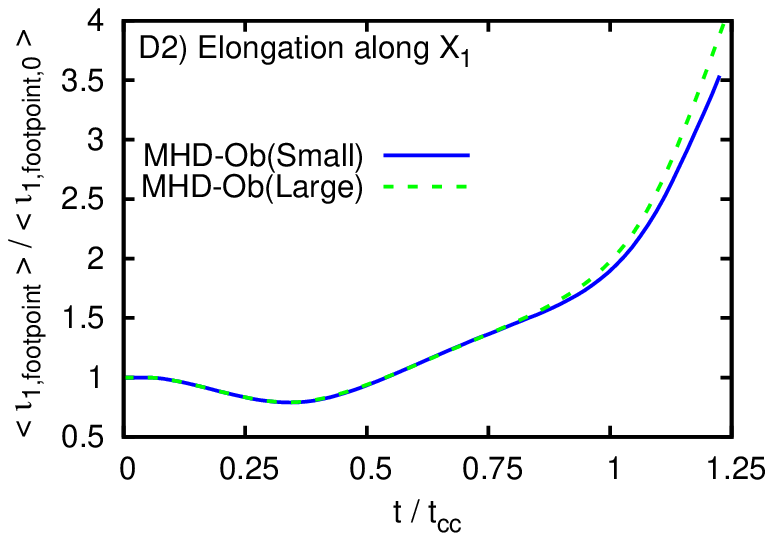}}\Tstrut\\
  \end{tabular}
  \caption{Comparison between the time evolution of the quantities shown in Figures \ref{Figure6} and \ref{Figure11} for two models with oblique magnetic fields (MHD-Ob) at the same resolution ($R_{64}$), but different domain sizes.} 
  \label{FigureA1}
\end{center}
\end{figure*}

Figure \ref{FigureA1} shows the evolution of the parameters presented in Figures \ref{Figure3}, \ref{Figure6}, and \ref{Figure11} in Sections \ref{sec:CloudDisruption} and \ref{sec:FilamentFormation}. Divergence between the curves for tail material starts at $t/t_{\rm cc}=0.2$, while for footpoint material it starts at $t/t_{\rm cc}=0.6$. These are the times at which material of either component commences to flow out of the smaller simulation grid. Panels A1 and A2 show that the aspect ratio is the only parameter affected by the simulation domain size significantly, with differences for tail material being as large as $6$ in units of aspect ratio. The absolute numbers for the aspect ratios presented in Section \ref{sec:FilamentFormation} above should therefore be considered by the reader solely as either lower limits or reference numbers if comparing different models with one another. Despite this bias, we notice that the curves in these panels show the trend expected for this parameter, displaying the tail formation, tail erosion, and footpoint dispersion phases clearly. Note, for instance, how after the break-up of the cloud at $t/t_{\rm cc}=1.0$, the aspect ratio decreases in both tail and footpoint, in a manner that coincides with an increase of the lateral size of the cloud (presented in Panels D1 and D2). Panels B1, C1, and D1 show that the other parameters, namely the transverse velocity dispersion, the mixing fraction, and the elongation along the $X_1$ direction for tail material are converged to e.g., within $10\,\%$ at $t/t_{\rm cc}=1.0$. Similarly, Panels B2, C2, and D2 show that footpoint parameters are unaffected by the domain size.\par

\begin{figure*}
\begin{center}
  \begin{tabular}{c c}
      \resizebox{80mm}{!}{\includegraphics{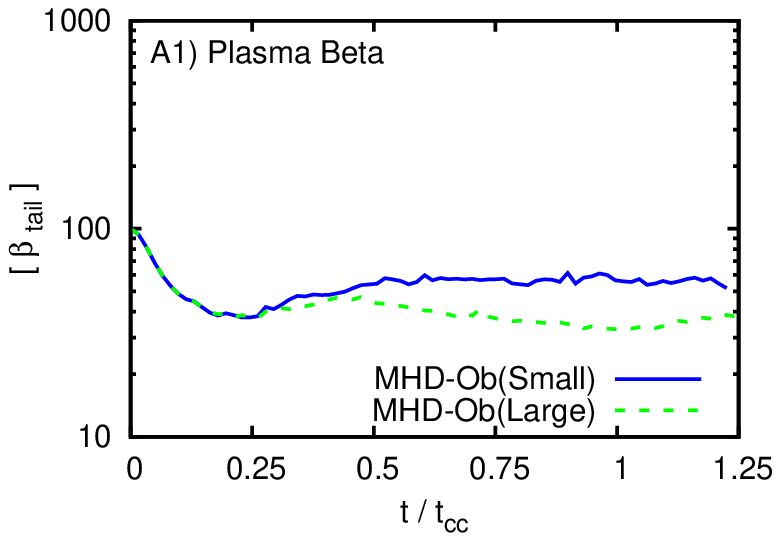}} & \resizebox{80mm}{!}{\includegraphics{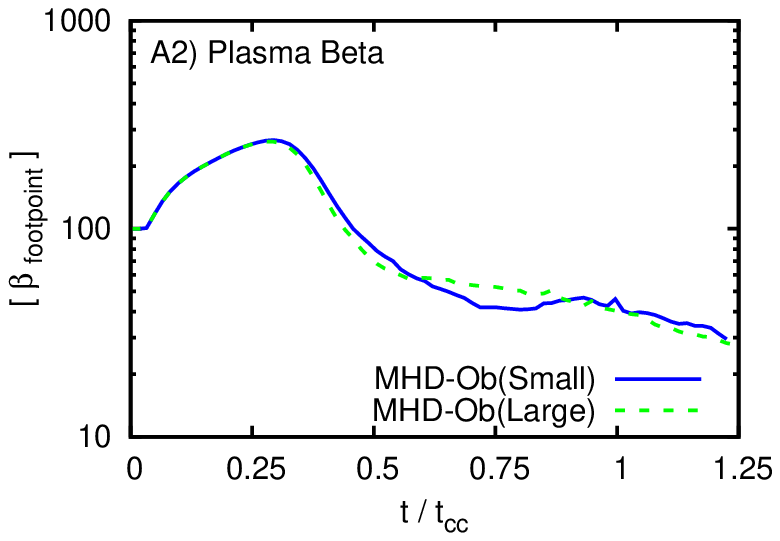}}\\
      \resizebox{80mm}{!}{\includegraphics{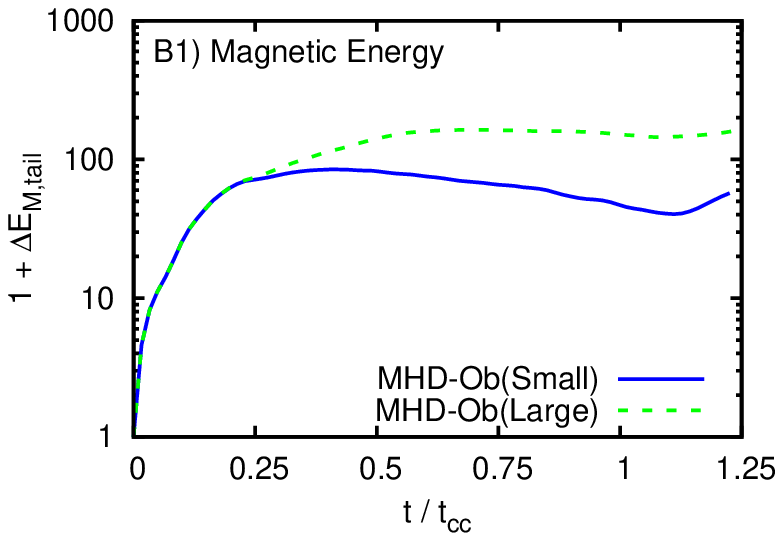}} & \resizebox{80mm}{!}{\includegraphics{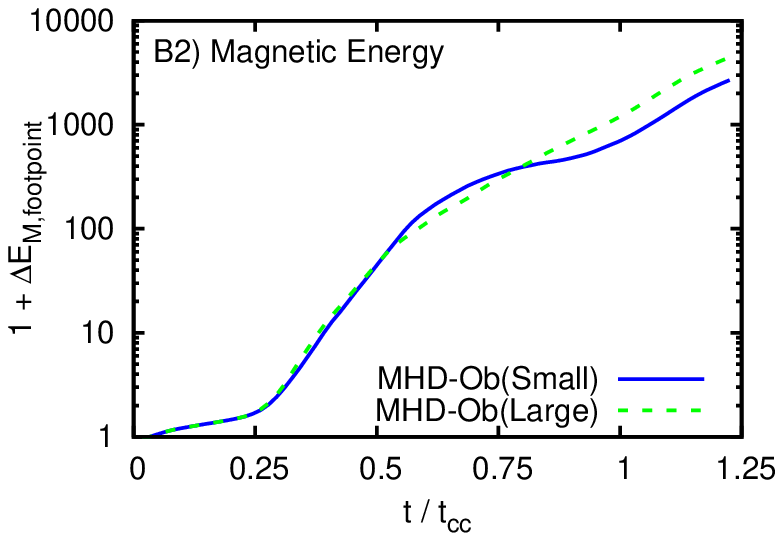}}\\
  \end{tabular}
  \caption{Comparison between the time evolution of the quantities shown in Figure \ref{Figure8} for two models with oblique magnetic fields (MHD-Ob) at the same resolution ($R_{64}$), but different domain sizes.} 
  \label{FigureA2}
\end{center}
\end{figure*}

We note that only $\sim5\,\%$ of the original mass of the cloud is lost from the simulation domains until $t/t_{\rm cc}=1.0$ (and $\sim9\,\%$ until $t/t_{\rm cc}=1.2$) in all our small-domain models. In the case of tail material, $\sim14\,\%$ of its original mass is lost until $t/t_{\rm cc}=1.0$ (and $\sim18\,\%$ until $t/t_{\rm cc}=1.2$), and in the case of footpoint material, $\sim1\,\%$ of its original mass is lost until $t/t_{\rm cc}=1.0$ (and $\sim4\,\%$ until $t/t_{\rm cc}=1.2$). In model MHD-Ob(Large) these numbers drop by a factor of $4$. All the simulations are stopped at that time to ensure that the measurements are trustworthy. In small-domain simulations, however, tail material that leaves the simulation domain is magnetised, so the diagnostics for the plasma beta and magnetic energy enhancement presented in Panels A1 and B1 of Figure \ref{Figure8} should be regarded as upper and lower limits of these quantities, respectively. Our comparison with a large-domain simulation in Figure \ref{FigureA2} also shows that the trends for these diagnostics behave as expected. Panels A2 and B2 of Figure \ref{FigureA2} indicate that the measurements for the filament footpoint are unaffected by the finite size of the simulation domain.\par

\begin{figure*}
\begin{center}
  \begin{tabular}{c c}
      \resizebox{80mm}{!}{\includegraphics{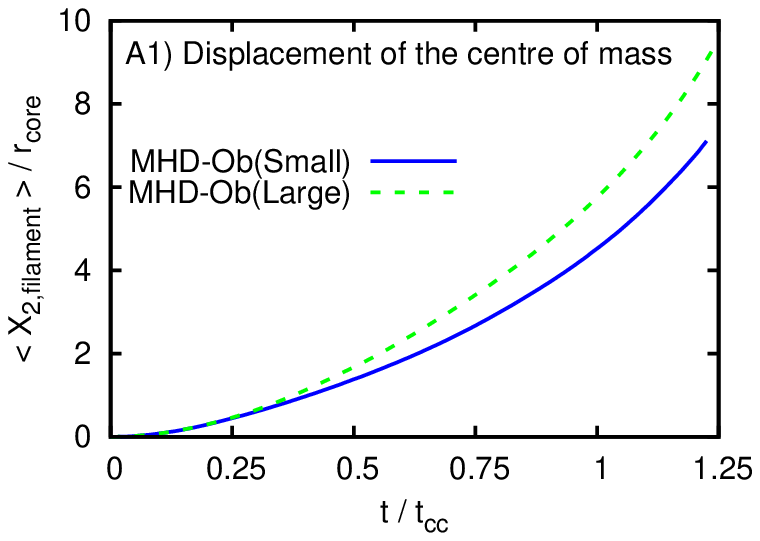}} & \resizebox{80mm}{!}{\includegraphics{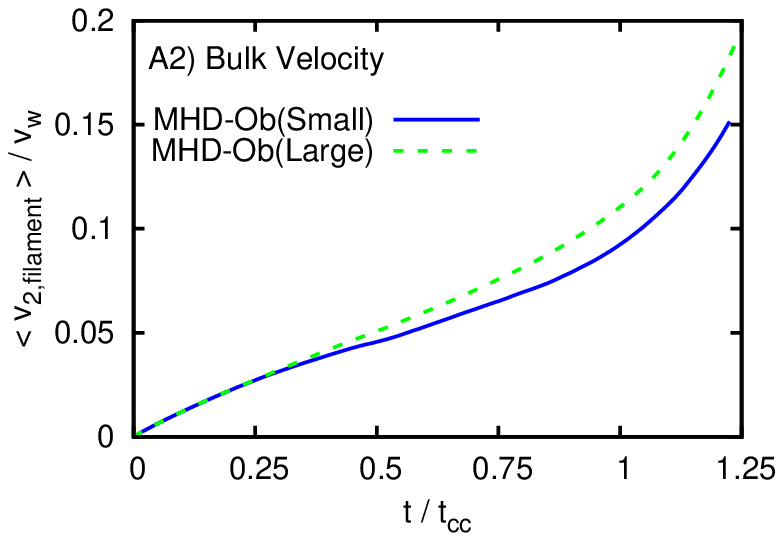}}\\
  \end{tabular}
  \caption{Comparison between the time evolution of the quantities shown in Figure \ref{FigureAA} for two models with oblique magnetic fields (MHD-Ob) at the same resolution ($R_{64}$), but different domain sizes.} 
  \label{FigureA3}
\end{center}
\end{figure*}

Figure \ref{FigureA3} presents a comparison between the parameters described in Figure \ref{FigureAA} in Section \ref{sec:FilamentFormation} for models MHD-Ob(Large) and MHD-Ob(Small). It shows that the trends observed for these parameters can be trusted throughout the entire simulation, with errors increasing as more filament material leaves the domain. The values reported in Section \ref{sec:Entrainment} for the distance travelled by the filaments' centre of mass and bulk velocity are conservative, but represent reliable lower limits for these quantities. Nonetheless, further numerical work with enlarged computational volumes is warranted if more precise measurements of these quantities are desired. For the purpose of this work, our Figures \ref{FigureA1}, \ref{FigureA2}, and \ref{FigureA3} show that the estimates of our diagnostics are reliable and can be used to asses relative differences in the properties of filaments arising from distinct native environments or as lower (or upper) limits.

\label{lastpage}

\end{document}